\documentclass[
    a4paper,
    aps,
    prb,
    twocolumn,
    superscriptaddress,
    floatfix,
    preprintnumbers,
    nobalancelastpage,
]{revtex4-2}

\usepackage{inputenc}
\usepackage{graphicx}
\usepackage{xcolor}
\usepackage{amsmath}
\usepackage{amssymb}
\usepackage{amsfonts}
\usepackage{physics}
\usepackage[colorlinks=true,allcolors=blue]{hyperref}
\usepackage{bm}

\newcommand{\EHU}{EHU Quantum Center and Department of Physical Chemistry, University of the Basque Country UPV/EHU, P.O. Box 644, 48080 Bilbao, Spain}
\newcommand{\MaxPlanck}{Max-Planck-Institut f\"ur Quantenoptik, Hans-Kopfermann-Str.~1, D-85748 Garching, Germany}
\newcommand{\MCQST}{Munich Center for Quantum Science and Technology (MCQST), Schellingstr. 4, D-80799 M\"unchen, Germany}
\newcommand{\CERN}{European Organization for Nuclear Research (CERN),1211 Geneva 23, Switzerland}
\newcommand{\DIPC}{Donostia International Physics Center, 20018 Donostia-San Sebastián, Spain}
\newcommand{\Ikerbasque}{IKERBASQUE, Basque Foundation for Science, Plaza Euskadi 5, 48009 Bilbao, Spain}

\newcommand{\Lattice}{\Lambda}
\newcommand{\DualLattice}{\Lambda^\ast}
\newcommand{\onehalf}{\frac{1}{2}}

\newcommand{\elec}{\sigma^x}
\newcommand{\gauge}{\sigma^z}

\newcommand{\dualE}{\mu^z}
\newcommand{\dualB}{\mu^x}

\newcommand{\interior}{\operatorname{int}}

\newcommand{\kinkmass}{m_{\text{k}}}

\newcommand{\CritPointEE}{0.570(4)}
\newcommand{\CritPointSuscep}{0.57(1)}
\newcommand{\CritPointTheor}{0.5731}

\newcommand{\critpoint}{g_{\text{c}}}
\newcommand{\roughpoint}{g_{\text{r}}}

\begin{document}

\title{Roughening and dynamics of an electric flux string  in a (2+1)D lattice gauge theory}

\author{Francesco Di Marcantonio}
\email[]{fra.di.marcantonio@gmail.com}
\affiliation{\EHU}

\author{Sunny Pradhan}
\email[]{sunny.pradhan@ehu.eus}
\affiliation{\EHU}

\author{Sofia Vallecorsa}
\email[]{sofia.vallecorsa@cern.ch}
\affiliation{\CERN}

\author{Mari Carmen Bañuls}
\email[]{banulsm@mpq.mpg.de}
\affiliation{\MaxPlanck}
\affiliation{\MCQST}

\author{Enrique Rico Ortega}
\email[]{enrique.rico.ortega@gmail.com}
\affiliation{\EHU}
\affiliation{\CERN}
\affiliation{\DIPC}
\affiliation{\Ikerbasque}

\preprint{CERN-TH-2025-105}

\begin{abstract}
We investigate the roughening transition in the pure $\mathbb{Z}_2$ lattice gauge theory in (2+1) dimensions.
Using numerical simulations with matrix product states, we explore the static and dynamical properties of an electric flux string between two static charges as the coupling is varied and approaches the deconfinement phase transition from the confined phase.
Within the roughening region, we obtain the universal L\"uscher correction to the confining potential and observe the expected restoration of rotational symmetry.
Our simulations of the out-of-equilibrium evolution of a string reveal that the growth of the entanglement entropy of the state and the string width exhibit qualitatively different behavior in the roughening region compared to the deeply confined one.
In particular, we find that the rate of entropy growth is consistent with an effective description of the string excitations by a bosonic model in the roughening phase.
\end{abstract}

\maketitle

\section{Introduction}
\label{section:intoduction}

Lattice gauge theories are at the center of the most fundamental theoretical physics and also appear in the effective description of some of the most interesting condensed matter phenomena.
Their relevance and complexity in solving them have motivated the interest in numerical and simulation approaches, and they have been at the heart of recent advances in the field of quantum simulation \cite{banuls2020simulating,davoudi2020,Funcke2023,bass2021qtech,bauer2023quantum,halimeh2023coldatom,di-meglio2024quantum}.
 Experimentally, some simple instances have already been realized on a variety of platforms \cite{martinez2016,schweizer2019,lu2019phot,yang2020observation,mil2020u1}. 
 
One of the most intriguing aspects of gauge theories is the confinement phenomenon, which appears as an attractive potential binding together pairs of charges and preventing their isolated observation.
Although the theoretical framework for confinement is well-established, its intricate dynamics, particularly in higher dimensions, remain a subject of intense research.
A simpler analogy can be found in $(1+1)$D spin chains, where mesonic bound states form because of the energy cost associated with the separation of domain walls \cite{mccoy1978two-dimensional, rutkevich2008energy, kormos2017confinement, lake2009confinement, tan2021domain-wall}.
However, the situation becomes more complex in higher spatial dimensions, such as $(2+1)$D, primarily due to the transverse degrees of freedom in the object responsible for confinement, the electric flux string.

The simple picture of confinement via the electric flux string is incomplete.
The flux tube is fundamentally unstable in a continuous space-time, so it becomes \emph{delocalized} for large separations.
This fact is signaled by its divergent width, as shown in \cite{luscher1981thick}.
This is no longer strictly true for a gauge theory defined on the lattice; breaking the translation symmetry into a discrete group makes it possible to support infinitely long strings for very strong couplings.
However, it has been observed that below some critical coupling $\roughpoint$ the electric string delocalizes nonetheless \cite{hasenfratz1981generalized, luscher1981symmetry-breaking, munster1981roughening, munster1981nonabelian, drouffe1981rougheningI, drouffe1981rougheningII}.
This signifies that a lattice gauge theory (LGT) goes through what is called a \emph{roughening transition} \cite{chui1976transition, fradkin1983roughening, hasenbusch1997roughening}.

The early numerical analyses of the roughening transition
\cite{caselle1996width, caselle1997string, caselle2003string, di-giacomo1990evidence, di-giacomo1990confinement, hasenbusch1997roughening, gliozzi2010width, luscher2002quark, fukugita1983distribution, agostini1997spectrum, athenodorou2023excitations, athenodorou2011closed, athenodorou2007closed},
were mainly based on the Euclidean action formulation of LGT and mainly Monte Carlo techniques, which means that only static properties could be investigated.
More recently, quantum and quantum-inspired simulations have been contemplated as potential alternatives that could overcome some of these difficulties.
This has motivated an interest in the experimental exploration of the confinement phenomenology, including string breaking, in experimental platforms~\cite{cochran2024visualizing, de2024observation, gonzalez-cuadra2024observation, surace2020lattice,mildenberger2025confinement, meth2025simulating}.
From the side of classical simulation, tensor networks (TN)
\cite{verstraete2008matrix, schollwock2011dmrg, orus2014practical,Bridgeman_2017, silvi2019tensor,Ran2020tncontr,Cirac2021rmp,Okunishi2022,Banuls2023}
have indeed been employed for LGT problems, and found especial success for equilibrium (1+1)D setups (see~ 
\cite{banuls2020review,banuls2020simulating} for reviews), but the number of their applications in this field is ever expanding~\cite{meurice2022trg,Kadoh:2022Ia,felser2020u1,robaina2021simulating,magnifico2021qed3d,emonts2023z2,kelman2024gpeps,Akiyama_2024}, including real time evolution scenarios~\cite{pichler2016real-time, chanda2020quenches,notarnicola2020ryd,rigobello2021entanglement,belyansky2023high,barata2025real,papaefstathiou2025,banuls2025pdf}.

\begin{figure}[t]
    \centering
    \includegraphics{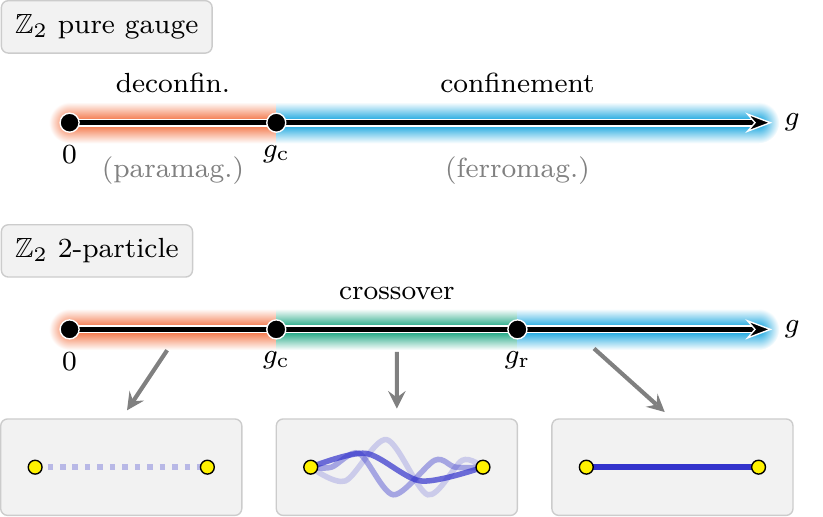}
    \caption{Phase diagram of the $\mathbb{Z}_2$ LGT in the vacuum and the 2-particle sector.
    In the vacuum sector, there are two different phases (the deconfined phase dual to the paramagnetic one and the confined phase dual to the ferromagnetic one), while in the 2-particle sector, an intermediate crossover region appears.}
    \label{fig:phase_diagram}
\end{figure}

Despite its phenomenological nature in high-energy physics, and in particular in quantum chromo-dynamics (QCD), confinement emerges independently in the strong coupling limit of pure LGTs. This striking observation grants us access to a plethora of simpler LGTs. One of the most basic yet insightful example is the pure gauge $\mathbb{Z}_2$ model. We already mentioned how increasing the spatial dimensions of our model gives rise to the roughening phenomenon, therefore we also expect it to occur in the $\mathbb{Z}_2$ case. 
In this work, we use TN methods to investigate both static and dynamical properties of the roughening transition in a 
pure gauge $\mathbb{Z}_2$ model in (2+1)D. 
The model has a confinement-deconfinement transition.
The presence of the roughening transition means that, when static charges are introduced, the confined phase is divided into two regions (see Fig.~\ref{fig:phase_diagram}).
The first occurs in strong coupling, where the electric string is smooth and rigid. On this side lies a strongly confined regime characterized by stiff string excitations, where the transverse fluctuations of the electric flux string are suppressed, resulting in a flux tube with a finite thickness.
The second, on the other hand, is a crossover region close to the deconfining transition.
This other side is a weakly confined regime with floppy string excitations, where these transverse fluctuations become significant.
The string becomes highly fluctuating and delocalized.
In this floppy regime, the width of the flux tube connecting two charges diverges logarithmically with the distance separating them.
This crossover region, where the roughening occurs, is significant because it indicates that the continuous space-time symmetries are effectively restored while remaining in a confining phase \cite{luscher1981symmetry-breaking, kogut1981string}.

The transition between these two behaviors is driven by the strength of the confining potential, which is intrinsically linked to the electric field in the system.
This qualitative change in the nature of the confining string has profound implications for various physical observables, including the entanglement entropy of the system and the spatial extent of the flux tube itself.
Defining a simple, local order parameter that sharply distinguishes between the stiff and floppy string regimes has proven to be a non-trivial task.
This difficulty arises from the subtle nature of the transition, which is not associated with a conventional symmetry-breaking pattern characterized by a local order parameter acquiring a non-zero expectation value.
Recent research suggests that more sophisticated nonlocal measures, such as entanglement entropy and order parameters related to translational symmetry breaking, can serve as effective indicators of the roughening transition \cite{xu2025tensor}.
The absence of a straightforward local order parameter underscores the complexity of the roughening phenomenon and motivates the exploration of novel diagnostic tools rooted in quantum information theory and symmetry principles.

The $\mathbb{Z}_2$ LGT is well-suited for TN simulations~\cite{sugihara2005matrix, tagliacozzo2011entanglement,wu2025gauge_invariant_tn, emonts2023z2, liu2013blocking_formulas,borla2025breaking}. 
We use a \emph{Hamiltonian formulation} of LGTs, based on the Kogut-Susskind approach \cite{kogut1975hamiltonian, horn1979hamiltonian}, and approximate the states of the system using the \emph{matrix product state} (MPS) ansatz \cite{PerezGarcia2007,verstraete2008matrix, schollwock2011dmrg}, which also allows the simulation of \emph{real-time evolution} \cite{Paeckel2019tevol}.
We study the ground state of the pure gauge theory across the phase diagram, with and without external charges. This gives us access to different observables, including entanglement entropy, string width, and confining potential, which we use to determine the roughening region.
We also perform dynamical simulations, in which we quench the system by introducing a pair of external charges on the interacting vacuum. 
We then follow the evolution of entropy and string width as a function of time in the various regimes. In particular, for the entanglement entropy, we find qualitatively different dynamics within the roughening region, which can be connected to the effective model describing the string excitations in such a case.
Our approach is complementary to other recent TN studies of the roughening problem in this model~\cite{xu2025tensor, krinitsin2024roughening}.

The manuscript is structured as follows.
In Sect.~\ref {sec:model_and_setup} we describe the $\mathbb{Z}_2$ LGT in 2+1 dimensions, its duality with the quantum Ising model (Sect.~\ref{sub:dual_ising_model}), and the matrix product state setup for the numerics (Sect.~\ref{sub:numerical_setup}).
After that, in Sect.~\ref{sec:the_roughening_transition} we first review the roughening transition, and we proceed with the analysis of four aspects: string width (Sect.~\ref{sub:string_width}), entanglement entropy (Sect.~\ref{sub:entanglement_entropy}), confining potential (Sect.~\ref{sub:confining_potential}), and restoration of rotational symmetry (Sect.~\ref{sub:restoration_of_rotational_symmetry}).
Subsequently, in Sect.~\ref{sec:dynamical_studies} we study the dynamical properties of the flux string in the roughening region by observing the growth patterns of the entanglement entropy (Sect.~\ref{sub:entanglement_entropy_dynamics}) and the string width (Sect.~\ref{sub:string_width_dynamics}).
Final remarks and outlook are given in Sect.~\ref{sec:conclusions_and_outlook}.

\section{Model and setup}
\label{sec:model_and_setup}

We consider a $\mathbb{Z}_2$ lattice gauge theory (LGT) on a square lattice $\Lattice$ with $L$ plaquettes in the horizontal direction $\hat{1}$ and $N$ plaquettes in the vertical direction $\hat{2}$.
We impose periodic boundary conditions in the vertical direction.
We denote the sites with $v = (x,y)$ ($x = 1, \dots, L+1$ and $y = 1,\dots,N$) and the links with the pair $(v, \hat{\mu})$ where $\hat{\mu}$ is the direction ($\mu = 1,2$).
The expression $(v, -\hat{\mu})$ has to be understood as the link $(v - \hat{\mu}, \hat{\mu})$.

The $\mathbb{Z}_2$ LGT has spin-$\onehalf$ degrees of freedom on the links.
We identify $\gauge$ as the gauge field and $\elec$ as the electric field.
With this in mind, we can write the Hamiltonian in the spirit of Kogut and Susskind \cite{kogut1975hamiltonian}:
\begin{equation}
    H_{\text{LGT}}(g) = -g \sum_{(v, \hat{\mu})} \elec_{(v, \hat{\mu})} - \frac{1}{g} \sum_{\square} (\gauge \gauge \gauge \gauge)_{\square}.
    \label{eq:LGT_hamiltonian}
\end{equation}
The first term can be interpreted as the electric field energy, and the second as the magnetic field energy.

In a LGT, together with the Hamiltonian, it is also important to define \emph{gauge transformations}, which ``rotate'' the gauge degrees of freedom attached to a site.
For this reason, given a site $v \in \Lattice$, the gauge transformation is defined as
\begin{equation}
    G_v = \elec_{v, \hat{1}}  \elec_{v, \hat{2}}  \elec_{v, -\hat{1}}  \elec_{v, -\hat{2}}.
    \label{eq:gauge_operator}
\end{equation}
Consider now a general situation where we have a distribution of static charges around the lattice $\Lattice$, which can be denoted with $\{q_v = \pm 1\}_{v \in \Lattice}$, with $q_v = -1$ meaning that a charge is present in $v$.
The \emph{physical states} are those that satisfy
\begin{equation}
    G_v \ket{\phi_\text{phys}} = q_v \ket{\phi_\text{phys}}
    \qquad \forall v \in \Lattice,
    \label{eq:gauss_law}
\end{equation}
which can be understood as a discretized version of Gauss' law.

\begin{figure}[t]
    \centering
    \includegraphics{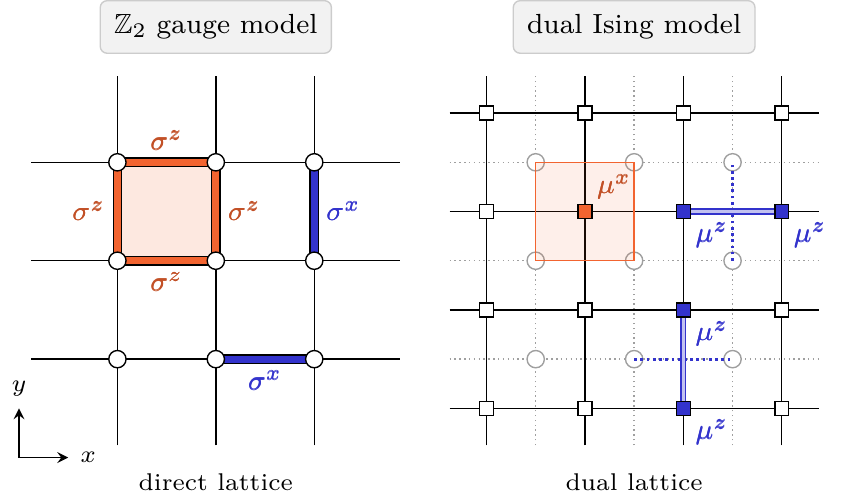}

    \caption{Picture of plaquette and link operators in the $\mathbb{Z}_2$ gauge model (\emph{left}) and the corresponding dual operators in the Ising model (\emph{right}).}
    \label{fig:operators}
\end{figure}

\subsection{Dual Ising model}
\label{sub:dual_ising_model}

It is well known that a \emph{pure gauge} $\mathbb{Z}_2$ model in (2+1)D can be mapped to a (2+1)D quantum Ising model (or equivalently a 3D classical Ising model) \cite{fradkin1978order, kogut1979introduction, wegner1971duality, cobanera2011bond}.
The dual model is defined on the dual lattice $\DualLattice$, whose sites lie at the center of the plaquettes of $\Lattice$.
We label each plaquette $\square$ with the site $v$ in the bottom left corner, which also acts as a label for the corresponding dual site.
Then, we introduce the new operators
\begin{equation}
    \dualB_v = (\gauge \gauge \gauge \gauge)_{\square_v}
    \label{eq:def_dualB}
\end{equation}
and
\begin{equation}
    \dualE_{(x,y)} = \prod_{j=1}^{x} \elec_{(j, y), \hat{2}}
    \label{eq:def_dualE}
\end{equation}
that act on the dual sites of $\DualLattice$.
These new operators satisfy the same algebra of $\gauge$ and $\elec$ and can be used to map the gauge Hamiltonian \eqref{eq:LGT_hamiltonian} into a transverse field Ising model.
To express $H_{\text{LGT}}$ in terms of the new variables \eqref{eq:def_dualB} and \eqref{eq:def_dualE}, one has to take into account the Gauss law \eqref{eq:gauss_law}.
Consequently, this means that the duality is sensitive to static charges.
Furthermore, the non-local nature of \eqref{eq:def_dualE} also makes it dependent on the boundary conditions.

We find that the dual Hamiltonian can be written
\begin{multline}
    H_{\text{dual}}(g, \{q_v\}) =
    - \frac{1}{g} \sum_{\Lattice} \dualB_v
    - g \sum_{\nu = 1,2} \sum_{\interior \Lattice} \omega(v, \hat{\nu}) \dualE_v \dualE_{v - \hat{\nu}} \\
    - g \sum_{y} \qty( \dualE_{(1,y)} + \dualE_{(L,y)} \dualE_A ),
    \label{eq:dual_ising_model}
\end{multline}
where $\Lattice$ is the lattice, $\interior \Lattice$ its interior ($x \neq 1, L+1$), and $\omega(v, \hat{\nu})$ a phase factor required for the correct implementation of the Gauss law, defined as
\begin{equation}
    \omega(v, \hat{\nu}) =
    \begin{cases}
        1 & \hat{\nu} = 1, \\
        \prod_{j = 1}^{x} q_{(j, y)} & \hat{\nu} = 2.
    \end{cases}
    \label{eq:duality_phase_factor}
\end{equation}
The last term in brackets in \eqref{eq:dual_ising_model} are boundary terms, and $\dualE_{A}$ is an ancillary spin that takes into account the different topological sectors in the case of PBC.
In Appendix \ref{app:duality_transformation_with_static_charges}, we show the full steps necessary to build \eqref{eq:dual_ising_model}.
We see that \eqref{eq:dual_ising_model} is a 2D quantum Ising model with \emph{defects} introduced by the static charges.

Given that we will consider at most two charges, we will use the notation $H(g; v_1, v_2)$ to denote $H_{\text{dual}}(g, \{q_v\})$ when two charges are put in the positions $v_1$ and $v_2$,
\begin{equation}
    H(g; v_1, v_2) = H_{\text{dual}}(g, \{q_{v_1} = -1, q_{v_2} = -1\}),
    \label{eq:hamiltonian_2p}
\end{equation}
whereas in the case of no charges (empty charge sector), we will use instead the notation
\begin{equation}
    H(g; \varnothing) = H_{\text{dual}}(g, \{q_v = +1, \; \forall  v\}).
    \label{eq:hamiltonian_vac}
\end{equation}
For later convenience, we will introduce the symbol $\Gamma$ to denote the path connecting two static charges.

\subsection{Numerical setup}
\label{sub:numerical_setup}

We simulate $\mathbb{Z}_2$ LGT using matrix product states (MPS).
Importantly, instead of working directly with the gauge Hamiltonian \eqref{eq:LGT_hamiltonian}, we use the dual Ising Hamiltonian \eqref{eq:dual_ising_model} in both the vacuum sector and the two-particle sector.
In particular, the Hamiltonian \eqref{eq:hamiltonian_vac} for the former and \eqref{eq:hamiltonian_2p} for the latter.

A generic state $\ket{\Psi}$ of $n$ sites, each with (physical) dimension $d$ can be written in a basis $\{\ket{i_1 \cdots i_n}\}$, where each $i_k=1,\ldots d$, as
\begin{equation*}
    \ket{\psi} = \sum_{i_1 \dots i_n} \psi_{i_1 \dots i_n} \ket{i_1 \dots i_n},
\end{equation*}
which requires $d^n$ coefficients.
The main idea behind MPS is to approximate $\psi_{i_1 \dots i_n}$ with a contraction of lower-rank tensors:
\begin{equation}
    \psi_{i_1 \dots i_n}^{\text{(MPS)}} =
    \sum_{\sigma_1 \dots \sigma_n}
    A[1]^{i_1}_{\sigma_1}
    A[2]^{i_2}_{\sigma_1 \sigma_2}
    \cdots
    A[n-1]^{i_{n-1}}_{\sigma_{n-2} \sigma_{n-1}}
    A[n]^{i_n}_{\sigma_{n-1}},
    \label{eq:MPS_def}
\end{equation}
where each $A[k]$ (except for the first and last ones) is a 3-rank tensor. Each index $\sigma_k$, called virtual, connects two neighboring tensors and takes values in the range $\{1,\ldots,\chi\}$,
where $\chi$ is called \emph{bond dimension}.
This means that we only need $O(n d \chi^2)$ coefficients for \eqref{eq:MPS_def}, in other words, the memory complexity is now polynomial with the size of the system.

\begin{figure}[t]
    \centering
    \includegraphics{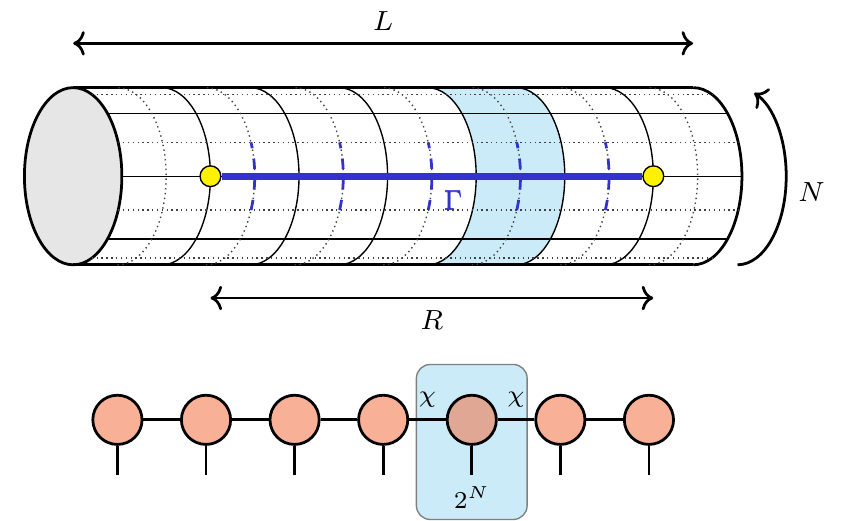}
    \caption{Cartoon picture of the setup.
    From the vacuum of a pure $\mathbb{Z}_2$ LGT, we create a string of length $R$ and path $\Gamma$ in a 2D system of dimensions $L \times N$, with periodic boundary conditions along the transversal direction.
    The edges of the direct lattice are shown with a solid line, while those of the dual lattice are shown with a dotted line.
    The dual links affected by the phase factor \eqref{eq:duality_phase_factor} are highlighted by blue dashed lines.
    Because we expect a logarithmic growth of the string in the transversal direction, this setup is still efficient for the simulation of the string dynamics with MPS methods.
    Each tensor of the MPS has bond dimension $\chi$ and physical dimension $2^N$ and represents a whole vertical strip of the 2D model.
    }
    \label{fig:cylinder_mps}
\end{figure}

Since in the roughening region the width of the string only grows logarithmically with its length, we do not need large transversal dimensions, and we restrict the study to narrow cylindrical systems of size $L\times N$, with $N \ll L$ (Fig.~\ref{fig:cylinder_mps}).
We represent the states of the system as MPS, where each tensor corresponds to a vertical strip of the lattice and thus has a physical dimension $d^N$.
The whole setup is depicted in Fig.~\ref{fig:cylinder_mps}.
This representation has exponential memory cost in transverse size of the lattice, which is thus limited to small values of $N$, while the cost is only polynomial in the longitudinal size.

We simulate the dual Hamiltonian \eqref{eq:dual_ising_model} with MPS for a variety of parameters.
That is, different lattice sizes $L \times N$, couplings $g$, and configurations of charges with bond dimensions up to $\chi = 256$.
We chose to work with the dual Hamiltonian because it is much more efficient, as it involves fewer degrees of freedom and at most two-body interaction terms.
For the static properties, we compute the ground state via variational search over the MPS family, while for the dynamics, the time-evolving block decimation (TEBD) algorithm is employed for the time evolution~\cite{verstraete2008matrix,schollwock2011dmrg}.
An error analysis of the numerical simulations is presented in Appendix \ref{app:numerical_methods_analysis}.

\section{The roughening transition}
\label{sec:the_roughening_transition}

The typical picture of confinement involves the formation of a flux tube, or string, that binds two charges.
In a (2+1)D gauge theory, the flux string is effectively a (1+1)D quantum system, as it can only oscillate along one transversal direction.
These fluctuations, depending on the coupling $g$, can be of two types: massive or massless \cite{hasenfratz1981generalized, kogut1981string}.
In the massive case, there is a finite non-zero energy cost for each deformation from the equilibrium position.
Hence, the string can be considered as \emph{rigid} and \emph{smooth}.
This happens deep inside the strong coupling region $g \gg \critpoint$, where the magnetic fluctuations induced by the plaquette terms are \textit{de facto} suppressed, forcing the string into its equilibrium configuration.
On the other hand, decreasing the coupling makes the magnetic fluctuations more prominent, which lowers the cost of these deformations.
When this penalty disappears, we reach a massless (or critical) regime, which leads to a \emph{highly fluctuating string}.
Or, in other words, a \emph{rough string}.
Not only that, it has been shown that the string delocalizes and its \emph{width diverges} logarithmically with $R$, while keeping a non-zero string tension, meaning that it is not a deconfining phase \cite{luscher1981thick}.
Therefore, in the massless case, an infinitely long string is fundamentally unstable \cite{luscher1981symmetry-breaking}.

From the above discussion, the flux string manifests different phases within the confinement phase.
In particular, it undergoes a \emph{roughening transition} at some coupling $\roughpoint$, distinct from the deconfining transition $\critpoint$, that leads to a new \emph{crossover region} $\critpoint < g \leq \roughpoint$ appearing inside the confinement phase (see Fig.~\ref{fig:phase_diagram}).
We will also refer to this region sometimes as the weak coupling region (not to be confused with the deconfining phase) or roughening region.
It can be shown that the roughening transition is a Berezinskii-Kosterlitz-Thouless (BKT) transition \cite{chui1976transition}.
It is not a second-order phase transition but an infinite-order one, and it cannot be characterized by the symmetry-breaking of a local order parameter \cite{berezinskii1971long-range, kosterlitz1973ordering, kosterlitz1974critical, frohlich1981kosterlitz-thouless}.

In the rest of this section, we will explore and analyze four key phenomena associated with roughening:
(i) the divergence of the string width; (ii) the entanglement entropy scaling with $R$; (iii) sub-leading corrections to the confining potential; (iv) the restoration of rotational symmetry.
Before continuing, we want to remark on two important facts.
First, in the crossover region, the ``background'' gauge theory is still in a gapped phase (confinement), only the flux string as a new effective (1+1)D quantum system has to be considered gapless.
Second, with dynamical matter, strings of an arbitrary large length $R$ are not allowed.
After a certain separation, the string-breaking mechanism would enter into action, producing bound states of the charges (mesons).
In our case, we are substantially considering a pure gauge theory with at most two external charges that have no dynamics, frozen into their positions, which are coupled to the gauge fields only via the Gauss law.

The roughening region is limited to the confined phase, and ends at the deconfining transition $\critpoint = \CritPointTheor$ \cite{blote2002cluster}.
We can locate this transition from our simulations, analyzing the ground state of the vacuum sector.
We find that both the dual magnetization and the entanglement entropy in this case (see Appendix \ref{app:finite_size_scaling_for_the_deconfining_transition}) yield values for the deconfining point consistent with the exact one.

In this section, to investigate the roughening phenomenology, we compute the variational ground state of the model on a $L \times N$ lattice with periodic boundary conditions along the transversal direction ($N$ is the transversal size).
We consider both the empty charge sector and the one with two charges, which are always positioned in the bulk such that the string is centered along the $x$-axis (see Fig.~\ref{fig:cylinder_mps}).
The properties of the string are systematically studied as a function of the coupling $g$, the transversal size $N$, and its length $R$.

\begin{figure}[t]
    \centering
    \includegraphics[width=\columnwidth]{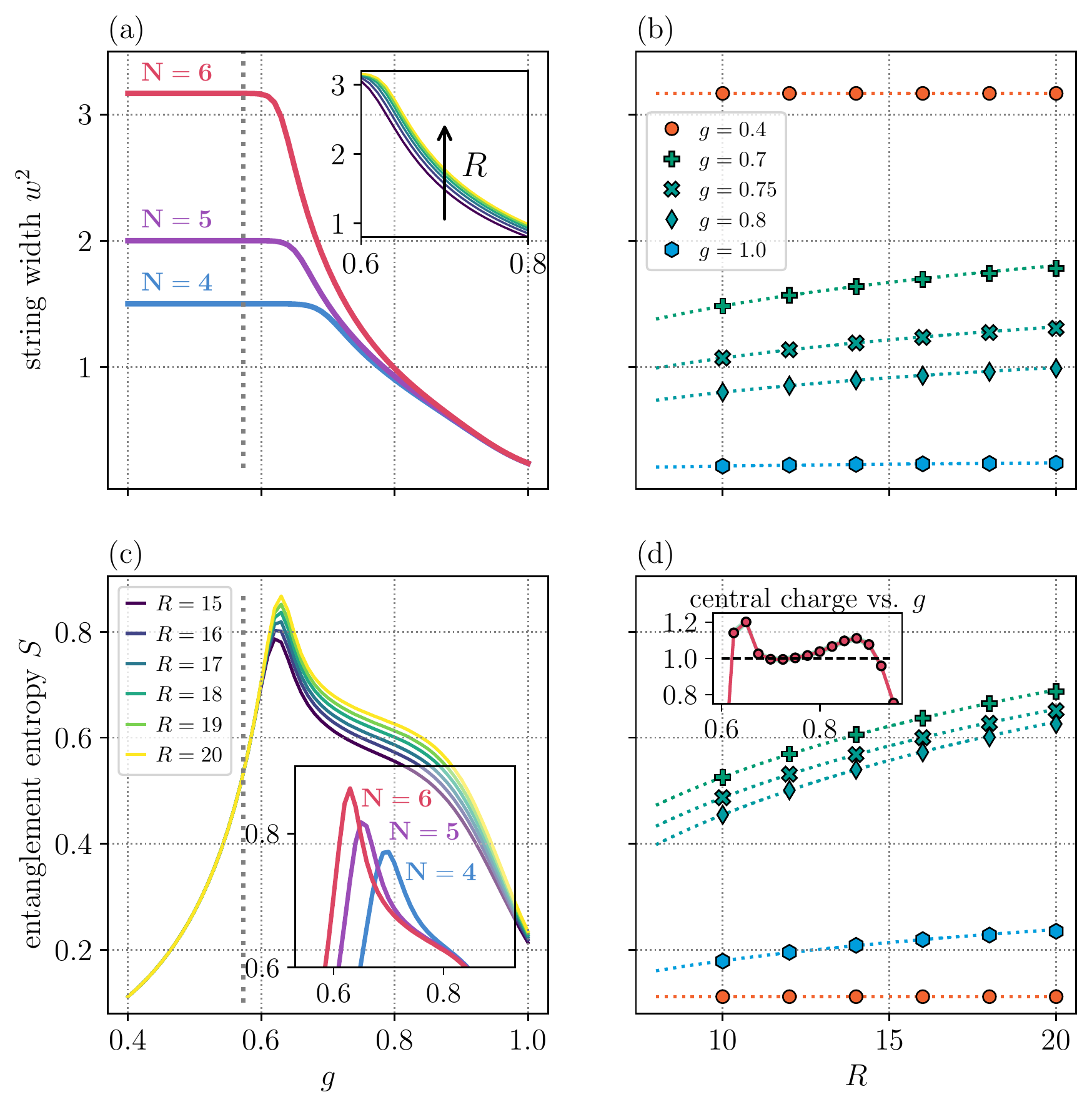}
    \caption{
        Numerical results on the string width $w^2$ and entanglement entropy $S$ for a system size in the longitudinal direction of the string, $L = 30$, and in the transversal direction of the string $N=6$, if not specified otherwise, with coupling step $\Delta g = 0.01$ together with a maximum bond dimension $\chi_{\text{max}} = 128$.
        \textbf{(a)}~String width $w^2$ \eqref{eq:string_width_def} vs. the coupling constant $g$ with fixed string length $R = 20$ and different transversal sizes $N$;
        The gray dotted line indicates the deconfining transition point $\critpoint$. The inset shows the dependence on $R \in [10,20]$ within the roughening region for $N=6$.
        \textbf{(b)}~String width $w^2$ \emph{vs} $R$ for different couplings;
        The dotted lines indicate the logarithmic fit in $R$ for each $g$.
        \textbf{(c)}~Entanglement entropy $S$ vs.~$g$ for different lengths $R$; The inset shows the dependence of the maximal curve on $N$.
        \textbf{(d)}~Entanglement entropy $S$ vs.~$R$ for different couplings $g$ across the phase diagram.
        \emph{Inset}: extrapolation of the central charge, with $\chi = 256$ and $\Delta g = 0.025$.
    }
    \label{fig:statics_sw_ee}
\end{figure}

\begin{figure*}[t]
    \centering
s    \includegraphics[width=0.9\textwidth]{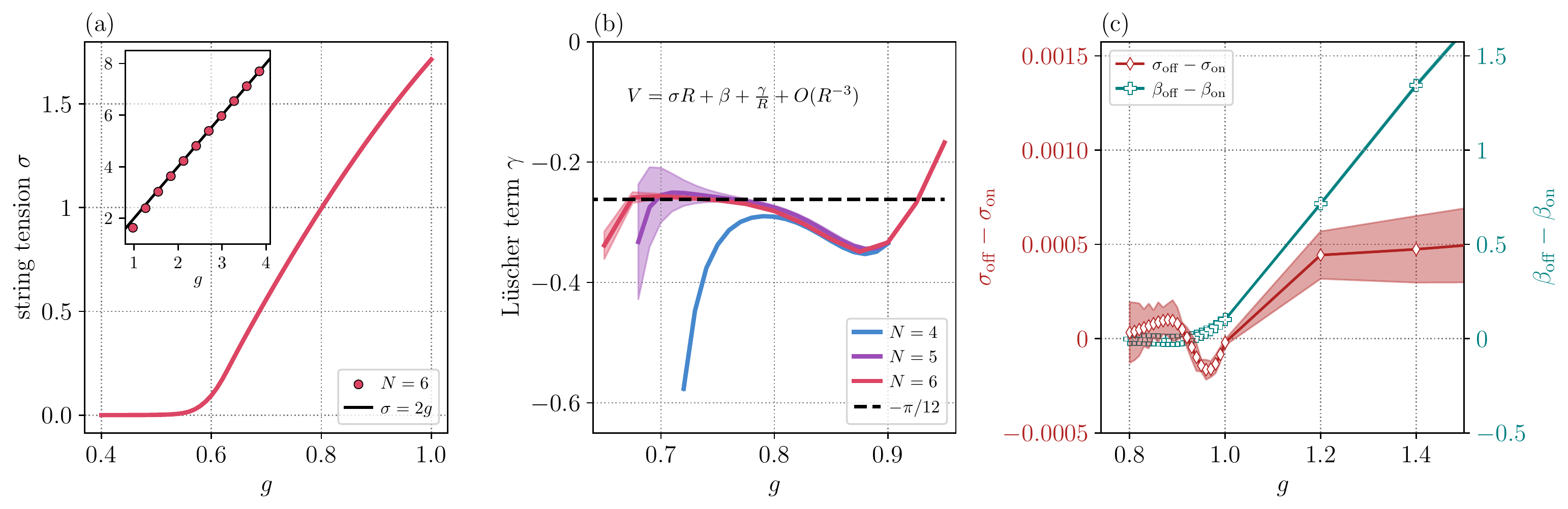}
    \caption{Numerical results on the confining potential \eqref{eq:potential_luscher} across the roughening transition for a system size in the longitudinal direction of the string, $L = 30$, and in the transversal direction of the string $N=6$, if not explicitly stated.
        \textbf{(a)}: String tension $\sigma$ vs. coupling constant $g$ (coupling step $\Delta g = 0.01$ and $\chi_{\text{max}} = 128$);
        For large coupling $\sigma \sim 2g$ as shown by the inset.
        \textbf{(b)}: Universal L\"uscher term $\gamma$ for $N=4,5$ and $6$ in the crossover region ($\Delta g = 0.025$ and $\chi_{\text{max}} = 256$);
        We have reported the values of $\gamma$ only when the string width is not saturated at the given $N$ (see Fig.~\ref{fig:statics_sw_ee}a).
        \textbf{(c)}: Difference of the on and off-axis potential terms $\sigma$ and $\beta$ in \eqref{eq:potential_luscher}; In this case a $50 \times 5$ lattice to allow for larger strings.
    }
    \label{fig:potential}
\end{figure*}

\subsection{String width}
\label{sub:string_width}

One of the main characteristics of the roughening transition, and consequently of the entire crossover region, is the logarithmic divergence of the string width $w^2$ as a function of charge separation $R$, i.e., $ w^2 \sim \log R$.
To observe this phenomenon, we follow \cite{luscher1981thick} and define first the electric field energy above the vacuum,
which measures the variation in the electric field caused by the string,
\begin{equation}
    \mathcal{E}(x, y) = -\frac{1}{2}\qty(
    \mel{\Omega_{v_1, v_2}}{\elec_{(x,y), \hat{1}}}{\Omega_{v_1, v_2}} -
    \mel{\Omega_{\varnothing}}{\elec_{(x,y), \hat{1}}}{\Omega_{\varnothing}}
    ),
\end{equation}
where $\ket{\Omega_{v_1, v_2}}$ and $\ket{\Omega_{\varnothing}}$ are the ground states of $H(g; v_1, v_2)$ and $H(g; \varnothing)$, respectively.
We consider only the horizontal links $(\mu = 1)$ for simplicity.
Then, we define the string width $w^2$ as the variance of $\mathcal{E}$ in transversal direction, taken in the middle of the string:
\begin{equation}
    w^2 = \frac{\sum_y (y-y_0)^2 \mathcal{E}(x_0, y)}{\sum_y \mathcal{E}(x_0, y) },
    \label{eq:string_width_def}
\end{equation}
where $(x_0,y_0)$ are the coordinates of the central point between both charges, and the sum runs over all the transversal range.

In Figs.~\ref{fig:statics_sw_ee}a--b, we have computed the string width $w^2$ in a lattice of longitudinal size $L = 30$.
In the left panel, we kept $R=20$ and varied both the transversal size ($N = 4,5,6$) and the coupling ($g \in [0.4, 1.0]$).
As it can be seen, $w^2$ grows in the crossover region as $g \to \critpoint$ (the critical point indicated by the gray dotted line) but saturates at different values of $g$ depending on $N$.
It is a clear sign that $w^2$ is limited by the transversal size $N$.
In the right panel, the transversal size is kept fixed ($N = 6$) and the width is plotted as a function of $R$ for different values of $g$ across the phase diagram.
$w^2$ grows slowly with $R$ inside the crossover region, while being constant outside of it.
Indeed, both in the strong confinement and in the deconfined region, the results are independent of $g$ but with different features.
In the strongly confined region, we expect $w^2 \rightarrow 0$ since the string is rigid, while in the deconfined region it will saturate because of the limited transversal size
\footnote{The saturation values will be $w^2 \rightarrow (2k^2 + 1)/6$ if $N = 2k$ and $w^2 \rightarrow (k(k + 1))/3$ if $N = 2k+1$.}.
The growth with $R$ in the roughening region is compatible with a logarithmic function, as shown by the fit in dotted line, which confirms the general prediction of \cite{luscher1981thick}.

\subsection{Entanglement entropy}
\label{sub:entanglement_entropy}

Given a bipartition $A \cup B$ of the whole system and a state $\rho$, the entanglement entropy (EE) is defined as
\begin{equation}
    S =
    - \Tr (\rho_A \log \rho_A) =
    - \sum_{i} p_i \log p_i
    \label{eq:entanglement_entropy}
\end{equation}
where $\rho_A$ is the reduced density matrix $\Tr_B(\rho)$ and $p_i$ are the Schmidt values.
Through the EE, it is possible to obtain a more quantum-informational perspective on the crossover region.
We consider a transverse cut of the cylinder in two equal parts of length $L/2$ and compute the EE for this bipartition, which corresponds to the middle bond in our MPS representation.

The flux string is effectively a one-dimensional system, and it is known that in (1+1)D the EE scales differently with the subsystem size depending on the phase \cite{amico2008entanglement}.
In a critical phase, it grows logarithmically due to the diverging correlation length, while it is capped in a gapped phase due to the short-range correlations.
Therefore, when the string is highly fluctuating, meaning that it is critical, we expect the EE to show a dependence on $R$.
We verify that this is indeed the case, as illustrated
in Fig.~\ref{fig:statics_sw_ee}c--d, where we show the EE
in a $30 \times 6$ lattice for $g \in [0.4, 1.0]$ and $R = 15, \dots, 20$.
We see that in the crossover region $0.6 \lesssim g \lesssim 1.0$, the EE develops a dependence on $R$ (Fig.~\ref{fig:statics_sw_ee}c) that is compatible with a logarithmic growth (Fig.~\ref{fig:statics_sw_ee}d).
This behavior then dies out when moving away from the crossover region.

Furthermore, it has been shown in \cite{holzhey1994entropy, calabrese2004entanglement, korepin2004universality, vidal2003entanglement} that the scaling coefficient of the EE in a 1D critical state depends on $c$, the central charge, as
\begin{equation}
    S \sim \frac{c}{6} \log(R) + \text{const.}
    \label{eq:chain_ee}
\end{equation}
From the logarithmic fits of the EE for different $g$, as the ones shown in Fig.~\ref{fig:statics_sw_ee}d, we find that the extracted central charge plateaus around $c \simeq 1$ in the crossover region, as shown in the inset of the panel, consistent with an effective description of the fluctuating string at low energies by a massless boson theory, as discussed in the next subsection.

On a final note, we also observe a dependence on the transverse size of the system $N$, as shown in the inset of Fig.~\ref{fig:statics_sw_ee}c, especially when close to the deconfining point $\critpoint$.
We regard this as an effect of the finite-size dependence of the EE near a second-order phase transition.

\subsection{Confining potential}
\label{sub:confining_potential}

The energy stored in a flux string connecting two static charges produces a \emph{confining potential} that corresponds to the energy cost to create such a string state from the vacuum of the theory.
In the Hamiltonian picture, the confining potential can be measured as the difference between the energies of the ground states with and without charges.
\begin{multline}
    V(g, \vec{r})
    = \mel{\Omega_{v_1, v_2}}{H(g; v_1, v_2)}{\Omega_{v_1, v_2}} -\\
    - \mel{\Omega_{\varnothing}}{H(g; \varnothing)}{\Omega_{\varnothing}},
    \label{eq:confining_potential_def}
\end{multline}
where $H(g; v_1, v_2)$ and $H(g, \varnothing)$ are defined respectively in \eqref{eq:hamiltonian_2p} and \eqref{eq:hamiltonian_vac}, while $\ket{\Omega_{v_1, v_2}}$ and $\ket{\Omega_{\varnothing}}$ are their respective ground states.
The charges are positions in $v_1$ and $v_2$ with the vector $\vec{r}$ connecting them.
As can be expected, the energy of a string in the confined region grows linearly with its length, meaning that at the zeroth order $V(g, \vec{r}) \simeq \sigma(g) r$, where $\sigma(g)$ is the \emph{string tension} and $r = \abs{\vec{r}}$.
The presence of subleading terms in the potential is a key difference between the weakly confined and strongly confined phases and will be discussed shortly.
In the deconfining phase ($g < \critpoint$), the notion of a binding string is not well defined, or even nonsensical, since the charges are not bound to each other.

Predictions about the confining potential can be obtained from a proper modeling of the quantized string.
In particular, one can predict that when the string is massless, the potential acquires extra \emph{universal subleading terms} for each transverse direction to the string \cite{luscher1980anomalies, luscher1981symmetry-breaking, kuti2006lattice, luscher2002quark, caselle1997string, caselle2003string, caselle2013quantisation}.
These terms are universal in the sense that they are independent of the gauge group, hence they are expected also for a $\mathbb{Z}_2$ LGT \cite{munster1981roughening, drouffe1981rougheningI, hasenfratz1981generalized}.

The current understanding \cite{luscher2002quark, caselle1997string, caselle2003string, caselle2013quantisation} of these correction terms is based on the following argument.
In a $D$-dimensional space-time, the string can fluctuate only in $D-2$ transversal directions.
At first approximation, these fluctuations can be modeled as $D-2$ uncoupled Gaussian fields in 1+1D, one for each transversal direction.
From models like the Coulomb gas or the planar XY model, it is known that a Gaussian model in 1+1D has a BKT phase transition that separates a massive phase from a critical one \cite{berezinskii1971long-range, kosterlitz1973ordering, kosterlitz1974critical, frohlich1981kosterlitz-thouless}, which in our language corresponds precisely to the roughening transition \cite{chui1976transition}.
Therefore, in the critical phase, the fluctuations are described by massless bosons, and at finite size their free energy has a $1/r$ correction due to the conformal anomaly \cite{blote1986conformal, affleck1988universal, difrancesco2012conformal}.

This first correction to the potential is the so-called \emph{L\"uscher term} \cite{luscher1980anomalies, luscher1981symmetry-breaking}, which corresponds to the sum of the Casimir energies of the bosons.
In the case of open strings we have $E_{\text{Casimir}} = - \pi / 24 r$, while for closed strings $E_{\text{Casimir}} = - \pi / 6 r$ (with central charge $c=1$) \cite{blote1986conformal, affleck1988universal, difrancesco2012conformal}.
Summing up the contributions from each mode, one obtains that
\begin{equation}
    V(g, \vec{r}) = \sigma(g) r + \beta + \frac{\gamma}{r} + O(r^{-3})
    \label{eq:potential_luscher}
\end{equation}
where $\sigma(g)$ is the string tension, $\beta$ a constant that depends at most on $g$ and the $1/r$ coefficient term is given by
\begin{equation}
    \gamma = - \frac{\pi}{24} (D-2).
    \label{eq:luscher_term}
\end{equation}
In our case $D - 2 = 1$.
It is important to point out that the value \eqref{eq:luscher_term} of $\gamma$ is valid within a (discrete) Lorentz-invariant \footnote{Lorentz-invariance on a lattice has to be understood as isotropy between the time and space directions} formulation of LGTs, which is not the case for a Hamiltonian framework \cite{luscher1981symmetry-breaking}.
When going to the Hamiltonian formalism in the lattice, the Casimir energy has to be rescaled by a (non-universal) speed of sound factor $v_s$, related to the relative scale of time and space discretizations.

In Fig.~\ref{fig:potential}a--b, we have calculated the potential and extracted the values of string tension $\sigma$ and the L\"uscher term $\gamma$ for $N = 4,5,6$ by fitting \eqref{eq:potential_luscher}.
For the fit, we use a set of charge configurations, where both external charges are along the horizontal axis, separated by distances $R=7,8,\dots,21$.
In Appendix \ref{app:numerical_methods_analysis}, we show how the finiteness of the string lengths influences the determination of the subleading terms of the potential.
The string tension is confirmed to be non-vanishing in the crossover region, meaning that we are, in fact, not observing a deconfining transition.
Meanwhile, the behavior of the L\"uscher term is much more interesting.
It reaches a plateau at $- \pi/12$ and not at $- \pi / 24$ in the crossover region.
Considering that we have only one transversal direction in \eqref{eq:luscher_term},  this suggests that the speed of sound in the effective string model is $v_s \simeq 2$ across the crossover region in our system.
Higher-order corrections to \eqref{eq:potential_luscher} are expected from a more accurate modeling of the flux string, which is a topic discussed in Appendix \ref{app:higher_order_corrections_to_the_potential}.

\subsection{Restoration of rotational symmetry}
\label{sub:restoration_of_rotational_symmetry}

\begin{figure*}[t]
    \centering
    \begin{minipage}{5cm}
        \centering
        \includegraphics{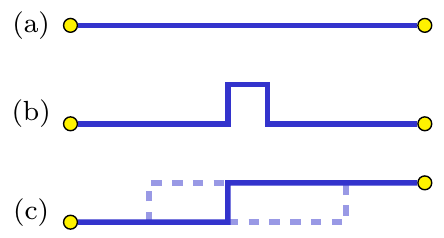}
    \end{minipage}
    \hspace{0.5cm}
    \begin{minipage}{9cm}
        \centering
        \includegraphics[width=9cm]{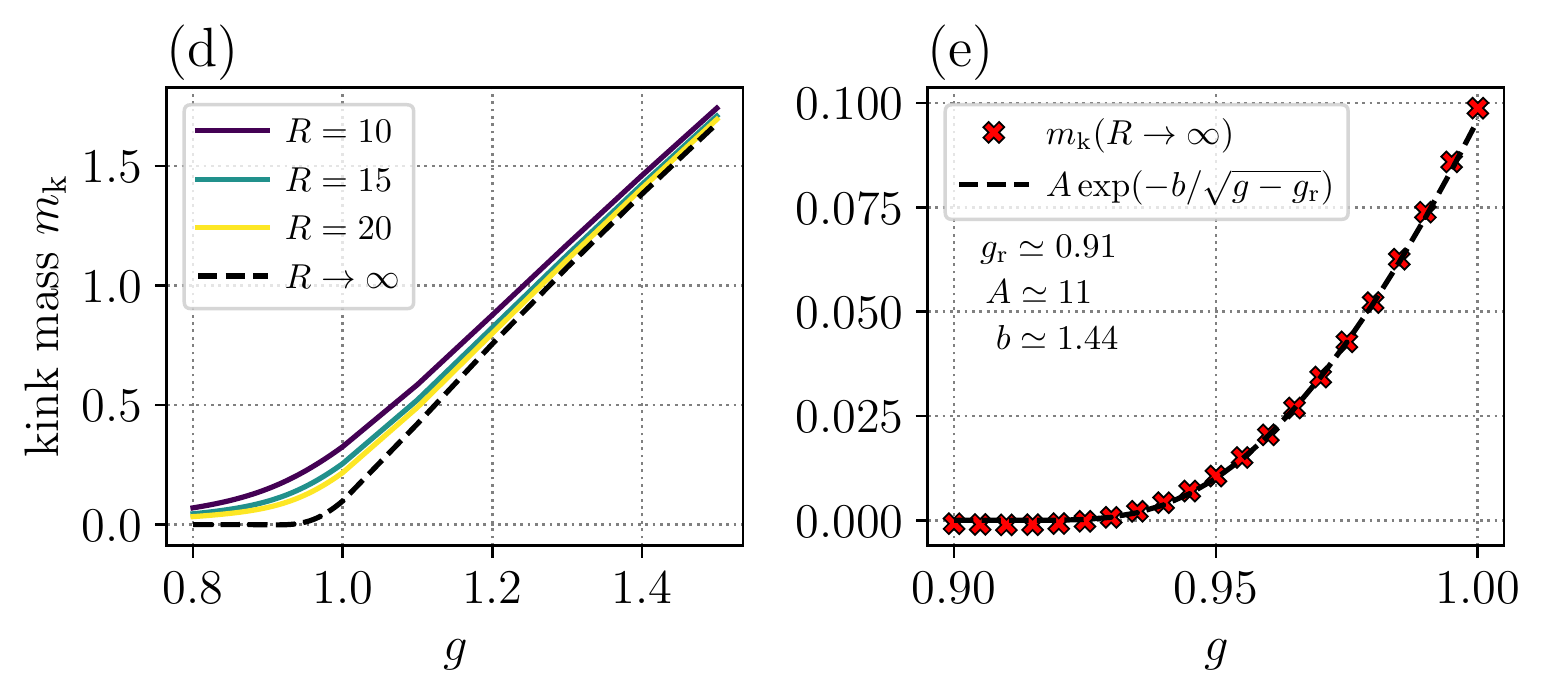}
    \end{minipage}
    \caption{Different string configurations.
    \textbf{(a)} On-axis string.
    \textbf{(b)} Lowest excited configuration of on-axis string (kink-antikink pair).
    \textbf{(c)} Off-axis string with a single kink; off-axis charges have multiple minimal paths connecting them.
    \textbf{(d)} Behavior of the kink mass $\kinkmass$ \eqref{eq:kink_mass_def} in the confined phase,
    at fixed $R$ and the extrapolation to $R \to \infty$; for clarity only $R=10, 15, 20$ are shown even though the extrapolation has been done from the set $R = 10, 11, \dots, 20$.
    \textbf{(e)} Vanishing of the kink mass fitted against the gap function of a BKT transition; we obtain a transition point $\roughpoint \simeq 0.91$.
    Both (d) and (e) are computed for a $30 \times 5$ lattice.
    We have used $41$ points in the interval $[0.8, 1.0]$ and $11$ points in $[1.0, 2.0]$.
    }
    \label{fig:string_confs_and_kink_mass}
\end{figure*}

An interesting feature of the roughening transition in LGTs is the \emph{restoration of rotational symmetry} \cite{kogut1981string}.
On a lattice, a flux string behaves very differently from its continuum counterpart, especially at strong coupling.
Not only does the string's spectrum become discrete, but it can also strongly depend on its orientation, meaning that the rotational invariance of the potential is broken.

To visualize this fact, we may compare the scenarios of two charges on- and off-axis.
In the first case, both charges lie along the same lattice axis (in our case along a horizontal leg of the lattice), in the second, both charges have different vertical coordinates (see left panel of Fig.~\ref{fig:string_confs_and_kink_mass}).
The main contribution to the potential comes from the electric field on the links along the shortest lattice path connecting the two charges.
In the on-axis case, this path is unique,
and each deformation of it can have a finite cost in energy.
Meanwhile, in the off-axis case, there is a degeneracy in the lowest energy configuration of the string, as there are multiple paths of the same minimal length connecting the charges (see Fig.~\ref{fig:string_confs_and_kink_mass}c).
This means that the off-axis string cannot be rigid but is already fluctuating, and it has been shown that it is already ``rough'' \cite{kogut1981analyticity}.
Therefore, the confining potential does not simply depend on the distance between the charges, i.e., on the length of the vector connecting their positions, but also on its orientation, hence breaking rotational invariance.
As transverse excitations of the string become massless, the difference between on- and off-axis strings vanishes, and this restoration of rotational symmetry is a signature of the roughening point \cite{kogut1981string, kogut1981analyticity}.

Let $\vec{r} = (\Delta x, \Delta y)$ be the vector connecting the positions of the two charges, the off-axis case corresponding to $|\Delta y|>0$.
In the strong coupling regime $g > \roughpoint$, the leading order contribution to the potential is given by the electric field on the links of the minimal lattice path connecting the charges.
There are multiple minimal paths, but they all involve
$\abs{\Delta x}$ and $\abs{\Delta y}$ links in the $\hat{1}$ and $\hat{2}$ directions, respectively, meaning that $V(\vec{r}) \propto \abs{\Delta x} + \abs{\Delta y}$.
This leads us to the definition of the \emph{kink mass} $\kinkmass$ as the energy contribution of a kink in an off-axis string.
Following \cite{kogut1981string},
in the strong coupling regime $g \gg  \roughpoint$, the potential of a slightly off-axis string will behave asymptotically as
\begin{equation}
    V(\vec{r}) \simeq \sigma \abs{\Delta x} + \kinkmass \abs{\Delta y}
    \qquad \abs{\Delta x} \gg \abs{\Delta y},
    \label{eq:potential_strong}
\end{equation}
where $\sigma$ is the string tension.
With this definition, we expect the kink mass $m_k$ to have a dependence on the string tension $\sigma$.
When the rotational symmetry gets restored at weak coupling $\critpoint < g \leq \roughpoint$, the potential should depend only on the distance between the charges, meaning that
\begin{equation}
    V(\vec{r})=V(r) \simeq \sigma \sqrt{(\Delta x)^2 + (\Delta y)^2}
    \label{eq:potential_weak}
\end{equation}
at the leading order.

The transition from \eqref{eq:potential_strong} to \eqref{eq:potential_weak} is expected to be non-analytical \cite{kogut1981string} and can be directly observed.
Given \eqref{eq:potential_strong}, we can extract the kink mass  from our simulations by comparing the energies of two ground states with a pair of static charges placed either on a horizontal lattice axis, at a (large) distance $R$, or off-axis, at the same longitudinal distance but with the minimal transversal displacement, via
\begin{equation}
    \kinkmass(R) = V(\vec{r}_{\text{off}}) - V(\vec{r}_{\text{on}}),
    \label{eq:kink_mass_def}
\end{equation}
where $\vec{r}_{\text{on}} = (R, 0)$ and $\vec{r}_{\text{off}} = (R, 1)$ (see Fig.~\ref{fig:string_confs_and_kink_mass}a and c).
In other words, we introduce a single kink in the string and measure the difference in the potential.
At very strong coupling $g \gg \roughpoint$ we expect the kink mass to be practically equal to the string tension, i.e.~$\kinkmass / \sigma \simeq 1$.
In contrast, we expect $\kinkmass / \sigma \simeq (2 R)^{-1}$ in the crossover region, which we obtain by substituting \eqref{eq:potential_weak} in \eqref{eq:kink_mass_def} and expanding for large $R$.
There is indeed a qualitatively different behavior when crossing the roughening transition, in particular the kink mass vanishes for $R \to \infty$ in the crossover region, while staying finite in the strong coupling region.

One may be tempted to consider $\kinkmass$ as equivalent to the excitation gap of the string.
However, this interpretation is valid only for the on-axis string in the strong-coupling regime.
This is because the lowest excited configurations are represented by a kink and ``antikink'' pair, as shown in Fig.~\ref{fig:string_confs_and_kink_mass}b, whose energy gap is $2\kinkmass$.
In the weak coupling region, the kink mass in \eqref{eq:kink_mass_def} only represents the difference in energy between a string of length $R$ and $\sqrt{R^2 + 1}$, which does not reflect the excitation gap.
With this warning in mind, we can try to compare $\kinkmass$ for $g > \roughpoint$ against the gap $\Delta$ of a BKT transition \cite{kosterlitz1974critical}
\begin{equation}
    \Delta(g) \sim A \exp \qty( - \frac{b}{\sqrt{g - \roughpoint}} ),
    \label{eq:bkt_gap}
\end{equation}
where $A$ and $b$ are to constant to be determined.
To get an estimation of the roughening point, the kink mass $\kinkmass$ in the limit $R \to \infty$ has been fitted against \eqref{eq:bkt_gap}.
This yields a value $\roughpoint \simeq 0.91$, which is comparable with the recent literature \cite{xu2025tensor, krinitsin2024roughening}, and the details are presented in Appendix \ref{app:scaling_of_the_kink_mass}.

In Fig.~\ref{fig:string_confs_and_kink_mass}d--e, we have computed the kink mass in a $30 \times 5$ lattice and for separation $R = 10, 11, \dots, 20$ and extrapolated the values for $R \to \infty$.
We observe a vanishing of the kink mass in the region $g \lesssim 1.0$, which is in agreement with the picture of rotational invariance restoration.
To further corroborate this point, we also compare the potential for on-axis and off-axis charges.
Meaning, that in both cases the potential is computed and fitted against the expression \eqref{eq:potential_luscher} and then the leading parameters, $\sigma$ and  $\beta$, are compared.
We used a $50 \times 5$ lattice with horizontal separations $R = 25, 26, \dots, 30$.
In the case of rotational invariance we expect the differences $\sigma_{\text{off}} - \sigma_{\text{on}}$ and $\beta_{\text{off}} - \beta_{\text{on}}$ to vanish.
Indeed, the results in Fig.~\ref{fig:potential}c are compatible with this expectation, with the differences vanishing around $\roughpoint \simeq 0.91$. Further results are in Appendix~\ref {app:numerical_methods_analysis}.

\section{Dynamical studies}
\label{sec:dynamical_studies}

A quantum quench involves a sudden change in the Hamiltonian of the system, driving it out of its initial equilibrium state.
Understanding the system's response to such sudden perturbations provides valuable information about its excitation spectrum and the nature of interactions, offering a complementary perspective to the static analysis of the roughening transition.

To isolate the roughening phenomenology under dynamics as much as possible from the properties of the bulk, we simulate quenches by applying a string operator $\Pi_{m=0}^{R-1}\sigma_{(x_0+m,y_0), \hat{1}}^z$ on the interacting vacuum state, variationally obtained in the sector of no charges for a certain value of the coupling $g$, and letting the system evolve with the same Hamiltonian.
Because the bulk is in the corresponding ground state, the quench has a relatively local character, restricted to the string.
For efficiency reasons, we perform these simulations in the dual model, where the effect of applying the string operator is that of actually changing the Hamiltonian \eqref{eq:dual_ising_model}, from \eqref{eq:hamiltonian_vac} to \eqref{eq:hamiltonian_2p}, as the insertion of new static charges changes the sector. 
This translates into a modification of the nearest-neighbor interaction between the dual electric field variables along the path of the string operator.
Thus, in our simulations, we prepare the initial state at $t=0$ as
the ground state $\ket{\Omega_{\varnothing}}$ of the vacuum Hamiltonian $H(g; \varnothing)$.
Then, for $t > 0$, we evolve the ground state with the two-particle Hamiltonian $H(g; v_1, v_2)$.
The difference between the two Hamiltonians is
\begin{equation}
    H(g; v_1, v_2) - H(g; \varnothing) =
    2 g \sum_{v \in \Gamma} \dualB_{v} \dualB_{v - \hat{2}},
\end{equation}
where $\Gamma$ is the lattice path connecting the two charges, which are at a distance $R$ between each other.
Thus, while $H(g; \varnothing)$ has only ferromagnetic interactions, our quench introduces antiferromagnetic defects along the path $\Gamma$.
From the point of view of the dual Ising model, considering that the confined phase corresponds to a ferromagnetic phase, this quench protocol introduces a domain wall of length $R$ inside the ordered region.
We have simulated quenches as described above at different values of the coupling $g$, for varying string length $R$.
Our TEBD simulations allow us to keep a faithful approximation of the evolved state at short and intermediate times, before the truncation error becomes too large.

A similar study was conducted in \cite{krinitsin2024roughening}, although the situation analyzed in it is quite different.
Since the initial state in \cite{krinitsin2024roughening} is a product state with two differently polarized regions, the quench is global.
In our case, we take as the initial state the ground state of the pure gauge theory, which is not necessarily a product state.
In this way, we isolate the string dynamics from the ``background'', which remains mostly stationary.

\subsection{Entanglement entropy dynamics}
\label{sub:entanglement_entropy_dynamics}

The behavior of entanglement entropy after a quench can reveal substantial information about the system.
In general, for (1+1)D cases, the entropy grows linearly with time.
In integrable models, this is traced back to the propagation of quasiparticle excitations~\cite{calabrese2005evolution,dechiara2006heisenberg, Alba2017}, whereas its origin in more generic systems is still investigated~\cite{Kim2013ballistic, Ho2017ent,Nahum2017rand}.
Alterations of this generic growth can occur, for instance, in the presence of strong disorder~\cite{Znidaric2008mbl, Bardarson2012mbl}, or in systems with confinement, where the formation of bound states constrains the dynamics of quasiparticles \cite{kormos2017confinement}.

Because our quenches explore the dynamics of the string in regimes where its effective description has drastically different properties, we can expect that the behavior of the entanglement entropy is sensitive to such phases.
In particular, we find that the rate of entanglement growth exhibits qualitatively different features across the confining region, which distinguish the rigid and fluctuating phases of the string.

Within our MPS setup, we can determine the EE as a function of time.
Fig.~\ref{fig:dynamics_sw_ee}a shows its evolution after performing the quench with a string of fixed length $R=10$, on a cylinder of width $N=5$
for the coupling sweeping from the weak confinement ($g = 0.75, 0.8, 0.9$ and $\delta t = 0.01$) to the strong confinement ($g = 1.0, 1.2$ and $\delta t = 0.02$).
After a transient at short times, we can observe a period of approximate linear growth of the EE for intermediate times, marked with a gray shaded area in the figure.
In the strongly confined region, the rate of this growth seems to decrease as $g$ increases.
Instead, within the roughening region, the rate becomes essentially independent of the coupling $g$, with a slope around $0.97(5)$.
This observation is consistent with the general expectation of a linear growth in critical 1D systems \cite{calabrese2005evolution}.

Fig.~\ref{fig:dynamics_sw_ee}b shows, on the other hand, the time-evolution as a function of the string length.
We have considered string lengths $R = 10, 11, 15, 20$, with the coupling fixed in the crossover region ($g = 0.8$ and $\delta t = 0.05$) on a lattice with $N=6$ and maximum bond dimension $\chi=256$.
We observe that the initial short-time transient and the consequent period of linear growth are unaffected by the change of $R$.
The rate of growth appears to be independent of the string length.
Instead, we notice that for different $R$'s the EE saturates at different values, each at its time $t^*(R)$.

\subsection{String width dynamics}
\label{sub:string_width_dynamics}

\begin{figure}[t]
    \centering
    \includegraphics[width=\columnwidth]{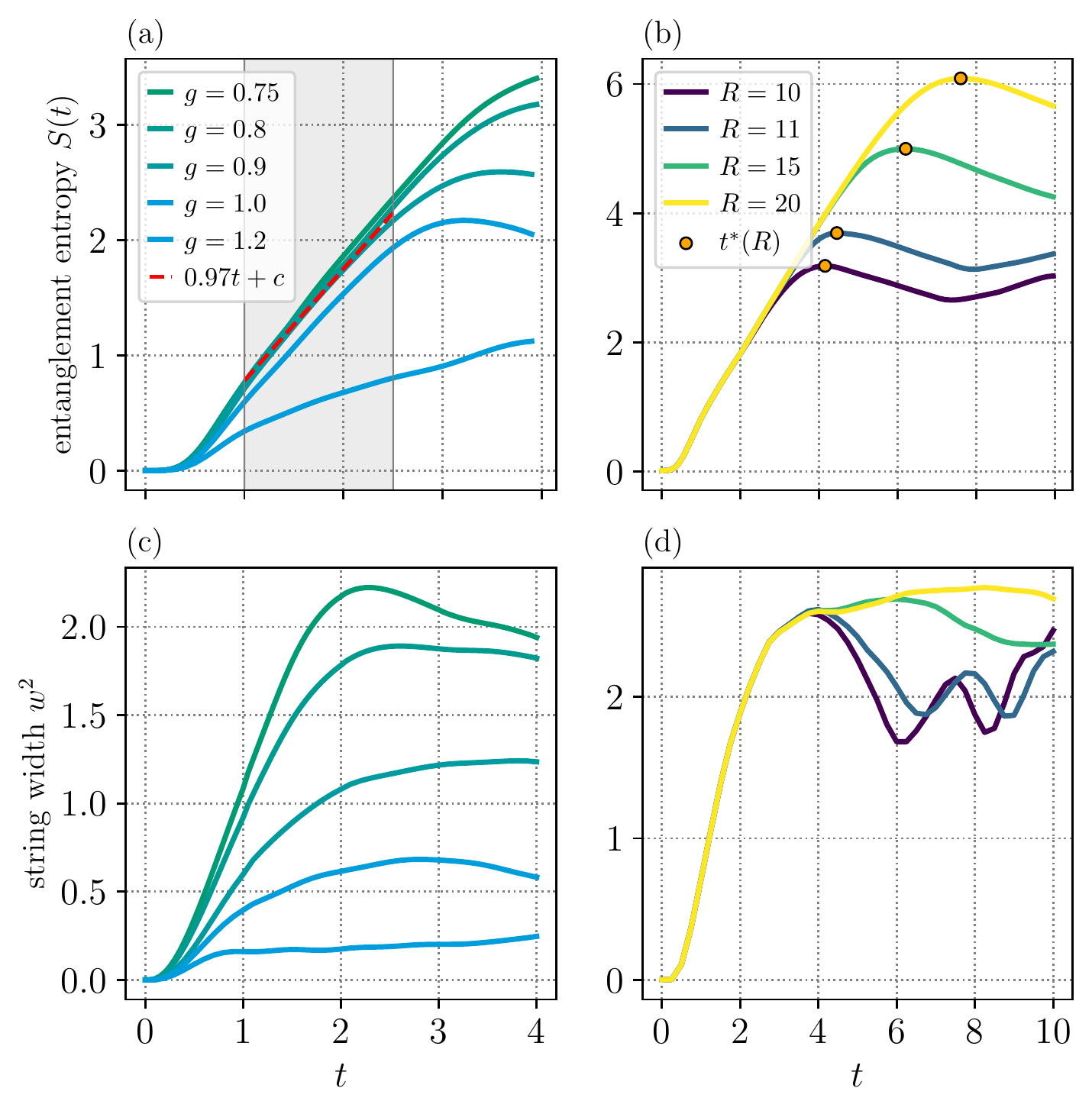}
    \caption{Time evolution of EE $S(t)$ (upper panels) and string width $w^2(t)$ (lower panels).
    In the panels (a) and (c), we consider a $30 \times 5$ lattice with variable $g$ and fixed string length ($R = 10$); the maximum bond dimension is $\chi_{\text{max}} = 128$.
    Conversely, in (b) and (d), a $30 \times 6$ lattice is considered instead with a variable $R$ and coupling $g=0.8$, fixed in the crossover region; the bond dimension is increased to $\chi_{\text{max}} = 256$.
    The shaded area in (a) displays the period of linear growth of the EE, which is approximated by the red dashed line with a slope of $0.97(5)$.
    The markers in (b) show the saturation times $t^*$ for each length $R$.
    }
    \label{fig:dynamics_sw_ee}
\end{figure}

In the same dynamical setup, in addition to the EE, we have also measured the width of the string $w^2$ as a function of time.
By looking at its evolution, we can observe how the electric field spreads around the string after a quench.
In Fig.~\ref{fig:dynamics_sw_ee}c--d, we show the time dependence of $w^2$, varying both the coupling $g$ and the string length $R$, using the same simulation parameters of the previous section.

What we see is a stark difference compared to the EE.
First, in Fig.~\ref{fig:dynamics_sw_ee}c, where the string length is fixed ($R=10$) while the coupling is varied, we observe a strong dependence of the growth rate on the coupling, even in the crossover region.
Second, in Fig.~\ref{fig:dynamics_sw_ee}d, where instead the coupling is fixed to $g=0.8$ and $R$ varied, the evolution of $w^2$ is uniform in $R$ up to the saturation point, which also does not depend on $R$.
Further numerical details on the dynamics are given in the Appendix~\ref{app:numerical_methods_analysis}.

\subsection{Electric field dynamics}
\label{sub:electric_field_dynamics}

\begin{figure}[t]
    \centering
    \includegraphics[width=0.9\columnwidth]{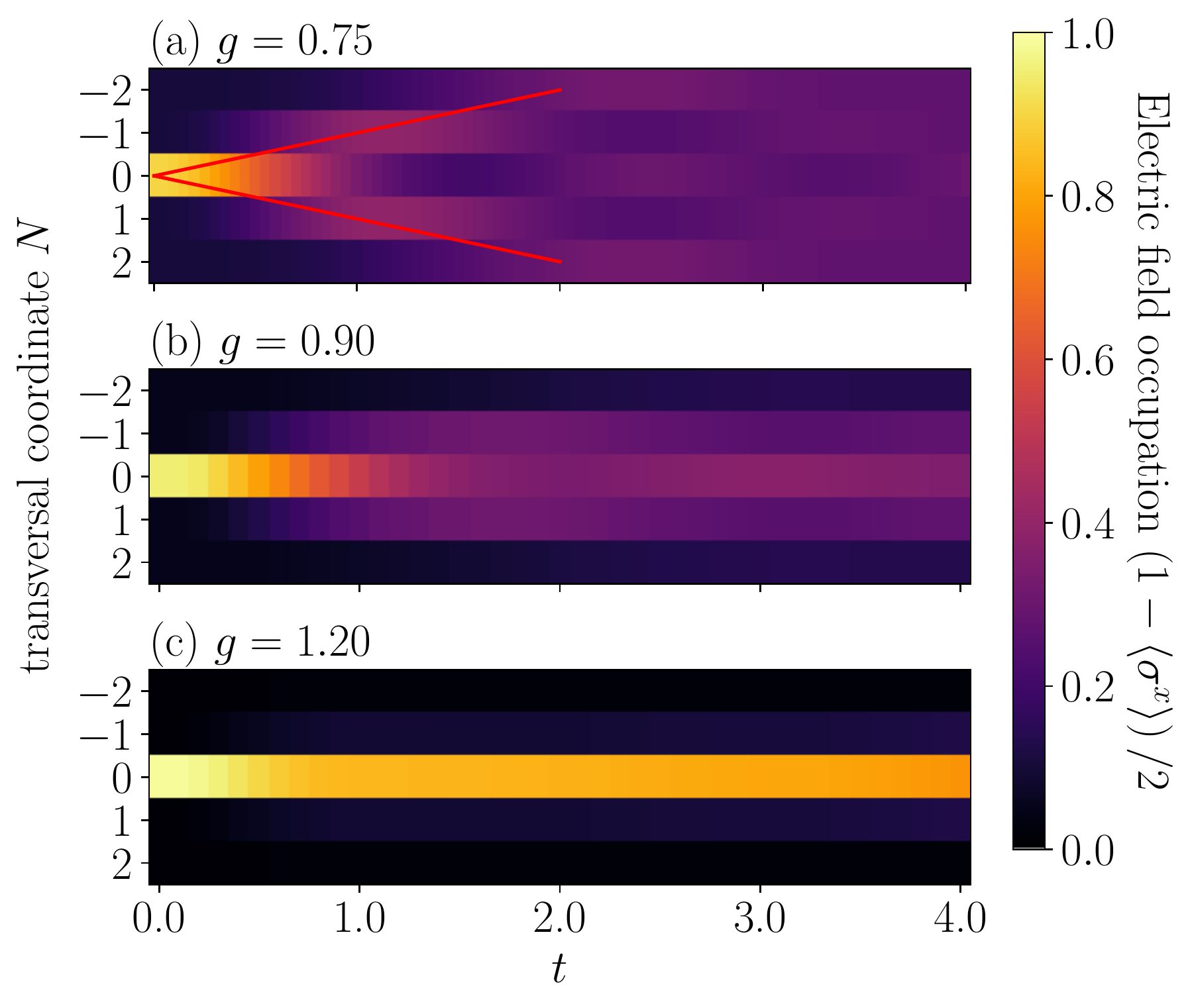}
    \caption{Electric field in time along a slice of the cylinder:
    \textbf{(a)} inside the crossover region ($g = 0.75$ and $\delta t = 0.01$);
    \textbf{(b)} near the roughening point ($g = 0.9$ and $\delta t = 0.02$);
    \textbf{(c)} strongly confined ($g = 1.2$ and $\delta t = 0.02$).
    As a measure of the electric field, we have used the expectation value of the occupation number $(1 - \elec_{v, \hat{1}}) / 2$, computed on the horizontal links in the middle of the cylinder.
    We have highlighted the linear spreading of the electric field in (a) with solid red lines (with slope equal to one).
    }
    \label{fig:dynamics_electric_fields}
\end{figure}

To further enrich our picture of the dynamics, we depict in Fig.~\ref{fig:dynamics_electric_fields} the time evolution on the electric basis of a slice of the cylinder.
In this visualization, we are going back to the original gauge model on the direct lattice.
We consider the electric field on the middle horizontal links of a $30 \times 5$ lattice and display their dynamics for three key parameters: inside the crossover region ($g = 0.75$), near the roughening transition ($g=0.9$), and in the strongly confined phase ($g=1.2$).

For $g < \roughpoint$, a spread of the fluctuations is noticeable, with a rate consistent with the one observed in Fig.~\ref{fig:dynamics_sw_ee}c.
Closer to the deconfining point ($g = 0.75$), the spread is much wider.
On the other hand, when $g > \roughpoint$, the string is visibly rigid in time with its fluctuations suppressed, meaning that the system is only slightly perturbed by the introduction of the string.

\section{Conclusions and outlook}
\label{sec:conclusions_and_outlook}

The research carried out in this paper demonstrates the power of tensor network methods to quantitatively analyze the roughening transition and the consequent crossover region.
We have directly observed different aspects of this transition, including: the logarithmic divergence of string width; the critical scaling of the entanglement entropy; universal corrections in the confining potential; and the restoration of rotational symmetry on the lattice.
The resulting scenario matches the description of the critical string via a (1+1)D massless boson.

We also performed a qualitative study of the string dynamics by exploring the local quench where the string is created in the vacuum of gauge theory, comparing the weak and strong confinement regions.
The behavior of the entanglement is again in agreement with the prediction of 1D critical systems.
Furthermore, we also offer a complementary view by displaying the string width in time, which measures the spread of the electric field.

Nevertheless, to develop a more complete picture of the string dynamics, we require a more thorough study with longer times.
Moreover, developing theoretical models to describe the time evolution of string width will further enhance our understanding.
Additionally, extending these investigations to lattice gauge theories with different gauge groups and matter content will help establish the universality of the observed phenomena.
The study of off-axis string dynamics could also provide relevant insights since their nature is rough by definition.

Finally, the exciting progress in quantum simulation opens the door to experimentally probing the dynamical aspects of the roughening transition, potentially providing a direct validation of the theoretical predictions discussed in this manuscript.
The continued exploration of dynamical phenomena in the roughening region promises to deepen our understanding of fundamental aspects of quantum field theory and its applications in various areas of physics.
The field of quantum simulation has witnessed significant advancements in recent years, with experimental platforms increasingly capable of simulating the dynamics of lattice gauge theories.
These experimental efforts, often utilizing systems of trapped ions, superconducting qubits, or neutral atoms, offer the exciting prospect of directly observing the dynamical behavior of electric flux strings and potentially studying the roughening transition in a controlled laboratory setting.

\begin{acknowledgments}

We acknowledge inputs and feedback at several project stages in discussions with J.~Halimeh, K.~Jansen, M.~Knap, S.~Kuhn, L.~Tagliacozzo, and T.~Zache.

S.P. and E.R. acknowledge the financial support received from the IKUR Strategy under the collaboration agreement between the Ikerbasque Foundation and UPV/EHU on behalf of the Department of Education of the Basque Government.

E.R. acknowledges support from the BasQ strategy of the Department of Science, Universities, and Innovation of the Basque Government. E.R. is supported by the grant PID2021-126273NB-I00 funded by MCIN/AEI/10.13039/501100011033 and by “ERDF A way of making Europe” and the Basque Government through Grant No. IT1470-22. This work was supported by the EU via QuantERA project T-NiSQ grant PCI2022-132984 funded by MCIN/AEI/10.13039/501100011033 and by the European Union “NextGenerationEU”/PRTR. This work has been financially supported by the Ministry of Economic Affairs and Digital Transformation of the Spanish Government through the QUANTUM ENIA project, called Quantum Spain project, and by the European Union through the Recovery, Transformation, and Resilience Plan – NextGenerationEU within the framework of the Digital Spain 2026 Agenda.

M.C.B. acknowledges support from the Deutsche Forschungsgemeinschaft (DFG, German Research Foundation) under Germany's Excellence Strategy -- EXC-2111 -- 390814868 and Research Unit FOR 5522 (grant nr. 499180199);
and the EU-QUANTERA project TNiSQ (BA 6059/1-1).

This work has been partially funded by the Eric \& Wendy Schmidt Fund for Strategic Innovation through the CERN Next Generation Triggers project under grant agreement number SIF-2023-004.

The authors developed the tensor network code from scratch in Python, which is available at \cite{qs-mps-repo}.
\end{acknowledgments}

\appendix

\section{Duality transformation with static charges}
\label{app:duality_transformation_with_static_charges}

We consider a lattice $\Lattice$ with $L$ plaquettes in the horizontal direction $\hat{1}$ and $N$ plaquettes in the vertical direction $\hat{2}$.
The sites are labeled as $v = (x, y)$, with integer coordinates $x$ and $y$ along $\hat{1}$ and $\hat{2}$, respectively.
With periodic boundary conditions in the vertical direction, we have $x = 1, \dots, L+1$ and $y = 1, \dots, N$, with $y + N \equiv y$.
From $\Lattice$ we define the dual lattice $\DualLattice$, where the plaquettes of the former are the sites of the latter.
Each plaquette $\square$ (hence dual site) is labeled with its direct site $v$ in the bottom left corner.

The main idea of the duality is to map the plaquette states into spin-$\onehalf$ states.
We first define the set of operators $\dualB_v$ and $\dualE_v$ for each site of the dual lattice.
They have the same operator algebra as $\gauge$ and $\elec$.
Then, we define $\dualB_v$ as the plaquette operator at the site $v$
\begin{equation}
    \dualB_v = (\gauge \gauge \gauge \gauge)_{\square_v}.
    \label{eq:def_dualB_app}
\end{equation}
Subsequently, we can define $\dualE$ at $v=(x,y)$ as a string of $\elec$ starting from the left boundary $x=1$:
\begin{equation}
    \dualE_{(x, y)} = \prod_{j=1}^{x} \elec_{(j, y), \hat{2}}.
    \label{eq:def_dualE_app}
\end{equation}
Notice that the definition of $\dualE_v$ involves only the $\elec$ operators on the vertical links.

To write down the dual Hamiltonian, we need to invert \eqref{eq:def_dualB_app} and \eqref{eq:def_dualE_app} and express the operators in \eqref{eq:LGT_hamiltonian} as a function of $\dualB$ and $\dualE$.
The plaquette term is given by \eqref{eq:def_dualB_app} by definition, while the electric terms require a bit more work.
On the vertical links, away from the boundary, we observe from \eqref{eq:def_dualE_app} that the electric field can be written down as an interaction term:
\begin{equation}
    \elec_{(x,y), \hat{2}} = \dualE_{(x,y)} \dualE_{(x-1, y)}.
    \label{eq:elec2_from_dualE_bulk}
\end{equation}
On the leftmost boundary $x=1$, $\dualE$ and $\elec_{\hat{2}}$ coincide, meaning
\begin{equation}
    \dualE_{(1, y)} = \elec_{(1, y), \hat{2}}.
    \label{eq:elec2_from_dualE_left}
\end{equation}
On the other hand, on the rightmost boundary $x=L+1$ it is necessary to introduce an ancillary spin $\dualE_A$:
\begin{equation}
    \elec_{(L+1, y), \hat{2}} = \dualE_{(L, y)} \dualE_A.
    \label{eq:elec2_from_dualE_right}
\end{equation}
The role of this ancilla needs a separate discussion, which will be dealt with later, about the topological sectors of the gauge model.

Now the goal is to find an expression for the electric fields $\elec_{\hat{1}}$ on the horizontal links, which can be achieved by consecutive applications of the Gauss law.
We start with the leftmost boundary $x=1$, by considering the action on physical states of the Gauss law at a site $(1, y)$:
\begin{equation}
    G_{(1, y)} = \elec_{(1, y), \hat{1}} \elec_{(1, y), \hat{2}} \elec_{(1, y), -\hat{2}} = q_{(1, y)}.
\end{equation}
We can substitute \eqref{eq:elec2_from_dualE_left} in place of $\elec_{(1,y), \hat{2}}$ and $\elec_{(1,y), -\hat{2}}$ in the above expression and obtain
\begin{equation}
    \elec_{(1,y), \hat{2}} = q_{(1,y)} \dualE_{(1, y)} \dualE_{(1, y-1)}.
    \label{eq:elec1_from_dualE_left}
\end{equation}
Notice that $\elec_{(1,y), \hat{2}}$ \emph{depends on the charge} $q_{(1, y)}$.
Then, we move to the left and consider the column $x=2$.
Here, the Gauss law at the site $(2, y)$ reads:
\begin{equation}
    G_{(2, y)}
    = \elec_{(2, y), \hat{1}} \elec_{(2, y), \hat{2}} \elec_{(2, y), -\hat{1}} \elec_{(2, y), -\hat{2}}
    = q_{(2, y)}.
\end{equation}
Now we can substitute both \eqref{eq:elec2_from_dualE_bulk} and \eqref{eq:elec1_from_dualE_left} and express $\elec_{(2, y), \hat{1}}$ in terms of dual operators:
\begin{equation}
    \elec_{(2, y)} = q_{(1, y)} q_{(2, y)} \dualE_{(2, y)} \dualE_{(2, y-1)}.
\end{equation}
We see that now it depends on the product of charges $q_{(1, y)} q_{(2, y)}$.
This process can then be iterated for all the other fields on the horizontal links, obtaining:
\begin{equation}
    \elec_{(x, y), \hat{1}} =
    \eta(x, y) \dualE_{(x, y)} \dualE_{(x, y-1)},
\end{equation}
with a phase factor
\begin{equation}
    \eta(x, y) = \prod_{j = 1}^{x} q_{(j, y)}
\end{equation}
that \emph{depends on a product string of charges}.
We can write down the mapping of the electric fields in a general way as
\begin{equation}
    \elec_{v, \hat{\mu}} \mapsto \omega(v, \hat{\nu}) \dualE_v \dualE_{v - \hat{\nu}},
    \label{eq:elec_from_dualE_general}
\end{equation}
where $\hat{\nu}$ is the direction perpendicular to $\hat{\mu}$.
In our case we have ($v = (x,y)$)
\begin{equation}
    \omega(v, \hat{\nu}) =
    \begin{cases}
        1 & \hat{\nu} = 1, \\
        \prod_{j = 1}^{x} q_{(j, y)} & \hat{\nu} = 2.
    \end{cases}
    \label{eq:general_phase_factor}
\end{equation}
From \eqref{eq:elec_from_dualE_general} and \eqref{eq:general_phase_factor}, we see that the Gauss law is automatically satisfied:
\begin{equation}
    G_v = \omega(v, \hat{2}) \omega(v, \hat{1}) \omega(v-\hat{1}, \hat{2}) \omega(v-\hat{2}, \hat{1})
    = q_v.
    \label{eq:dual_gauss_law}
\end{equation}
Putting together all these results, we obtain the final dual Hamiltonian
\begin{multline}
    H_{\text{dual}} = - \frac{1}{g} \sum_{v \in \Lattice} \dualB_v - g \sum_{\nu = 1,2} \sum_{v \in \interior \Lattice} \omega(v, \hat{\nu}) \dualE_v \dualE_{v - \hat{\nu}} \\
    -g \sum_{y} \qty( \dualE_{(1,y)} + \dualE_{(L,y)} \dualE_A )
\end{multline}
We remark on the following important fact: the information about the static charges is not encoded in the degrees of freedom of the dual model per se, but only in the couplings of the interaction term $\dualE \dualE$.
The extra phase factor that appears in this term carries the content of the Gauss law at each site, as shown by \eqref{eq:dual_gauss_law}.

The last point left up for discussion is the role of the ancillary qubit and the topological sectors of the gauge model.
The key insight is the fact that the plaquettes states do not actually fully specify a physical state, due to the possibility of non-contractible electric loops on the cylinder.
An electric string is a path along which $\elec = -1$, and a loop is just a closed string.
A non-contractible electric loop is an electric loop that goes around the cylinder and cannot be removed via plaquette operators.
It is possible to define an indicator for these non-contractible loops.
Consider the string operator at vertical position $y$:
\begin{equation}
    S_y = \qty[  \prod_{y^{\prime} = 0}^{y-1} \prod_{x=1}^{L+1} q_{(x, y^{\prime})} ] \prod_{j = 1}^{L+1} \elec_{(j,y), \hat{2}},
\end{equation}
The product in brackets is a sign factor that depends on the number of charges below the string in $y$.
The values of $S_y$ are either $+1$ or $-1$ in the vacuum sector, are independent of $y$, and are invariant under the application of gauge-invariant operators.
Hence, the different eigenspaces of $S_y$ are the different topological sectors of the gauge mode.
Using \eqref{eq:elec2_from_dualE_bulk}, \eqref{eq:elec2_from_dualE_left}, and \eqref{eq:elec2_from_dualE_right} we obtain
\begin{equation}
    S_y = \elec_A,
\end{equation}
meaning that the value of the ancilla depends on the topological sector.

\begin{figure}[t]
    \centering
    \includegraphics{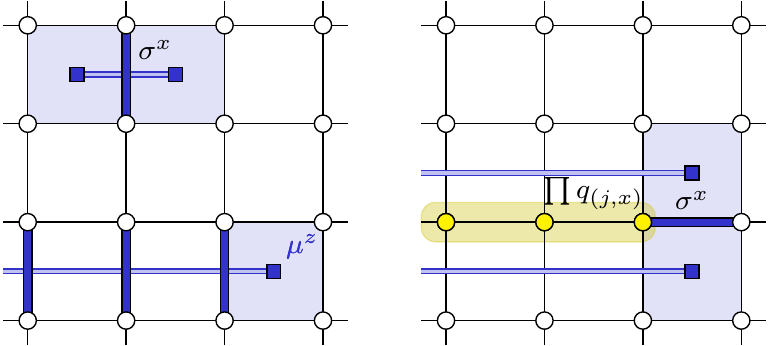}
    \caption{In a dual formulation of the $\mathbb{Z}_2$ LGT, initial plaquette interaction maps to local magnetic fields of the Ising model, and electric field terms map to nearest neighbor interaction in the transversal direction.
    In particular, electric fields on horizontal links (\emph{right}) depend on products of charges along a horizontal path (highlighted in yellow).
    }
\end{figure}

\section{Finite-size scaling for the deconfining transition}
\label{app:finite_size_scaling_for_the_deconfining_transition}

The gauge Hamiltonian \eqref{eq:LGT_hamiltonian} in the vacuum has two phases, a confining one for strong coupling $g \gg 1$ and a deconfining one for weak coupling $ g\ll 1$, with a second-order phase transition between them (see Fig.~\ref{fig:phase_diagram}).
In the confinement phase, the electric term dominates, and the excitations induced by the plaquettes are suppressed.
Conversely, in the deconfinement phase, the magnetic term dominates, which, combined with Gauss' law, realizes a ground state that is a superposition of all the possible gauge-invariant configurations.

The vacuum phase diagram of the gauge model can be mapped onto the phase diagram of the quantum Ising model.
The confining and deconfining phases, therefore, correspond to the ferromagnetic and paramagnetic phases, respectively.
The duality also extends to the phase transition point.
This means we can detect the critical point $\critpoint$ of the gauge model by finding its dual in the Ising model.
The order parameter of the dual Ising model is the magnetization
\begin{equation}
    M = \frac{1}{N} \sum_{x}  \ev*{\dualE_x},
\end{equation}
which has a second-order phase transition at the critical point.
Therefore, we expect the magnetic susceptibility $\chi_M = \dv*{M}{g}$ to diverge with a critical exponent $\gamma$:
\begin{equation}
    \chi_M \sim \abs{g - \critpoint}^{- \gamma}.
\end{equation}
This, however, is true only in the thermodynamic limit $L \to \infty$.
At finite size, we see finite peaks of height $\chi_c^{(L)}$ at positions $\critpoint^{(L)}$.
The latter can be considered as pseudo-critical points and shifted to the true value $\critpoint$.
We avoid boundary effects on the magnetization by considering only sites in the bulk of the system.

\begin{figure}[t]
    \centering
    \includegraphics[width=\columnwidth]{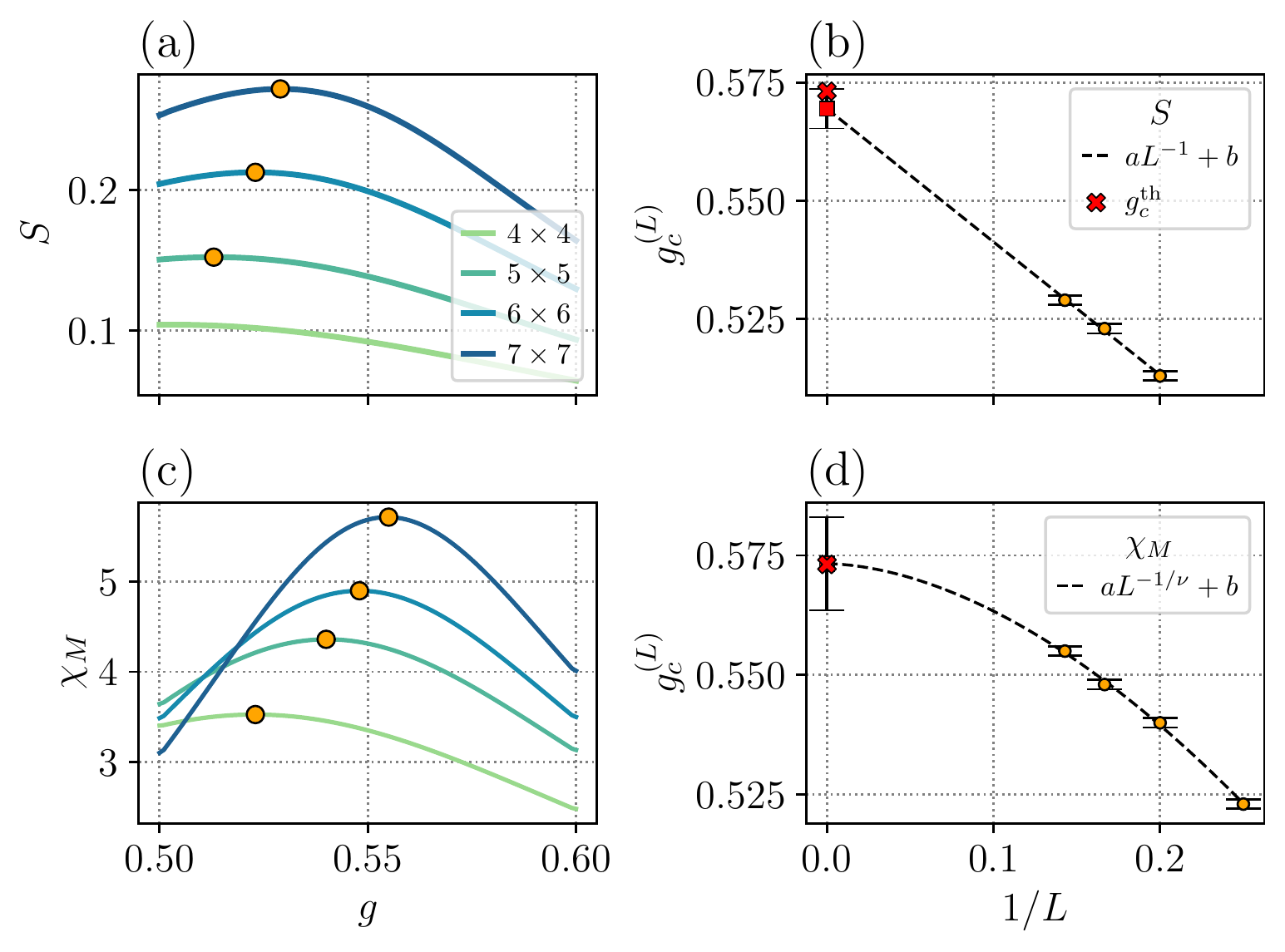}
    \caption{
        Finite-size scaling of entanglement entropy $S$ and magnetic susceptibility $\chi_M$.
        \emph{Left}: $S$ and $\chi_M$ ((a) and (c) respectively) around the critical point for $L = 4, \dots, 7$.
        The EE is computed for a half-lattice bipartition, while the magnetization is evaluated only in the bulk (square region with side $\lceil L/2 \rceil$).
        In both cases we used $101$ points for $g \in [0.5, 0.6]$.
        \emph{Right}: the extrapolation for $L \to \infty$ of the pseudo-critical points $\critpoint^{(L)}$, both for $S$ (b) and $\chi_M$ (d).
        From the EE we find $\critpoint^{\text{EE}} = \CritPointEE$, while from the susceptibility $\critpoint^{\text{M}} = \CritPointSuscep$.
        Both are compatible with the known value $\critpoint^{\text{th}} = \CritPointTheor$ that comes from the literature \cite{blote2002cluster}.
        }
        \label{fig:finite_size_scaling}
\end{figure}

We can apply the following reasoning to extract the true critical point.
The correlation length $\xi$ is also expected to diverge at the critical point with critical exponent $\nu$,
\begin{equation}
    \xi \sim \abs{g - \critpoint}^{- \nu}.
    \label{eq:corr_length_singular}
\end{equation}
But it is hindered by the finite size, meaning that $\xi \sim L$.
For this reason we replace $\xi$ with $L$ inside \eqref{eq:corr_length_singular} and invert the equation to obtain
$\abs{g - \critpoint} \sim L^{-1/\nu}$.
Therefore, we can assume that the pseudo-critical points $\critpoint^{(L)}$ follow the relation
\begin{equation}
    \critpoint^{(L)} = a L^{-1/\nu} + c,
\end{equation}
which we can then fit against our numerical data and extrapolate $\critpoint$ as
\begin{equation}
    \critpoint = \critpoint^{(\infty)} = \lim_{L \to \infty} \critpoint^{(L)}.
\end{equation}
The result of this procedure is shown in the bottom section of Fig.~\ref{fig:finite_size_scaling}, where we have obtained $\critpoint^{\text{M}} = \CritPointSuscep$.
This value is compatible with the known result $\critpoint^{\text{th}} = \CritPointTheor$ available in literature \cite{blote2002cluster}.

It is important to point out that in the gauge model, $M$ corresponds to the expectation values of 't Hooft strings, i.e., the product of strings of $\elec$ along paths in the dual lattice, which makes it a \emph{non-local} order parameter from the gauge model point of view.

Another method of finding the critical point is by measuring the entanglement entropy (EE), which is defined as
\begin{equation}
    S = - \Tr \rho_A \log \rho_A,
\end{equation}
where $\rho_A$ is the reduced density matrix $\rho_A = \Tr_B \rho$, with $A \cup B$ being the bipartition of the whole system.
In our case, we simply cut the system in the middle for the bipartition.
Both the ferromagnetic and paramagnetic phases are gapped, with a finite correlation length $\xi < L$, meaning that the entanglement saturates quickly with the system size.
On the other hand, at the critical point, due to $\xi \to \infty$, we expect the entanglement to grow, especially with increasing system size, and to saturate to a much higher value when compared to the gapped phases.
Therefore, we expect a maximum in $S$ at the critical point.

Similarly to before, with the positions $\critpoint^{(L)}$ of the EE maxima and finite-size scaling, we can extract the true value of $\critpoint$.
As before, we can extract $\critpoint$ by extrapolating for $L \to \infty$ with an appropriate scaling law, and the data suggest a simple relation
\begin{equation}
    \critpoint^{(L)} = \critpoint^{\infty} + \frac{a}{L}.
\end{equation}
We can see the outcome in the top panel of Fig.~\ref{fig:finite_size_scaling}, where we have obtained a value $\critpoint^{\text{EE}} = \CritPointEE$, again compatible with available literature.

We remark that we are not looking for an accurate estimation of the deconfining critical point, hence the rudimentary procedure.
We only need a value precise enough to distinguish the second-order phase transition of the gauge model from the roughening transition of the flux string.
Furthermore, the use of square lattices makes the extrapolation to the thermodynamic limit much more sensible than doing so with rectangular lattices.

\section{Higher-order corrections to the potential}
\label{app:higher_order_corrections_to_the_potential}

Besides the L\"uscher term \eqref{eq:luscher_term}, other higher-order corrections to the potential can be obtained through the proper modeling of the flux string.
But the latter has not been an easy task, especially if one intends to quantize the model.
For example, the simplest proposal is the Nambu-Goto action \cite{nambu1979qcd, goddard1973quantum} and it is Lorentz-invariant only at the critical dimension $D = 26$.
Another example is the Polyakov action \cite{polyakov1986string}, whose quantization leads to an extra oscillatory mode.
The most accepted proposal for a quantum model of a quantized string comes from the Polchinski-Strominger action \cite{polchinski1991effective}, which is an effective model valid only for long strings.

The spectrum for an open string with Dirichlet boundary conditions of the Polchinski-Strominger model is \cite{kuti2006lattice}
\begin{equation}
    E_n = \sigma r \sqrt{1 - \frac{\pi}{12 \sigma r^2} (D - 2) + \frac{2 \pi n}{\sigma r^2} }.
\end{equation}
By expanding the square root for $n = 0$ (the ground state), one can find the higher-order corrections to the potential.
At order $r^{-3}$ one obtains
\begin{equation}
    E_0(r) = \sigma r - \frac{\pi(D - 2)}{24 r}  - \frac{\pi^2 (D-2)^2}{1152 \sigma r^3} + O(r^{-5}).
\end{equation}
Notice that the L\"uscher term \eqref{eq:luscher_term} is recovered plus en extra term $\delta / r^3$, where
\begin{equation}
    \delta = - \frac{\pi^2 (D - 2)^2}{1152 \sigma}.
    \label{eq:delta_term}
\end{equation}

\begin{figure}[t]
    \centering
    \includegraphics[width=0.65\columnwidth]{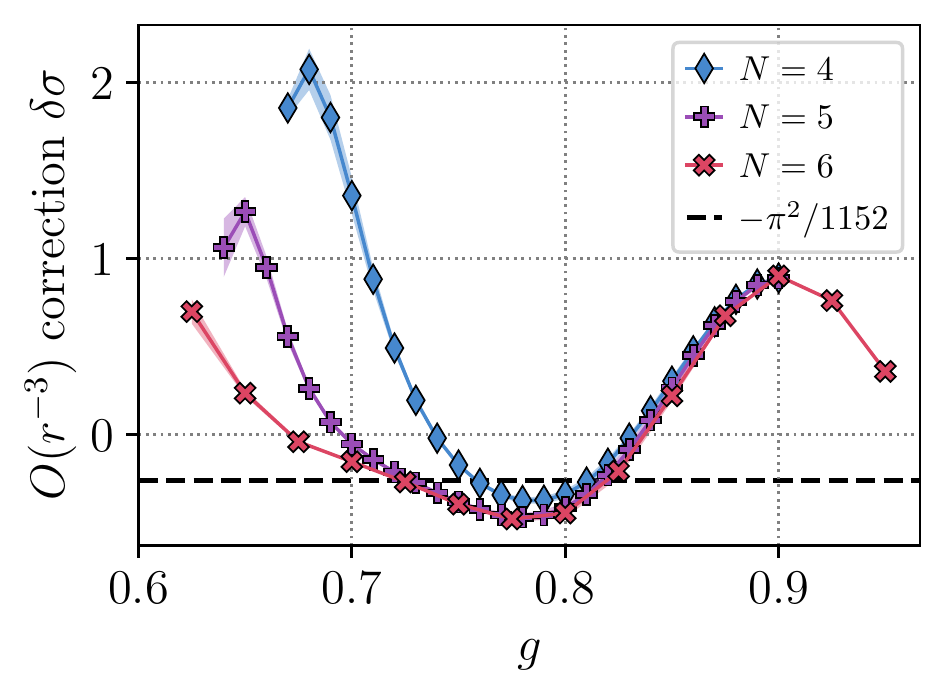}
    \caption{The $O(r^{-3})$ correction term $\delta$ times $\sigma$ of \eqref{eq:delta_term} on a lattice with $L = 30$.
    Data of the potential for $r = 7, 8, \dots, 21$ have been used for the fit. }
    \label{fig:deltasigma}

    \bigskip
    \includegraphics[width=0.65\columnwidth]{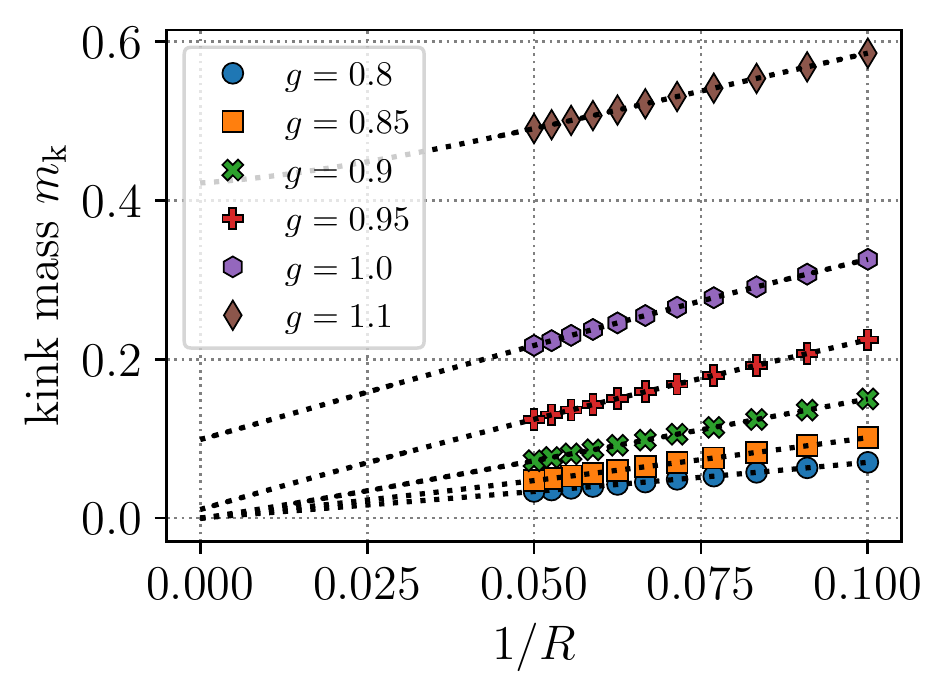}
    \caption{Scaling of the kink mass as a function of $1/R$ for different values of the coupling.
    The dotted lines represent the fit with a fifth-degree polynomial in $1/R$.}
    \label{fig:kink_mass_scaling}
\end{figure}

Using the same setup of Sect.~\ref{sub:confining_potential}, an attempt to fit the correction term \eqref{eq:delta_term} has been made, but the results have not been promising, as can be seen in Fig.~\ref{fig:deltasigma}.
In this regard, we recognize the following issue.
First, considering that \eqref{eq:delta_term} is obtained from the effective action of a long continuous string, we suspect that the finite-size and lattice effects in the present setup are still very strong for such a high-order term.
Second, similarly to the L\"uscher term in Sect.~\ref{sub:confining_potential}, the effective speed of sound $v_s$ of the string must enter the Hamiltonian description and therefore also in \eqref{eq:delta_term} in some, yet to be determined, way.

\section{Scaling of the kink mass and estimation of \texorpdfstring{$\roughpoint$}{gr}}
\label{app:scaling_of_the_kink_mass}

Following the definition \eqref{eq:kink_mass_def}, we have computed the kink mass $\kinkmass$ in the confined phase in a $30 \times 5$ lattice, for separations $R = 10, 11, \dots, 20$ and for different couplings.
Based on the literature \cite{hasenfratz1981generalized, kogut1981string, krinitsin2024roughening, xu2025tensor}, we expect the roughening point to be in the region $0.8 \lesssim g \lesssim 1.0$.
Hence, we have used 41 points in interval $[0.8, 1.0]$, while only 11 points in $[1.0, 2.0]$.
The kink mass as a function of $R$ is shown in Fig.~\ref{fig:kink_mass_scaling} only for some values of the coupling.
From observing the data, we have assumed that at finite sizes, $\kinkmass$ scales as
\begin{equation}
    \kinkmass(R) = \kinkmass(\infty) + \frac{a_1}{R} + \frac{a_2}{R^2} + \dots
    \label{eq:kink_mass_fit}
\end{equation}
For this reason, it has been fitted against a high-order polynomial in $1/R$ (in this case, fifth-order).
Then, an estimation of the limit value of $m_k(R \to \infty)$ is simply given by the constant term of the polynomial.
The results of this fit are shown in Fig.~\ref{fig:kink_mass_scaling} with a dotted line for some values of $g$, while the overall shape of $\kinkmass(R \to \infty)$ is depicted in Fig.~\ref{fig:string_confs_and_kink_mass}e.
The errors on $\kinkmass$ from this extrapolation are, on average, in the order of $10^{-5}$.

The gap $\Delta$ near a BKT transition $g_{\text{BKT}}$ is known to exponentially close as \cite{kosterlitz1974critical}
\begin{equation}
    \Delta \sim A \exp \qty( - \frac{b}{\sqrt{g - g_{\text{BKT}}}} ),
    \label{eq:bkt_gap_app}
\end{equation}
when coming from the massive phase ($A$ and $b$ are constant to be determined), whereas $\Delta$ vanishes in the critical phase in thermodynamic limit.
If we suppose that $\kinkmass$ indeed quantifies the gap \emph{only in the massive phase}, as argued in Sec.~\ref{sub:restoration_of_rotational_symmetry}, then from comparing $\kinkmass(R \to \infty)$ to \eqref{eq:bkt_gap_app} we should be able to get an estimation of the roughening point $\roughpoint = g_{\text{BKT}}$.
In order to do so, we need to be close enough to the roughening point, but without falling into the critical phase.
For this exact reason that we select only the points in the interval $g \in [0.9, 1.0]$.
The result of the fit is shown in Fig.~\ref{fig:string_confs_and_kink_mass}f, from we have found we find $\roughpoint = 0.907(1)$, $A = 11(2)$, and $b = 1.44(6)$

We warn that the precision about the roughening point $\roughpoint$ should be taken with a grain of salt.
Due to the nebulous relation between the kink mass and the real excitation gap of the string in the vicinity of the transition point, it is still unclear what the proper finite-size scaling procedure should be applied to $\kinkmass$.
For example, we have been unable to apply the techniques developed in \cite{mishra2011phase, dalmonte2015gap}, since $\kinkmass$ does not reflect the gap in the critical phase.
Second, it should be noted that the extrapolation of $\kinkmass(R \to \infty)$ yields in the interval $[0.8, 0.925]$ slightly negative values in the order of $-10^{-3}$ to $-10^{-4}$ with an error in the order of $10^{-5}$.
The magnitude of this error is sensitive to the order of the polynomial used in \eqref{eq:kink_mass_fit}.
We suspect that the issue lies with logarithmic corrections to the gap at the BKT point \cite{woynarovich1987finite-size}.

\section{Numerical methods analysis}
\label{app:numerical_methods_analysis}

In MPS simulations, a typical, problem-agnostic source of error is the approximation of the bond dimension $\chi$.
Nevertheless, our specific problem under study requires additional considerations.
The finiteness of the system considerably influences the outcome of the experiments, even if performed in ideal settings.
For this reason, we need to take into account finite size effects (as already explained in Appendix~\ref{app:finite_size_scaling_for_the_deconfining_transition}).
Since we embed our system in a $2D$ lattice, we have both transversal and longitudinal finite-size effects.
The scaling in those two dimensions allows us to retrieve what would be the correct thermodynamic limit.
Eventually, when studying the dynamics of the model, we have another source of numerical error due to the approximation of the evolution operator, that is, the Trotter error.
In this section, we show all the techniques adopted to obtain the static and dynamic results with a consistent degree of reliability.

\subsection{Vacuum sector analysis}
\label{sub_app:vacuum_sector_analysis}

\begin{figure}[t]
    \centering
    \includegraphics[width=\columnwidth]{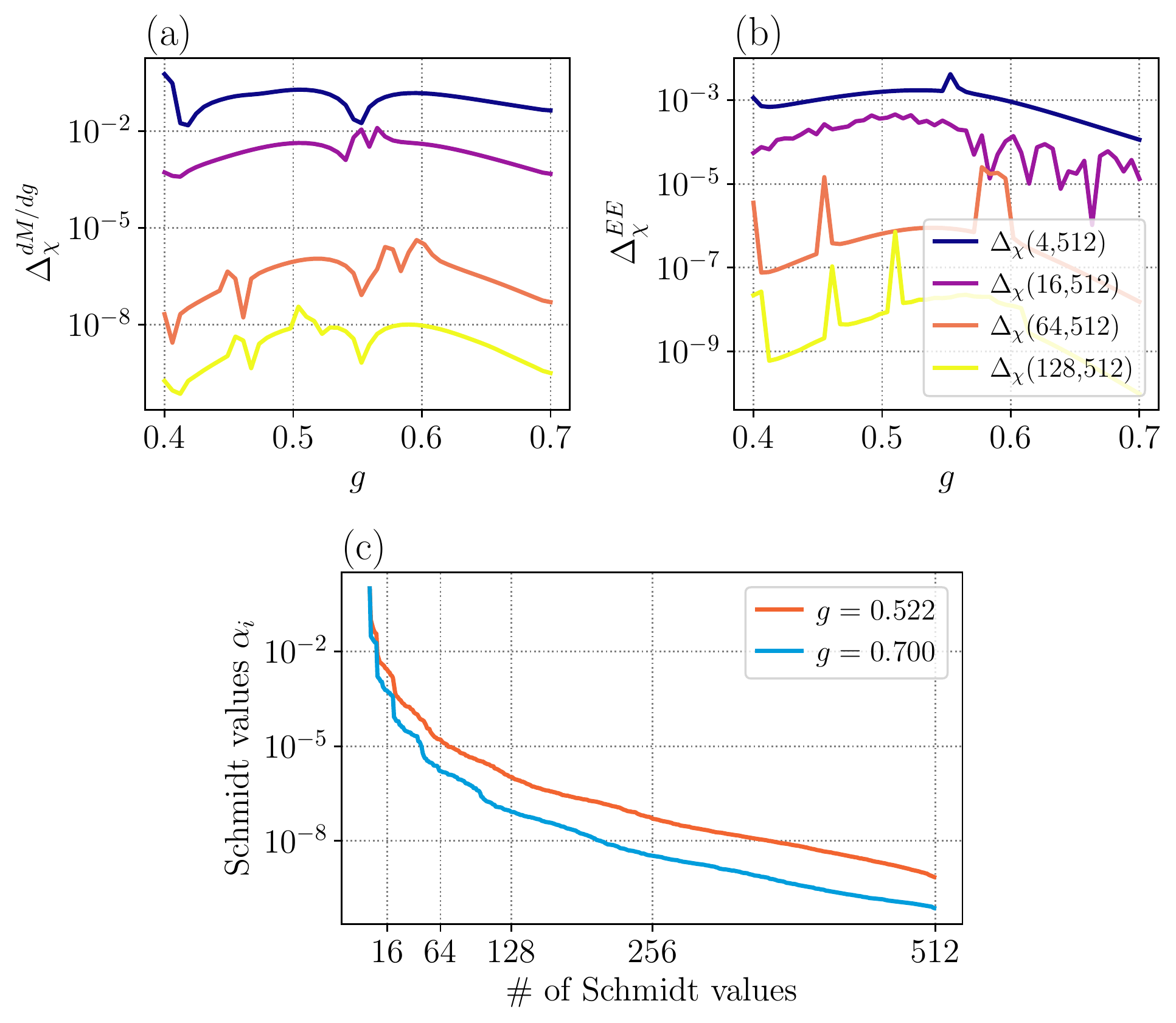}
    \caption{
        Convergence in bond dimension of 't Hooft and EE for $6 \times 6$ lattice.
        \emph{Top}: The difference between the maximum bond dimension $512$ and smaller ones for the Magnetic Susceptibility (left) and Entanglement Entropy (right).
        At $\chi=64$, we have in both quantities a significant convergence of $10^{-5}$ - $10^{-7}$.
        We explain this difference in the order of magnitude because of the phase transition that we are crossing.
        \emph{Bottom}: Schmidt values decay at two different electric couplings $g$.
        At the pseudo-critical point $\critpoint^{(L=6)}$, the system is ``harder'' to approximate than for other $g$ and the tail of Schmidt values decay slower than, e.g., $g=0.7$.}
        \label{fig:err_vac}
\end{figure}

For the results of the vacuum sector presented in the previous Appendix~\ref{app:finite_size_scaling_for_the_deconfining_transition}, we use square \emph{obc} lattices of side $N=L=4,5,6,7$ and maximum bond dimension of $\chi=64$.
As the physical dimension of our MPS is, as shown in Fig.~\ref{fig:cylinder_mps}, $2^N$, the larger the transversal size, the more entanglement will build up, and the larger should be the bond dimension to well represent the true ground state.
Thus, we take the largest lattice of $7 \times 7$ to show its convergence in $\chi$, implying this will be valid for smaller lattices. The convergence of the bond dimension depends on the observable we are analyzing, and can have different rates.
Here, we are interested in the convergence for the two quantities $\dv*{M}{g}$ and the entanglement entropy $S$.
To do so, we look at the difference between a maximum bond dimension and the others $\Delta_{\chi}^{\hat{O}} (\chi_{\text{max}}, \chi)$ for the observable $\hat{O}$ we are interested in.
The results in Fig.~\ref{fig:err_vac} show that this difference goes to $10^{-5}$ - $10^{-7}$ already for bond dimension $\chi=64$ that is the one used in the previous Appendix~\ref{app:finite_size_scaling_for_the_deconfining_transition} to find the critical point $\critpoint$ with great precision.

\begin{figure}[t]
    \centering
    \includegraphics[width=\columnwidth]{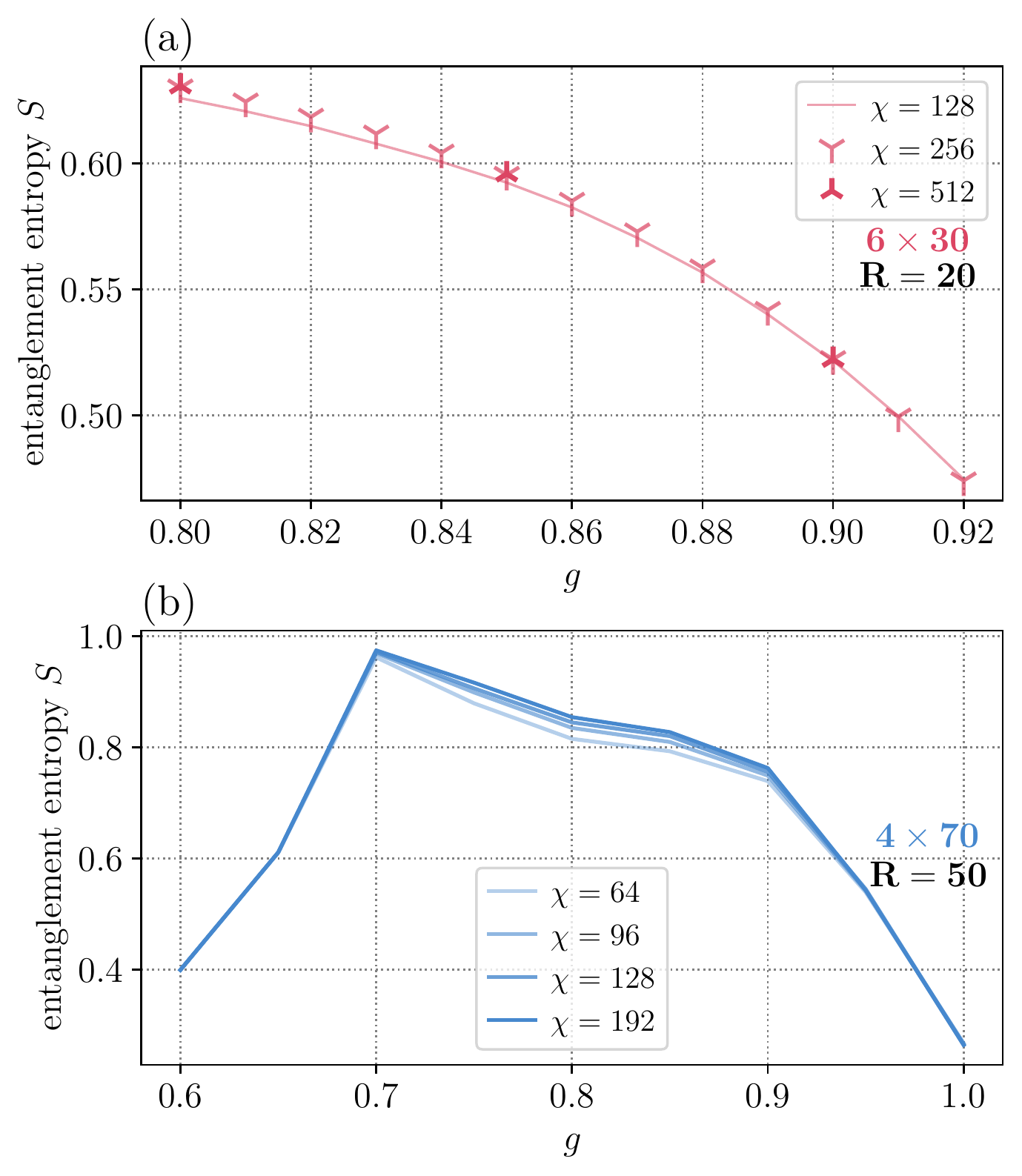}
    \caption{
        Convergence in bond dimension of EE for $6 \times 30$ and $4 \times 70$ lattice.
        \emph{Top}: The lattice with transversal size $N=6$ shows a convergence in bond dimension around $0.90$ for $\chi=128$. For $\chi=256$ we see convergence with $\chi=512$ for all the couplings $g > 0.8$. This is valid for $R=20$. \emph{Bottom}: The same procedure is applied for $N=4$, but we take a larger longitudinal size $L=70$ and maximum string of $R=50$. Here we can appreciate the complexity of the roughening region where the $\chi$ required to correctly represent the ground state of a large string grows.}
        \label{fig:ee_bond_dim}
\end{figure}

\begin{figure}[t]
    \centering
    \includegraphics[width=\columnwidth]{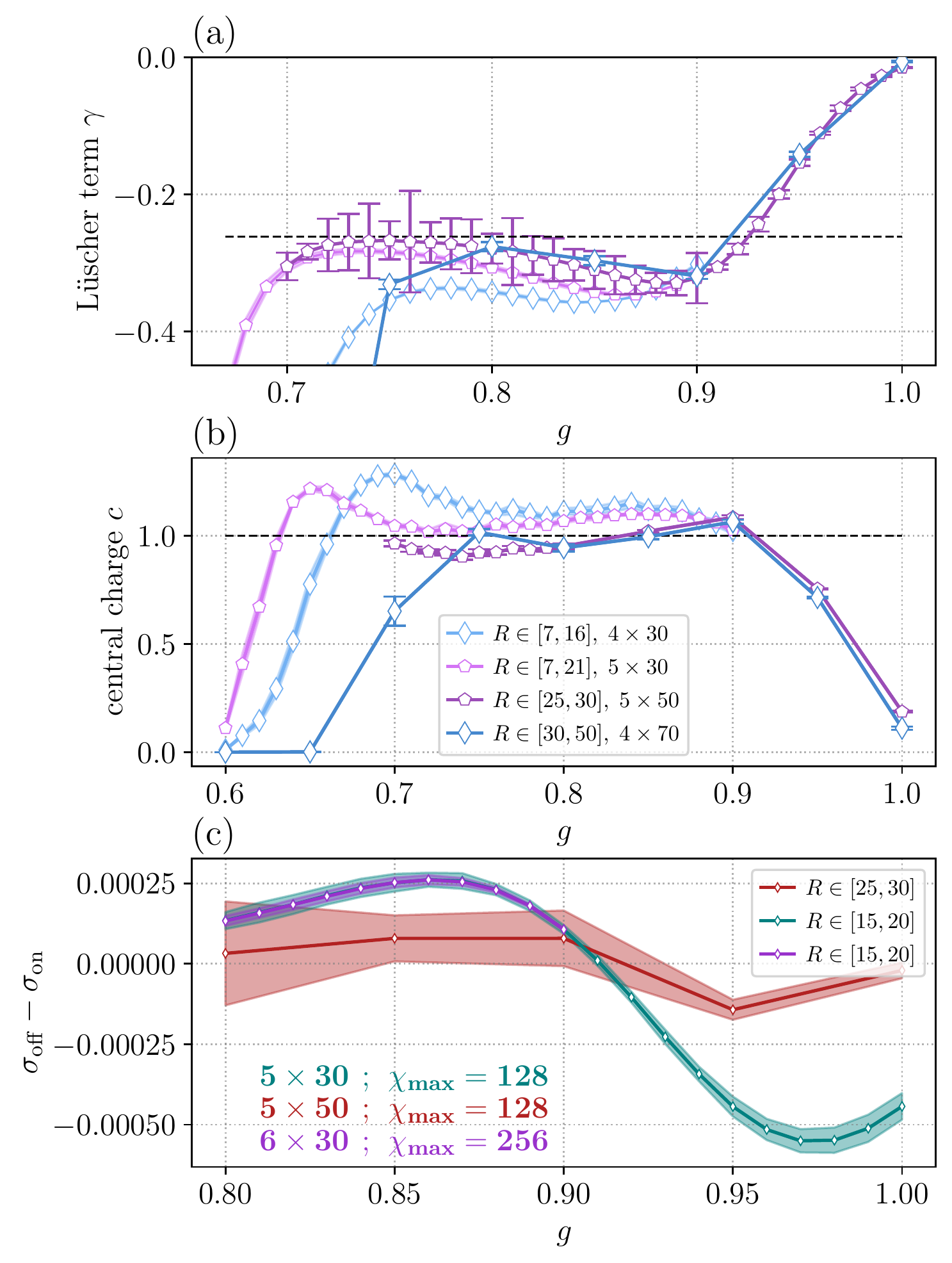}
    \caption{L\"uscher term $\gamma$ and central charge $c$ for lattices increasing in transversal $N$ and longitudinal $L$ size.
    \emph{Top}: The L\"uscher term $\gamma$ is affected by both $N$ and $L$. By increasing N, we increase the extent of the roughening region, by increasing $L$, and consequently $R$, the value of $\gamma(g)$ gets closer to the universal value (black dashed line).
    \emph{Middle}: The same results seem to apply also for the central charge $c$, which is extracted fitting for different $R$ following \eqref{eq:chain_ee}.
    The plateau gets closer to the bosonic string prediction of $c=1$ (black dashed line) as we use larger strings with many transversal degrees of freedom available for fluctuations.
    \emph{Bottom}: The difference in on-axis and off-axis potentials of the string tension $\sigma$.
    We change $N$ and $L$ and see that the onset of the roughening region depends more on the length $R$ of the strings than on the transversal size $N$.}
    \label{fig:gamma_cc_different_R_N}
\end{figure}

\subsection{Two-particle sector analysis}
\label{sub_app:2p_sector_analysis}

The numerics of the two-particle sector makes use of cylindrical lattices, which allows us to reduce finite size effects in the transversal direction.
The effective string symmetry is $U(1)$ in the thermodynamic limit, but for us the transversal fluctuations are truncated and effectively represented by a $\mathbb{Z}_N$ gauge group.
Thus, as already seen for the entropy (inset of Fig.\ref{fig:statics_sw_ee}c) and the string width (Fig.\ref{fig:statics_sw_ee}a), the transversal size $N$ highly influences the observed physics as $g \rightarrow g_c$, the deconfinement point. This is not true for all the coupling values $g$, as for some quantities the results are independent of $N$. Even though the results converge in $N$, most of the quantities we are interested in are valid for the limit of $R \rightarrow \infty$, and thus, need to converge in our longitudinal size $L$. Hence, fixing $N$, we show these two aspects in the following: convergence in bond dimension of the EE and $w^2$ for fixed charge separations $R$, and convergence to the asymptotic limit of the L\"uscher term, central charge, and on-axis vs. off-axis comparisons varying $R$.

Firstly, for a lattice $6 \times 30$ we find a maximum relative error between the string widths $w^2 (g=0.625, R=19, \chi=256)$ and $w^2 (g=0.625, R=19, \chi=128)$ of $4\times10^{-4}$.
Nonetheless, the entropy converges more slowly, and higher bond dimensions are needed (top panel of Fig.~\ref{fig:ee_bond_dim}). We show in the bottom panel of Fig.~\ref{fig:ee_bond_dim} the EE for a cylinder $4 \times 70$ with a string of $R=50$, and $\chi = 64, 96, 128, 192$. Here we can appreciate the computational hardness of the roughening region, that is, where the bond dimension struggles the most to converge.

\begin{figure*}[t]
    \centering
    \includegraphics[width=0.9\textwidth]{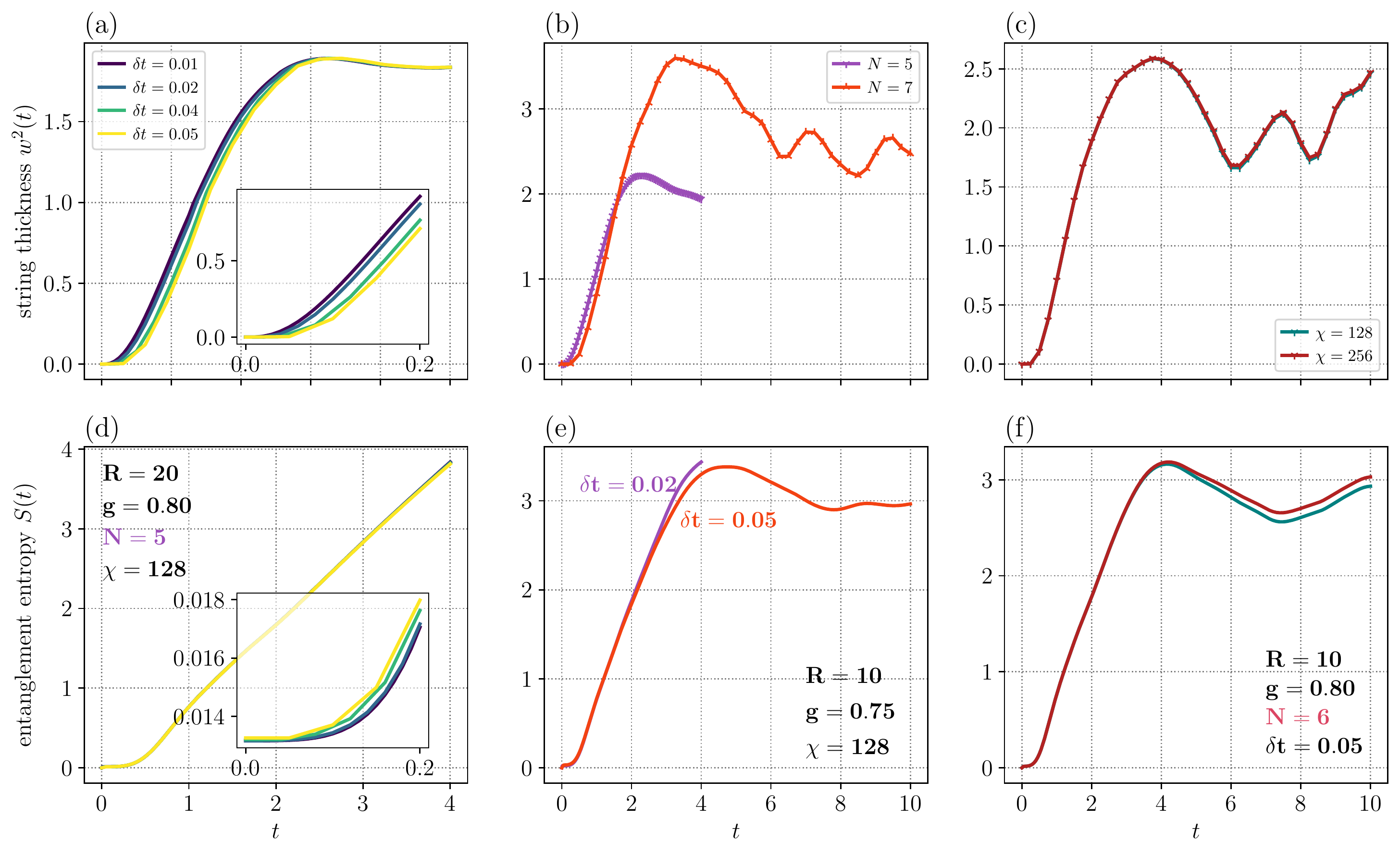}
    \caption{Error dynamics in EE and string width $w^2$ due to finite trotter step $\delta t$, transversal size $N$, and bond dimension $\chi$. On the upper part, we show the string thickness behavior for trotter steps $\delta t \in [0.01,0.05]$ (a), transversal size $N$ from $5$ to $7$ (b), and bond dimension $\chi$ from $128$ to $256$ (c). While for (a) and (b) we see that $w^2$ is affected, it seems to converge nicely with the bond dimension. On the lower part, we present the entanglement entropy in (d), (e), and (f) with the same parameters as above. Here, the entropy shows quite different behaviors. It is converging faster for changing $\delta t$ and $N$, but does not converge in bond dimension for $t \gtrsim 4$.}
    \label{fig:error_dynamics}
\end{figure*}

Then, we address the convergence to the L\"uscher term $\gamma$ and central charge $c$ that should be $-\pi/12$, and $1$, respectively, in the crossover region.
This analysis allows us to see that the extent of the roughening region depends on the transversal size $N$ as depicted in Fig.~\ref{fig:gamma_cc_different_R_N}. The larger the $N$, the more the plateaus in the L\"uscher term and the central charge sweep couplings up to the thermodynamical limit of the deconfining point $g_c$, the actual critical point of our model.
On the other hand, by increasing the length of the cylinder and the separations $R$ considered in the fits to extract $\gamma$ and $c$, we observe a flattening of the plateaus to the expected theoretical values. As a last remark, we also report the difference in the on-axis and off-axis string tensions $\sigma_{\mathrm{off}} - \sigma_{\mathrm{on}}$ at the onset of the critical region for the string. The oscillations present also in the right plot of Fig.~\ref{fig:potential} are not due to the transversal size $N$ (here it is clear that for $N=5$ and $N=6$ the result is unchanged), but they are caused by the short $R$ considered for the fit of the on-axis and off-axis potentials.

\subsection{Dynamical error analysis}
\label{sub_app:dynamical_error_analysis}

In this section, we see the main limitations of our TEBD algorithm. The Trotterization procedure can greatly affect the expectation values of our observables. From Fig.~\ref{fig:error_dynamics}a--d, we see that the string width varies by changing the $\delta t$, with a great influence at short times as shown in the inset, but the entanglement entropy converges faster.
The transversal size also plays a big role in the saturation of the quantum fluctuations of the string. Letting $N$ grow to $7$, we see that the spreading of the string does not seem to stop for a value in the roughening region. This suggests that more studies on the transversal size are needed to understand when the string is saturating because of the coupling or for the finite transversal size.
This is clear in Fig.~\ref{fig:error_dynamics}b--e where $w^2$ seems to saturate around $t=2$ for $N=5$, but the result for larger $N$ clearly shows that this was a finite size effect. Contrarily, the entropy seems to follow a bit better the trend of larger lattices even though they are not converging.
Finally, in Fig.~\ref{fig:error_dynamics}c--f, the result of changing the bond dimension, fixing the $\delta t = 0.05$ and the lattice to $30 \times 6$ shows no particular difference in the trends for a string of length $R=10$. 

\clearpage


\begin{thebibliography}{122}%
\makeatletter
\providecommand \@ifxundefined [1]{%
 \@ifx{#1\undefined}
}%
\providecommand \@ifnum [1]{%
 \ifnum #1\expandafter \@firstoftwo
 \else \expandafter \@secondoftwo
 \fi
}%
\providecommand \@ifx [1]{%
 \ifx #1\expandafter \@firstoftwo
 \else \expandafter \@secondoftwo
 \fi
}%
\providecommand \natexlab [1]{#1}%
\providecommand \enquote  [1]{``#1''}%
\providecommand \bibnamefont  [1]{#1}%
\providecommand \bibfnamefont [1]{#1}%
\providecommand \citenamefont [1]{#1}%
\providecommand \href@noop [0]{\@secondoftwo}%
\providecommand \href [0]{\begingroup \@sanitize@url \@href}%
\providecommand \@href[1]{\@@startlink{#1}\@@href}%
\providecommand \@@href[1]{\endgroup#1\@@endlink}%
\providecommand \@sanitize@url [0]{\catcode `\\12\catcode `\$12\catcode
  `\&12\catcode `\#12\catcode `\^12\catcode `\_12\catcode `\%12\relax}%
\providecommand \@@startlink[1]{}%
\providecommand \@@endlink[0]{}%
\providecommand \url  [0]{\begingroup\@sanitize@url \@url }%
\providecommand \@url [1]{\endgroup\@href {#1}{\urlprefix }}%
\providecommand \urlprefix  [0]{URL }%
\providecommand \Eprint [0]{\href }%
\providecommand \doibase [0]{https://doi.org/}%
\providecommand \selectlanguage [0]{\@gobble}%
\providecommand \bibinfo  [0]{\@secondoftwo}%
\providecommand \bibfield  [0]{\@secondoftwo}%
\providecommand \translation [1]{[#1]}%
\providecommand \BibitemOpen [0]{}%
\providecommand \bibitemStop [0]{}%
\providecommand \bibitemNoStop [0]{.\EOS\space}%
\providecommand \EOS [0]{\spacefactor3000\relax}%
\providecommand \BibitemShut  [1]{\csname bibitem#1\endcsname}%
\let\auto@bib@innerbib\@empty
\bibitem [{\citenamefont {Ba{\~n}uls}\ \emph {et~al.}(2020)\citenamefont
  {Ba{\~n}uls}, \citenamefont {Blatt}, \citenamefont {Catani}, \citenamefont
  {Celi}, \citenamefont {Cirac}, \citenamefont {Dalmonte}, \citenamefont
  {Fallani}, \citenamefont {Jansen}, \citenamefont {Lewenstein}, \citenamefont
  {Montangero} \emph {et~al.}}]{banuls2020simulating}%
  \BibitemOpen
  \bibfield  {author} {\bibinfo {author} {\bibfnamefont {M.~C.}\ \bibnamefont
  {Ba{\~n}uls}}, \bibinfo {author} {\bibfnamefont {R.}~\bibnamefont {Blatt}},
  \bibinfo {author} {\bibfnamefont {J.}~\bibnamefont {Catani}}, \bibinfo
  {author} {\bibfnamefont {A.}~\bibnamefont {Celi}}, \bibinfo {author}
  {\bibfnamefont {J.~I.}\ \bibnamefont {Cirac}}, \bibinfo {author}
  {\bibfnamefont {M.}~\bibnamefont {Dalmonte}}, \bibinfo {author}
  {\bibfnamefont {L.}~\bibnamefont {Fallani}}, \bibinfo {author} {\bibfnamefont
  {K.}~\bibnamefont {Jansen}}, \bibinfo {author} {\bibfnamefont
  {M.}~\bibnamefont {Lewenstein}}, \bibinfo {author} {\bibfnamefont
  {S.}~\bibnamefont {Montangero}}, \emph {et~al.},\ }\bibfield  {title}
  {\bibinfo {title} {Simulating lattice gauge theories within quantum
  technologies},\ }\href {https://doi.org/10.1140/epjd/e2020-100571-8}
  {\bibfield  {journal} {\bibinfo  {journal} {The European Physical Journal D}\
  }\textbf {\bibinfo {volume} {74}},\ \bibinfo {pages} {1} (\bibinfo {year}
  {2020})}\BibitemShut {NoStop}%
\bibitem [{\citenamefont {Davoudi}\ \emph {et~al.}(2020)\citenamefont
  {Davoudi}, \citenamefont {Hafezi}, \citenamefont {Monroe}, \citenamefont
  {Pagano}, \citenamefont {Seif},\ and\ \citenamefont {Shaw}}]{davoudi2020}%
  \BibitemOpen
  \bibfield  {author} {\bibinfo {author} {\bibfnamefont {Z.}~\bibnamefont
  {Davoudi}}, \bibinfo {author} {\bibfnamefont {M.}~\bibnamefont {Hafezi}},
  \bibinfo {author} {\bibfnamefont {C.}~\bibnamefont {Monroe}}, \bibinfo
  {author} {\bibfnamefont {G.}~\bibnamefont {Pagano}}, \bibinfo {author}
  {\bibfnamefont {A.}~\bibnamefont {Seif}},\ and\ \bibinfo {author}
  {\bibfnamefont {A.}~\bibnamefont {Shaw}},\ }\bibfield  {title} {\bibinfo
  {title} {Towards analog quantum simulations of lattice gauge theories with
  trapped ions},\ }\href {https://doi.org/10.1103/PhysRevResearch.2.023015}
  {\bibfield  {journal} {\bibinfo  {journal} {Phys. Rev. Res.}\ }\textbf
  {\bibinfo {volume} {2}},\ \bibinfo {pages} {023015} (\bibinfo {year}
  {2020})},\ \Eprint {https://arxiv.org/abs/1908.03210} {arXiv:1908.03210
  [quant-ph]} \BibitemShut {NoStop}%
\bibitem [{\citenamefont {Funcke}\ \emph {et~al.}(2023)\citenamefont {Funcke},
  \citenamefont {Hartung}, \citenamefont {Jansen},\ and\ \citenamefont
  {Kühn}}]{Funcke2023}%
  \BibitemOpen
  \bibfield  {author} {\bibinfo {author} {\bibfnamefont {L.}~\bibnamefont
  {Funcke}}, \bibinfo {author} {\bibfnamefont {T.}~\bibnamefont {Hartung}},
  \bibinfo {author} {\bibfnamefont {K.}~\bibnamefont {Jansen}},\ and\ \bibinfo
  {author} {\bibfnamefont {S.}~\bibnamefont {Kühn}},\ }\bibfield  {title}
  {\bibinfo {title} {{Review on Quantum Computing for Lattice Field Theory}},\
  }\href {https://doi.org/10.22323/1.430.0228} {\bibfield  {journal} {\bibinfo
  {journal} {Proceedings of The 39th International Symposium on Lattice Field
  Theory {\textemdash} PoS(LATTICE2022)}\ }\textbf {\bibinfo {volume} {430}},\
  \bibinfo {pages} {228} (\bibinfo {year} {2023})},\ \Eprint
  {https://arxiv.org/abs/2302.00467} {arXiv:2302.00467 [hep-lat]} \BibitemShut
  {NoStop}%
\bibitem [{\citenamefont {Bass}\ and\ \citenamefont
  {Zohar}(2022)}]{bass2021qtech}%
  \BibitemOpen
  \bibfield  {author} {\bibinfo {author} {\bibfnamefont {S.~D.}\ \bibnamefont
  {Bass}}\ and\ \bibinfo {author} {\bibfnamefont {E.}~\bibnamefont {Zohar}},\
  }\bibfield  {title} {\bibinfo {title} {Quantum technologies in particle
  physics},\ }\href {https://doi.org/https://doi.org/10.1098/rsta.2021.0072}
  {\bibfield  {journal} {\bibinfo  {journal} {Phil. Trans. R. Soc. A}\ }\textbf
  {\bibinfo {volume} {380}},\ \bibinfo {pages} {20210072} (\bibinfo {year}
  {2022})}\BibitemShut {NoStop}%
\bibitem [{\citenamefont {Bauer}\ \emph {et~al.}(2023)\citenamefont {Bauer}
  \emph {et~al.}}]{bauer2023quantum}%
  \BibitemOpen
  \bibfield  {author} {\bibinfo {author} {\bibfnamefont {C.~W.}\ \bibnamefont
  {Bauer}} \emph {et~al.},\ }\bibfield  {title} {\bibinfo {title} {Quantum
  simulation for high-energy physics},\ }\bibfield  {journal} {\bibinfo
  {journal} {PRX Quantum}\ }\textbf {\bibinfo {volume} {4}},\ \href
  {https://doi.org/10.1103/prxquantum.4.027001} {10.1103/prxquantum.4.027001}
  (\bibinfo {year} {2023})\BibitemShut {NoStop}%
\bibitem [{\citenamefont {Halimeh}\ \emph {et~al.}(2023)\citenamefont
  {Halimeh}, \citenamefont {Aidelsburger}, \citenamefont {Grusdt},
  \citenamefont {Hauke},\ and\ \citenamefont {Yang}}]{halimeh2023coldatom}%
  \BibitemOpen
  \bibfield  {author} {\bibinfo {author} {\bibfnamefont {J.~C.}\ \bibnamefont
  {Halimeh}}, \bibinfo {author} {\bibfnamefont {M.}~\bibnamefont
  {Aidelsburger}}, \bibinfo {author} {\bibfnamefont {F.}~\bibnamefont
  {Grusdt}}, \bibinfo {author} {\bibfnamefont {P.}~\bibnamefont {Hauke}},\ and\
  \bibinfo {author} {\bibfnamefont {B.}~\bibnamefont {Yang}},\ }\href@noop {}
  {\bibinfo {title} {{Cold-atom quantum simulators of gauge theories}}}
  (\bibinfo {year} {2023}),\ \Eprint {https://arxiv.org/abs/2310.12201}
  {arXiv:2310.12201} \BibitemShut {NoStop}%
\bibitem [{\citenamefont {Di~Meglio}\ \emph {et~al.}(2024)\citenamefont
  {Di~Meglio} \emph {et~al.}}]{di-meglio2024quantum}%
  \BibitemOpen
  \bibfield  {author} {\bibinfo {author} {\bibfnamefont {A.}~\bibnamefont
  {Di~Meglio}} \emph {et~al.},\ }\bibfield  {title} {\bibinfo {title} {Quantum
  computing for high-energy physics: State of the art and challenges},\
  }\bibfield  {journal} {\bibinfo  {journal} {PRX Quantum}\ }\textbf {\bibinfo
  {volume} {5}},\ \href {https://doi.org/10.1103/prxquantum.5.037001}
  {10.1103/prxquantum.5.037001} (\bibinfo {year} {2024})\BibitemShut {NoStop}%
\bibitem [{\citenamefont {Martinez}\ \emph {et~al.}(2016)\citenamefont
  {Martinez}, \citenamefont {Muschik}, \citenamefont {Schindler}, \citenamefont
  {Nigg}, \citenamefont {Erhard}, \citenamefont {Heyl}, \citenamefont {Hauke},
  \citenamefont {Dalmonte}, \citenamefont {Monz}, \citenamefont {Zoller},\ and\
  \citenamefont {Blatt}}]{martinez2016}%
  \BibitemOpen
  \bibfield  {author} {\bibinfo {author} {\bibfnamefont {E.~A.}\ \bibnamefont
  {Martinez}}, \bibinfo {author} {\bibfnamefont {C.~A.}\ \bibnamefont
  {Muschik}}, \bibinfo {author} {\bibfnamefont {P.}~\bibnamefont {Schindler}},
  \bibinfo {author} {\bibfnamefont {D.}~\bibnamefont {Nigg}}, \bibinfo {author}
  {\bibfnamefont {A.}~\bibnamefont {Erhard}}, \bibinfo {author} {\bibfnamefont
  {M.}~\bibnamefont {Heyl}}, \bibinfo {author} {\bibfnamefont {P.}~\bibnamefont
  {Hauke}}, \bibinfo {author} {\bibfnamefont {M.}~\bibnamefont {Dalmonte}},
  \bibinfo {author} {\bibfnamefont {T.}~\bibnamefont {Monz}}, \bibinfo {author}
  {\bibfnamefont {P.}~\bibnamefont {Zoller}},\ and\ \bibinfo {author}
  {\bibfnamefont {R.}~\bibnamefont {Blatt}},\ }\bibfield  {title} {\bibinfo
  {title} {Real-time dynamics of lattice gauge theories with a few-qubit
  quantum computer},\ }\href@noop {} {\bibfield  {journal} {\bibinfo  {journal}
  {Nature}\ }\textbf {\bibinfo {volume} {534}},\ \bibinfo {pages} {516}
  (\bibinfo {year} {2016})}\BibitemShut {NoStop}%
\bibitem [{\citenamefont {Schweizer}\ \emph {et~al.}(2019)\citenamefont
  {Schweizer}, \citenamefont {Grusdt}, \citenamefont {Berngruber},
  \citenamefont {Barbiero}, \citenamefont {Demler}, \citenamefont {Goldman},
  \citenamefont {Bloch},\ and\ \citenamefont {Aidelsburger}}]{schweizer2019}%
  \BibitemOpen
  \bibfield  {author} {\bibinfo {author} {\bibfnamefont {C.}~\bibnamefont
  {Schweizer}}, \bibinfo {author} {\bibfnamefont {F.}~\bibnamefont {Grusdt}},
  \bibinfo {author} {\bibfnamefont {M.}~\bibnamefont {Berngruber}}, \bibinfo
  {author} {\bibfnamefont {L.}~\bibnamefont {Barbiero}}, \bibinfo {author}
  {\bibfnamefont {E.}~\bibnamefont {Demler}}, \bibinfo {author} {\bibfnamefont
  {N.}~\bibnamefont {Goldman}}, \bibinfo {author} {\bibfnamefont
  {I.}~\bibnamefont {Bloch}},\ and\ \bibinfo {author} {\bibfnamefont
  {M.}~\bibnamefont {Aidelsburger}},\ }\bibfield  {title} {\bibinfo {title}
  {Floquet approach to $\mathbb{Z}_2$ lattice gauge theories with ultracold
  atoms in optical lattices},\ }\href@noop {} {\bibfield  {journal} {\bibinfo
  {journal} {Nature Physics}\ }\textbf {\bibinfo {volume} {15}},\ \bibinfo
  {pages} {1168} (\bibinfo {year} {2019})}\BibitemShut {NoStop}%
\bibitem [{\citenamefont {Lu}\ \emph {et~al.}(2019)\citenamefont {Lu},
  \citenamefont {Klco}, \citenamefont {Lukens}, \citenamefont {Morris},
  \citenamefont {Bansal}, \citenamefont {Ekstr\"om}, \citenamefont {Hagen},
  \citenamefont {Papenbrock}, \citenamefont {Weiner}, \citenamefont {Savage},\
  and\ \citenamefont {Lougovski}}]{lu2019phot}%
  \BibitemOpen
  \bibfield  {author} {\bibinfo {author} {\bibfnamefont {H.-H.}\ \bibnamefont
  {Lu}}, \bibinfo {author} {\bibfnamefont {N.}~\bibnamefont {Klco}}, \bibinfo
  {author} {\bibfnamefont {J.~M.}\ \bibnamefont {Lukens}}, \bibinfo {author}
  {\bibfnamefont {T.~D.}\ \bibnamefont {Morris}}, \bibinfo {author}
  {\bibfnamefont {A.}~\bibnamefont {Bansal}}, \bibinfo {author} {\bibfnamefont
  {A.}~\bibnamefont {Ekstr\"om}}, \bibinfo {author} {\bibfnamefont
  {G.}~\bibnamefont {Hagen}}, \bibinfo {author} {\bibfnamefont
  {T.}~\bibnamefont {Papenbrock}}, \bibinfo {author} {\bibfnamefont {A.~M.}\
  \bibnamefont {Weiner}}, \bibinfo {author} {\bibfnamefont {M.~J.}\
  \bibnamefont {Savage}},\ and\ \bibinfo {author} {\bibfnamefont
  {P.}~\bibnamefont {Lougovski}},\ }\bibfield  {title} {\bibinfo {title}
  {Simulations of subatomic many-body physics on a quantum frequency
  processor},\ }\href {https://doi.org/10.1103/PhysRevA.100.012320} {\bibfield
  {journal} {\bibinfo  {journal} {Phys. Rev. A}\ }\textbf {\bibinfo {volume}
  {100}},\ \bibinfo {pages} {012320} (\bibinfo {year} {2019})}\BibitemShut
  {NoStop}%
\bibitem [{\citenamefont {Yang}\ \emph {et~al.}(2020)\citenamefont {Yang},
  \citenamefont {Sun}, \citenamefont {Ott}, \citenamefont {Wang}, \citenamefont
  {Zache}, \citenamefont {Halimeh}, \citenamefont {Yuan}, \citenamefont
  {Hauke},\ and\ \citenamefont {Pan}}]{yang2020observation}%
  \BibitemOpen
  \bibfield  {author} {\bibinfo {author} {\bibfnamefont {B.}~\bibnamefont
  {Yang}}, \bibinfo {author} {\bibfnamefont {H.}~\bibnamefont {Sun}}, \bibinfo
  {author} {\bibfnamefont {R.}~\bibnamefont {Ott}}, \bibinfo {author}
  {\bibfnamefont {H.-Y.}\ \bibnamefont {Wang}}, \bibinfo {author}
  {\bibfnamefont {T.~V.}\ \bibnamefont {Zache}}, \bibinfo {author}
  {\bibfnamefont {J.~C.}\ \bibnamefont {Halimeh}}, \bibinfo {author}
  {\bibfnamefont {Z.-S.}\ \bibnamefont {Yuan}}, \bibinfo {author}
  {\bibfnamefont {P.}~\bibnamefont {Hauke}},\ and\ \bibinfo {author}
  {\bibfnamefont {J.-W.}\ \bibnamefont {Pan}},\ }\bibfield  {title} {\bibinfo
  {title} {Observation of gauge invariance in a 71-site bose–hubbard quantum
  simulator},\ }\href {https://doi.org/10.1038/s41586-020-2910-8} {\bibfield
  {journal} {\bibinfo  {journal} {Nature}\ }\textbf {\bibinfo {volume} {587}},\
  \bibinfo {pages} {392–396} (\bibinfo {year} {2020})}\BibitemShut {NoStop}%
\bibitem [{\citenamefont {Mil}\ \emph {et~al.}(2020)\citenamefont {Mil},
  \citenamefont {Zache}, \citenamefont {Hegde}, \citenamefont {Xia},
  \citenamefont {Bhatt}, \citenamefont {Oberthaler}, \citenamefont {Hauke},
  \citenamefont {Berges},\ and\ \citenamefont {Jendrzejewski}}]{mil2020u1}%
  \BibitemOpen
  \bibfield  {author} {\bibinfo {author} {\bibfnamefont {A.}~\bibnamefont
  {Mil}}, \bibinfo {author} {\bibfnamefont {T.~V.}\ \bibnamefont {Zache}},
  \bibinfo {author} {\bibfnamefont {A.}~\bibnamefont {Hegde}}, \bibinfo
  {author} {\bibfnamefont {A.}~\bibnamefont {Xia}}, \bibinfo {author}
  {\bibfnamefont {R.~P.}\ \bibnamefont {Bhatt}}, \bibinfo {author}
  {\bibfnamefont {M.~K.}\ \bibnamefont {Oberthaler}}, \bibinfo {author}
  {\bibfnamefont {P.}~\bibnamefont {Hauke}}, \bibinfo {author} {\bibfnamefont
  {J.}~\bibnamefont {Berges}},\ and\ \bibinfo {author} {\bibfnamefont
  {F.}~\bibnamefont {Jendrzejewski}},\ }\bibfield  {title} {\bibinfo {title} {A
  scalable realization of local u(1) gauge invariance in cold atomic
  mixtures},\ }\href@noop {} {\bibfield  {journal} {\bibinfo  {journal}
  {Science}\ }\textbf {\bibinfo {volume} {367}},\ \bibinfo {pages} {1128}
  (\bibinfo {year} {2020})}\BibitemShut {NoStop}%
\bibitem [{\citenamefont {McCoy}\ and\ \citenamefont
  {Wu}(1978)}]{mccoy1978two-dimensional}%
  \BibitemOpen
  \bibfield  {author} {\bibinfo {author} {\bibfnamefont {B.~M.}\ \bibnamefont
  {McCoy}}\ and\ \bibinfo {author} {\bibfnamefont {T.~T.}\ \bibnamefont {Wu}},\
  }\bibfield  {title} {\bibinfo {title} {Two-dimensional ising field theory in
  a magnetic field: Breakup of the cut in the two-point function},\ }\href
  {https://doi.org/10.1103/physrevd.18.1259} {\bibfield  {journal} {\bibinfo
  {journal} {Physical Review D}\ }\textbf {\bibinfo {volume} {18}},\ \bibinfo
  {pages} {1259–1267} (\bibinfo {year} {1978})}\BibitemShut {NoStop}%
\bibitem [{\citenamefont {Rutkevich}(2008)}]{rutkevich2008energy}%
  \BibitemOpen
  \bibfield  {author} {\bibinfo {author} {\bibfnamefont {S.~B.}\ \bibnamefont
  {Rutkevich}},\ }\bibfield  {title} {\bibinfo {title} {Energy spectrum of
  bound-spinons in the quantum ising spin-chain ferromagnet},\ }\href
  {https://doi.org/10.1007/s10955-008-9495-1} {\bibfield  {journal} {\bibinfo
  {journal} {Journal of Statistical Physics}\ }\textbf {\bibinfo {volume}
  {131}},\ \bibinfo {pages} {917–939} (\bibinfo {year} {2008})}\BibitemShut
  {NoStop}%
\bibitem [{\citenamefont {Kormos}\ \emph {et~al.}(2017)\citenamefont {Kormos},
  \citenamefont {Collura}, \citenamefont {Tak{\'a}cs},\ and\ \citenamefont
  {Calabrese}}]{kormos2017confinement}%
  \BibitemOpen
  \bibfield  {author} {\bibinfo {author} {\bibfnamefont {M.}~\bibnamefont
  {Kormos}}, \bibinfo {author} {\bibfnamefont {M.}~\bibnamefont {Collura}},
  \bibinfo {author} {\bibfnamefont {G.}~\bibnamefont {Tak{\'a}cs}},\ and\
  \bibinfo {author} {\bibfnamefont {P.}~\bibnamefont {Calabrese}},\ }\bibfield
  {title} {\bibinfo {title} {Real-time confinement following a quantum quench
  to a non-integrable model},\ }\href {https://doi.org/10.1038/nphys3934}
  {\bibfield  {journal} {\bibinfo  {journal} {Nature Physics}\ }\textbf
  {\bibinfo {volume} {13}},\ \bibinfo {pages} {246} (\bibinfo {year}
  {2017})}\BibitemShut {NoStop}%
\bibitem [{\citenamefont {Lake}\ \emph {et~al.}(2009)\citenamefont {Lake},
  \citenamefont {Tsvelik}, \citenamefont {Notbohm}, \citenamefont
  {Alan~Tennant}, \citenamefont {Perring}, \citenamefont {Reehuis},
  \citenamefont {Sekar}, \citenamefont {Krabbes},\ and\ \citenamefont
  {Büchner}}]{lake2009confinement}%
  \BibitemOpen
  \bibfield  {author} {\bibinfo {author} {\bibfnamefont {B.}~\bibnamefont
  {Lake}}, \bibinfo {author} {\bibfnamefont {A.~M.}\ \bibnamefont {Tsvelik}},
  \bibinfo {author} {\bibfnamefont {S.}~\bibnamefont {Notbohm}}, \bibinfo
  {author} {\bibfnamefont {D.}~\bibnamefont {Alan~Tennant}}, \bibinfo {author}
  {\bibfnamefont {T.~G.}\ \bibnamefont {Perring}}, \bibinfo {author}
  {\bibfnamefont {M.}~\bibnamefont {Reehuis}}, \bibinfo {author} {\bibfnamefont
  {C.}~\bibnamefont {Sekar}}, \bibinfo {author} {\bibfnamefont
  {G.}~\bibnamefont {Krabbes}},\ and\ \bibinfo {author} {\bibfnamefont
  {B.}~\bibnamefont {Büchner}},\ }\bibfield  {title} {\bibinfo {title}
  {Confinement of fractional quantum number particles in a condensed-matter
  system},\ }\href {https://doi.org/10.1038/nphys1462} {\bibfield  {journal}
  {\bibinfo  {journal} {Nature Physics}\ }\textbf {\bibinfo {volume} {6}},\
  \bibinfo {pages} {50–55} (\bibinfo {year} {2009})}\BibitemShut {NoStop}%
\bibitem [{\citenamefont {Tan}\ \emph {et~al.}(2021)\citenamefont {Tan},
  \citenamefont {Becker}, \citenamefont {Liu}, \citenamefont {Pagano},
  \citenamefont {Collins}, \citenamefont {De}, \citenamefont {Feng},
  \citenamefont {Kaplan}, \citenamefont {Kyprianidis}, \citenamefont
  {Lundgren}, \citenamefont {Morong}, \citenamefont {Whitsitt}, \citenamefont
  {Gorshkov},\ and\ \citenamefont {Monroe}}]{tan2021domain-wall}%
  \BibitemOpen
  \bibfield  {author} {\bibinfo {author} {\bibfnamefont {W.~L.}\ \bibnamefont
  {Tan}}, \bibinfo {author} {\bibfnamefont {P.}~\bibnamefont {Becker}},
  \bibinfo {author} {\bibfnamefont {F.}~\bibnamefont {Liu}}, \bibinfo {author}
  {\bibfnamefont {G.}~\bibnamefont {Pagano}}, \bibinfo {author} {\bibfnamefont
  {K.~S.}\ \bibnamefont {Collins}}, \bibinfo {author} {\bibfnamefont
  {A.}~\bibnamefont {De}}, \bibinfo {author} {\bibfnamefont {L.}~\bibnamefont
  {Feng}}, \bibinfo {author} {\bibfnamefont {H.~B.}\ \bibnamefont {Kaplan}},
  \bibinfo {author} {\bibfnamefont {A.}~\bibnamefont {Kyprianidis}}, \bibinfo
  {author} {\bibfnamefont {R.}~\bibnamefont {Lundgren}}, \bibinfo {author}
  {\bibfnamefont {W.}~\bibnamefont {Morong}}, \bibinfo {author} {\bibfnamefont
  {S.}~\bibnamefont {Whitsitt}}, \bibinfo {author} {\bibfnamefont {A.~V.}\
  \bibnamefont {Gorshkov}},\ and\ \bibinfo {author} {\bibfnamefont
  {C.}~\bibnamefont {Monroe}},\ }\bibfield  {title} {\bibinfo {title}
  {Domain-wall confinement and dynamics in a quantum simulator},\ }\href
  {https://doi.org/10.1038/s41567-021-01194-3} {\bibfield  {journal} {\bibinfo
  {journal} {Nature Physics}\ }\textbf {\bibinfo {volume} {17}},\ \bibinfo
  {pages} {742–747} (\bibinfo {year} {2021})}\BibitemShut {NoStop}%
\bibitem [{\citenamefont {Lüscher}\ \emph {et~al.}(1981)\citenamefont
  {Lüscher}, \citenamefont {Münster},\ and\ \citenamefont
  {Weisz}}]{luscher1981thick}%
  \BibitemOpen
  \bibfield  {author} {\bibinfo {author} {\bibfnamefont {M.}~\bibnamefont
  {Lüscher}}, \bibinfo {author} {\bibfnamefont {G.}~\bibnamefont {Münster}},\
  and\ \bibinfo {author} {\bibfnamefont {P.}~\bibnamefont {Weisz}},\ }\bibfield
   {title} {\bibinfo {title} {How thick are chromo-electric flux tubes?},\
  }\href {https://doi.org/10.1016/0550-3213(81)90151-6} {\bibfield  {journal}
  {\bibinfo  {journal} {Nuclear Physics B}\ }\textbf {\bibinfo {volume}
  {180}},\ \bibinfo {pages} {1–12} (\bibinfo {year} {1981})}\BibitemShut
  {NoStop}%
\bibitem [{\citenamefont {Hasenfratz}\ \emph {et~al.}(1981)\citenamefont
  {Hasenfratz}, \citenamefont {Hasenfratz},\ and\ \citenamefont
  {Hasenfratz}}]{hasenfratz1981generalized}%
  \BibitemOpen
  \bibfield  {author} {\bibinfo {author} {\bibfnamefont {A.}~\bibnamefont
  {Hasenfratz}}, \bibinfo {author} {\bibfnamefont {E.}~\bibnamefont
  {Hasenfratz}},\ and\ \bibinfo {author} {\bibfnamefont {P.}~\bibnamefont
  {Hasenfratz}},\ }\bibfield  {title} {\bibinfo {title} {Generalized roughening
  transition and its effect on the string tension},\ }\href
  {https://doi.org/10.1016/0550-3213(81)90426-0} {\bibfield  {journal}
  {\bibinfo  {journal} {Nuclear Physics B}\ }\textbf {\bibinfo {volume}
  {180}},\ \bibinfo {pages} {353–367} (\bibinfo {year} {1981})}\BibitemShut
  {NoStop}%
\bibitem [{\citenamefont {Lüscher}(1981)}]{luscher1981symmetry-breaking}%
  \BibitemOpen
  \bibfield  {author} {\bibinfo {author} {\bibfnamefont {M.}~\bibnamefont
  {Lüscher}},\ }\bibfield  {title} {\bibinfo {title} {Symmetry-breaking
  aspects of the roughening transition in gauge theories},\ }\href
  {https://doi.org/10.1016/0550-3213(81)90423-5} {\bibfield  {journal}
  {\bibinfo  {journal} {Nuclear Physics B}\ }\textbf {\bibinfo {volume}
  {180}},\ \bibinfo {pages} {317–329} (\bibinfo {year} {1981})}\BibitemShut
  {NoStop}%
\bibitem [{\citenamefont {Münster}\ and\ \citenamefont
  {Weisz}(1981{\natexlab{a}})}]{munster1981roughening}%
  \BibitemOpen
  \bibfield  {author} {\bibinfo {author} {\bibfnamefont {G.}~\bibnamefont
  {Münster}}\ and\ \bibinfo {author} {\bibfnamefont {P.}~\bibnamefont
  {Weisz}},\ }\bibfield  {title} {\bibinfo {title} {On the roughening
  transition in abelian lattice gauge theories},\ }\href
  {https://doi.org/10.1016/0550-3213(81)90152-8} {\bibfield  {journal}
  {\bibinfo  {journal} {Nuclear Physics B}\ }\textbf {\bibinfo {volume}
  {180}},\ \bibinfo {pages} {13–22} (\bibinfo {year}
  {1981}{\natexlab{a}})}\BibitemShut {NoStop}%
\bibitem [{\citenamefont {Münster}\ and\ \citenamefont
  {Weisz}(1981{\natexlab{b}})}]{munster1981nonabelian}%
  \BibitemOpen
  \bibfield  {author} {\bibinfo {author} {\bibfnamefont {G.}~\bibnamefont
  {Münster}}\ and\ \bibinfo {author} {\bibfnamefont {P.}~\bibnamefont
  {Weisz}},\ }\bibfield  {title} {\bibinfo {title} {On the roughening
  transition in non-abelian lattice gauge theories},\ }\href
  {https://doi.org/10.1016/0550-3213(81)90424-7} {\bibfield  {journal}
  {\bibinfo  {journal} {Nuclear Physics B}\ }\textbf {\bibinfo {volume}
  {180}},\ \bibinfo {pages} {330–340} (\bibinfo {year}
  {1981}{\natexlab{b}})}\BibitemShut {NoStop}%
\bibitem [{\citenamefont {Drouffe}\ and\ \citenamefont
  {Zuber}(1981{\natexlab{a}})}]{drouffe1981rougheningI}%
  \BibitemOpen
  \bibfield  {author} {\bibinfo {author} {\bibfnamefont {J.}~\bibnamefont
  {Drouffe}}\ and\ \bibinfo {author} {\bibfnamefont {J.}~\bibnamefont
  {Zuber}},\ }\bibfield  {title} {\bibinfo {title} {Roughening transition in
  lattice gauge theories in arbitrary dimension: (i). the {$Z_2$} case},\
  }\href {https://doi.org/10.1016/0550-3213(81)90418-1} {\bibfield  {journal}
  {\bibinfo  {journal} {Nuclear Physics B}\ }\textbf {\bibinfo {volume}
  {180}},\ \bibinfo {pages} {253–263} (\bibinfo {year}
  {1981}{\natexlab{a}})}\BibitemShut {NoStop}%
\bibitem [{\citenamefont {Drouffe}\ and\ \citenamefont
  {Zuber}(1981{\natexlab{b}})}]{drouffe1981rougheningII}%
  \BibitemOpen
  \bibfield  {author} {\bibinfo {author} {\bibfnamefont {J.}~\bibnamefont
  {Drouffe}}\ and\ \bibinfo {author} {\bibfnamefont {J.}~\bibnamefont
  {Zuber}},\ }\bibfield  {title} {\bibinfo {title} {Roughening transition in
  lattice gauge theories in arbitrary dimension: (ii). the groups {$Z_3$},
  {$U(1)$}, {$SU(2)$}, {$SU(3)$}},\ }\href
  {https://doi.org/10.1016/0550-3213(81)90419-3} {\bibfield  {journal}
  {\bibinfo  {journal} {Nuclear Physics B}\ }\textbf {\bibinfo {volume}
  {180}},\ \bibinfo {pages} {264–274} (\bibinfo {year}
  {1981}{\natexlab{b}})}\BibitemShut {NoStop}%
\bibitem [{\citenamefont {Chui}\ and\ \citenamefont
  {Weeks}(1976)}]{chui1976transition}%
  \BibitemOpen
  \bibfield  {author} {\bibinfo {author} {\bibfnamefont {S.~T.}\ \bibnamefont
  {Chui}}\ and\ \bibinfo {author} {\bibfnamefont {J.~D.}\ \bibnamefont
  {Weeks}},\ }\bibfield  {title} {\bibinfo {title} {Phase transition in the
  two-dimensional coulomb gas, and the interfacial roughening transition},\
  }\href {https://doi.org/10.1103/physrevb.14.4978} {\bibfield  {journal}
  {\bibinfo  {journal} {Physical Review B}\ }\textbf {\bibinfo {volume} {14}},\
  \bibinfo {pages} {4978–4982} (\bibinfo {year} {1976})}\BibitemShut
  {NoStop}%
\bibitem [{\citenamefont {Fradkin}(1983)}]{fradkin1983roughening}%
  \BibitemOpen
  \bibfield  {author} {\bibinfo {author} {\bibfnamefont {E.}~\bibnamefont
  {Fradkin}},\ }\bibfield  {title} {\bibinfo {title} {Roughening transition in
  quantum interfaces},\ }\href {https://doi.org/10.1103/physrevb.28.5338}
  {\bibfield  {journal} {\bibinfo  {journal} {Physical Review B}\ }\textbf
  {\bibinfo {volume} {28}},\ \bibinfo {pages} {5338–5341} (\bibinfo {year}
  {1983})}\BibitemShut {NoStop}%
\bibitem [{\citenamefont {Hasenbusch}\ and\ \citenamefont
  {Pinn}(1997)}]{hasenbusch1997roughening}%
  \BibitemOpen
  \bibfield  {author} {\bibinfo {author} {\bibfnamefont {M.}~\bibnamefont
  {Hasenbusch}}\ and\ \bibinfo {author} {\bibfnamefont {K.}~\bibnamefont
  {Pinn}},\ }\bibfield  {title} {\bibinfo {title} {Computing the roughening
  transition of ising and solid-on-solid models by bcsos model matching},\
  }\href {https://doi.org/10.1088/0305-4470/30/1/006} {\bibfield  {journal}
  {\bibinfo  {journal} {Journal of Physics A: Mathematical and General}\
  }\textbf {\bibinfo {volume} {30}},\ \bibinfo {pages} {63–80} (\bibinfo
  {year} {1997})}\BibitemShut {NoStop}%
\bibitem [{\citenamefont {Caselle}\ \emph {et~al.}(1996)\citenamefont
  {Caselle}, \citenamefont {Gliozzi}, \citenamefont {Magnea},\ and\
  \citenamefont {Vinti}}]{caselle1996width}%
  \BibitemOpen
  \bibfield  {author} {\bibinfo {author} {\bibfnamefont {M.}~\bibnamefont
  {Caselle}}, \bibinfo {author} {\bibfnamefont {F.}~\bibnamefont {Gliozzi}},
  \bibinfo {author} {\bibfnamefont {U.}~\bibnamefont {Magnea}},\ and\ \bibinfo
  {author} {\bibfnamefont {S.}~\bibnamefont {Vinti}},\ }\bibfield  {title}
  {\bibinfo {title} {Width of long colour flux tubes in lattice gauge
  systems},\ }\href {https://doi.org/10.1016/0550-3213(95)00639-7} {\bibfield
  {journal} {\bibinfo  {journal} {Nuclear Physics B}\ }\textbf {\bibinfo
  {volume} {460}},\ \bibinfo {pages} {397–412} (\bibinfo {year}
  {1996})}\BibitemShut {NoStop}%
\bibitem [{\citenamefont {Caselle}\ \emph {et~al.}(1997)\citenamefont
  {Caselle}, \citenamefont {Fiore}, \citenamefont {Gliozzi}, \citenamefont
  {Hasenbusch},\ and\ \citenamefont {Provero}}]{caselle1997string}%
  \BibitemOpen
  \bibfield  {author} {\bibinfo {author} {\bibfnamefont {M.}~\bibnamefont
  {Caselle}}, \bibinfo {author} {\bibfnamefont {R.}~\bibnamefont {Fiore}},
  \bibinfo {author} {\bibfnamefont {F.}~\bibnamefont {Gliozzi}}, \bibinfo
  {author} {\bibfnamefont {M.}~\bibnamefont {Hasenbusch}},\ and\ \bibinfo
  {author} {\bibfnamefont {P.}~\bibnamefont {Provero}},\ }\bibfield  {title}
  {\bibinfo {title} {String effects in the wilson loop: a high precision
  numerical test},\ }\href {https://doi.org/10.1016/s0550-3213(96)00672-4}
  {\bibfield  {journal} {\bibinfo  {journal} {Nuclear Physics B}\ }\textbf
  {\bibinfo {volume} {486}},\ \bibinfo {pages} {245–260} (\bibinfo {year}
  {1997})}\BibitemShut {NoStop}%
\bibitem [{\citenamefont {Caselle}\ \emph {et~al.}(2003)\citenamefont
  {Caselle}, \citenamefont {Hasenbusch},\ and\ \citenamefont
  {Panero}}]{caselle2003string}%
  \BibitemOpen
  \bibfield  {author} {\bibinfo {author} {\bibfnamefont {M.}~\bibnamefont
  {Caselle}}, \bibinfo {author} {\bibfnamefont {M.}~\bibnamefont
  {Hasenbusch}},\ and\ \bibinfo {author} {\bibfnamefont {M.}~\bibnamefont
  {Panero}},\ }\bibfield  {title} {\bibinfo {title} {String effects in the 3d
  gauge ising model},\ }\href {https://doi.org/10.1088/1126-6708/2003/01/057}
  {\bibfield  {journal} {\bibinfo  {journal} {Journal of High Energy Physics}\
  }\textbf {\bibinfo {volume} {2003}},\ \bibinfo {pages} {057–057} (\bibinfo
  {year} {2003})}\BibitemShut {NoStop}%
\bibitem [{\citenamefont {Di~Giacomo}\ \emph
  {et~al.}(1990{\natexlab{a}})\citenamefont {Di~Giacomo}, \citenamefont
  {Maggiore},\ and\ \citenamefont {{Olejn\'ik}}}]{di-giacomo1990evidence}%
  \BibitemOpen
  \bibfield  {author} {\bibinfo {author} {\bibfnamefont {A.}~\bibnamefont
  {Di~Giacomo}}, \bibinfo {author} {\bibfnamefont {M.}~\bibnamefont
  {Maggiore}},\ and\ \bibinfo {author} {\bibfnamefont
  {{\v{S}tefan}.}~\bibnamefont {{Olejn\'ik}}},\ }\bibfield  {title} {\bibinfo
  {title} {Evidence for flux tubes from cooled qcd configurations},\ }\href
  {https://doi.org/10.1016/0370-2693(90)90828-t} {\bibfield  {journal}
  {\bibinfo  {journal} {Physics Letters B}\ }\textbf {\bibinfo {volume}
  {236}},\ \bibinfo {pages} {199–202} (\bibinfo {year}
  {1990}{\natexlab{a}})}\BibitemShut {NoStop}%
\bibitem [{\citenamefont {Di~Giacomo}\ \emph
  {et~al.}(1990{\natexlab{b}})\citenamefont {Di~Giacomo}, \citenamefont
  {Maggiore},\ and\ \citenamefont {{Olejn\'ik}}}]{di-giacomo1990confinement}%
  \BibitemOpen
  \bibfield  {author} {\bibinfo {author} {\bibfnamefont {A.}~\bibnamefont
  {Di~Giacomo}}, \bibinfo {author} {\bibfnamefont {M.}~\bibnamefont
  {Maggiore}},\ and\ \bibinfo {author} {\bibfnamefont
  {{\v{S}tefan}.}~\bibnamefont {{Olejn\'ik}}},\ }\bibfield  {title} {\bibinfo
  {title} {Confinement and chromoelectric flux tubes in lattice qcd},\ }\href
  {https://doi.org/10.1016/0550-3213(90)90567-w} {\bibfield  {journal}
  {\bibinfo  {journal} {Nuclear Physics B}\ }\textbf {\bibinfo {volume}
  {347}},\ \bibinfo {pages} {441–460} (\bibinfo {year}
  {1990}{\natexlab{b}})}\BibitemShut {NoStop}%
\bibitem [{\citenamefont {Gliozzi}\ \emph {et~al.}(2010)\citenamefont
  {Gliozzi}, \citenamefont {Pepe},\ and\ \citenamefont
  {Wiese}}]{gliozzi2010width}%
  \BibitemOpen
  \bibfield  {author} {\bibinfo {author} {\bibfnamefont {F.}~\bibnamefont
  {Gliozzi}}, \bibinfo {author} {\bibfnamefont {M.}~\bibnamefont {Pepe}},\ and\
  \bibinfo {author} {\bibfnamefont {U.-J.}\ \bibnamefont {Wiese}},\ }\bibfield
  {title} {\bibinfo {title} {Width of the confining string in yang-mills
  theory},\ }\bibfield  {journal} {\bibinfo  {journal} {Physical Review
  Letters}\ }\textbf {\bibinfo {volume} {104}},\ \href
  {https://doi.org/10.1103/physrevlett.104.232001}
  {10.1103/physrevlett.104.232001} (\bibinfo {year} {2010})\BibitemShut
  {NoStop}%
\bibitem [{\citenamefont {Lüscher}\ and\ \citenamefont
  {Weisz}(2002)}]{luscher2002quark}%
  \BibitemOpen
  \bibfield  {author} {\bibinfo {author} {\bibfnamefont {M.}~\bibnamefont
  {Lüscher}}\ and\ \bibinfo {author} {\bibfnamefont {P.}~\bibnamefont
  {Weisz}},\ }\bibfield  {title} {\bibinfo {title} {Quark confinement and the
  bosonic string},\ }\href {https://doi.org/10.1088/1126-6708/2002/07/049}
  {\bibfield  {journal} {\bibinfo  {journal} {Journal of High Energy Physics}\
  }\textbf {\bibinfo {volume} {2002}},\ \bibinfo {pages} {049–049} (\bibinfo
  {year} {2002})}\BibitemShut {NoStop}%
\bibitem [{\citenamefont {Fukugita}\ and\ \citenamefont
  {Niuya}(1983)}]{fukugita1983distribution}%
  \BibitemOpen
  \bibfield  {author} {\bibinfo {author} {\bibfnamefont {M.}~\bibnamefont
  {Fukugita}}\ and\ \bibinfo {author} {\bibfnamefont {T.}~\bibnamefont
  {Niuya}},\ }\bibfield  {title} {\bibinfo {title} {The distribution of
  chromoelectric flux in {SU(2)} lattice gauge theory},\ }\href
  {https://doi.org/10.1016/0370-2693(83)90329-5} {\bibfield  {journal}
  {\bibinfo  {journal} {Physics Letters B}\ }\textbf {\bibinfo {volume}
  {132}},\ \bibinfo {pages} {374–378} (\bibinfo {year} {1983})}\BibitemShut
  {NoStop}%
\bibitem [{\citenamefont {Agostini}\ \emph {et~al.}(1997)\citenamefont
  {Agostini}, \citenamefont {Carlino}, \citenamefont {Caselle},\ and\
  \citenamefont {Hasenbusch}}]{agostini1997spectrum}%
  \BibitemOpen
  \bibfield  {author} {\bibinfo {author} {\bibfnamefont {V.}~\bibnamefont
  {Agostini}}, \bibinfo {author} {\bibfnamefont {G.}~\bibnamefont {Carlino}},
  \bibinfo {author} {\bibfnamefont {M.}~\bibnamefont {Caselle}},\ and\ \bibinfo
  {author} {\bibfnamefont {M.}~\bibnamefont {Hasenbusch}},\ }\bibfield  {title}
  {\bibinfo {title} {The spectrum of the 2 + 1-dimensional gauge ising model},\
  }\href {https://doi.org/10.1016/s0550-3213(96)00539-1} {\bibfield  {journal}
  {\bibinfo  {journal} {Nuclear Physics B}\ }\textbf {\bibinfo {volume}
  {484}},\ \bibinfo {pages} {331–352} (\bibinfo {year} {1997})}\BibitemShut
  {NoStop}%
\bibitem [{\citenamefont {Athenodorou}\ \emph {et~al.}(2023)\citenamefont
  {Athenodorou}, \citenamefont {Dubovsky}, \citenamefont {Luo},\ and\
  \citenamefont {Teper}}]{athenodorou2023excitations}%
  \BibitemOpen
  \bibfield  {author} {\bibinfo {author} {\bibfnamefont {A.}~\bibnamefont
  {Athenodorou}}, \bibinfo {author} {\bibfnamefont {S.}~\bibnamefont
  {Dubovsky}}, \bibinfo {author} {\bibfnamefont {C.}~\bibnamefont {Luo}},\ and\
  \bibinfo {author} {\bibfnamefont {M.}~\bibnamefont {Teper}},\ }\bibfield
  {title} {\bibinfo {title} {Excitations of ising strings on a lattice},\
  }\bibfield  {journal} {\bibinfo  {journal} {Journal of High Energy Physics}\
  }\textbf {\bibinfo {volume} {2023}},\ \href
  {https://doi.org/10.1007/jhep05(2023)082} {10.1007/jhep05(2023)082} (\bibinfo
  {year} {2023})\BibitemShut {NoStop}%
\bibitem [{\citenamefont {Athenodorou}\ \emph {et~al.}(2011)\citenamefont
  {Athenodorou}, \citenamefont {Bringoltz},\ and\ \citenamefont
  {Teper}}]{athenodorou2011closed}%
  \BibitemOpen
  \bibfield  {author} {\bibinfo {author} {\bibfnamefont {A.}~\bibnamefont
  {Athenodorou}}, \bibinfo {author} {\bibfnamefont {B.}~\bibnamefont
  {Bringoltz}},\ and\ \bibinfo {author} {\bibfnamefont {M.}~\bibnamefont
  {Teper}},\ }\bibfield  {title} {\bibinfo {title} {Closed flux tubes and their
  string description in {D=2+1} {SU(N)} gauge theories},\ }\bibfield  {journal}
  {\bibinfo  {journal} {Journal of High Energy Physics}\ }\textbf {\bibinfo
  {volume} {2011}},\ \href {https://doi.org/10.1007/jhep05(2011)042}
  {10.1007/jhep05(2011)042} (\bibinfo {year} {2011})\BibitemShut {NoStop}%
\bibitem [{\citenamefont {Athenodorou}\ \emph {et~al.}(2007)\citenamefont
  {Athenodorou}, \citenamefont {Bringoltz},\ and\ \citenamefont
  {Teper}}]{athenodorou2007closed}%
  \BibitemOpen
  \bibfield  {author} {\bibinfo {author} {\bibfnamefont {A.}~\bibnamefont
  {Athenodorou}}, \bibinfo {author} {\bibfnamefont {B.}~\bibnamefont
  {Bringoltz}},\ and\ \bibinfo {author} {\bibfnamefont {M.}~\bibnamefont
  {Teper}},\ }\bibfield  {title} {\bibinfo {title} {The closed string spectrum
  of {SU(N)} gauge theories in 2+1 dimensions},\ }\href
  {https://doi.org/10.1016/j.physletb.2007.09.045} {\bibfield  {journal}
  {\bibinfo  {journal} {Physics Letters B}\ }\textbf {\bibinfo {volume}
  {656}},\ \bibinfo {pages} {132–140} (\bibinfo {year} {2007})}\BibitemShut
  {NoStop}%
\bibitem [{\citenamefont {Cochran}\ \emph {et~al.}(2024)\citenamefont {Cochran}
  \emph {et~al.}}]{cochran2024visualizing}%
  \BibitemOpen
  \bibfield  {author} {\bibinfo {author} {\bibfnamefont {T.~A.}\ \bibnamefont
  {Cochran}} \emph {et~al.},\ }\href {https://arxiv.org/abs/2409.17142}
  {\bibinfo {title} {Visualizing dynamics of charges and strings in {(2+1)D}
  lattice gauge theories}} (\bibinfo {year} {2024}),\ \Eprint
  {https://arxiv.org/abs/2409.17142} {arXiv:2409.17142 [quant-ph]} \BibitemShut
  {NoStop}%
\bibitem [{\citenamefont {De}\ \emph {et~al.}(2024)\citenamefont {De},
  \citenamefont {Lerose}, \citenamefont {Luo}, \citenamefont {Surace},
  \citenamefont {Schuckert}, \citenamefont {Bennewitz}, \citenamefont {Ware},
  \citenamefont {Morong}, \citenamefont {Collins}, \citenamefont {Davoudi},
  \citenamefont {Gorshkov}, \citenamefont {Katz},\ and\ \citenamefont
  {Monroe}}]{de2024observation}%
  \BibitemOpen
  \bibfield  {author} {\bibinfo {author} {\bibfnamefont {A.}~\bibnamefont
  {De}}, \bibinfo {author} {\bibfnamefont {A.}~\bibnamefont {Lerose}}, \bibinfo
  {author} {\bibfnamefont {D.}~\bibnamefont {Luo}}, \bibinfo {author}
  {\bibfnamefont {F.~M.}\ \bibnamefont {Surace}}, \bibinfo {author}
  {\bibfnamefont {A.}~\bibnamefont {Schuckert}}, \bibinfo {author}
  {\bibfnamefont {E.~R.}\ \bibnamefont {Bennewitz}}, \bibinfo {author}
  {\bibfnamefont {B.}~\bibnamefont {Ware}}, \bibinfo {author} {\bibfnamefont
  {W.}~\bibnamefont {Morong}}, \bibinfo {author} {\bibfnamefont {K.~S.}\
  \bibnamefont {Collins}}, \bibinfo {author} {\bibfnamefont {Z.}~\bibnamefont
  {Davoudi}}, \bibinfo {author} {\bibfnamefont {A.~V.}\ \bibnamefont
  {Gorshkov}}, \bibinfo {author} {\bibfnamefont {O.}~\bibnamefont {Katz}},\
  and\ \bibinfo {author} {\bibfnamefont {C.}~\bibnamefont {Monroe}},\ }\href
  {https://arxiv.org/abs/2410.13815} {\bibinfo {title} {Observation of
  string-breaking dynamics in a quantum simulator}} (\bibinfo {year} {2024}),\
  \Eprint {https://arxiv.org/abs/2410.13815} {arXiv:2410.13815 [quant-ph]}
  \BibitemShut {NoStop}%
\bibitem [{\citenamefont {Cuadra}\ \emph {et~al.}(2024)\citenamefont {Cuadra},
  \citenamefont {Hamdan}, \citenamefont {Zache}, \citenamefont {Braverman},
  \citenamefont {Kornjaca}, \citenamefont {Lukin}, \citenamefont {Cantu},
  \citenamefont {Liu}, \citenamefont {Wang}, \citenamefont {A.}, \citenamefont
  {Lukin}, \citenamefont {Zoller},\ and\ \citenamefont
  {Bylinskii}}]{gonzalez-cuadra2024observation}%
  \BibitemOpen
  \bibfield  {author} {\bibinfo {author} {\bibfnamefont {D.~G.}\ \bibnamefont
  {Cuadra}}, \bibinfo {author} {\bibfnamefont {M.}~\bibnamefont {Hamdan}},
  \bibinfo {author} {\bibfnamefont {T.~V.}\ \bibnamefont {Zache}}, \bibinfo
  {author} {\bibfnamefont {B.}~\bibnamefont {Braverman}}, \bibinfo {author}
  {\bibfnamefont {M.}~\bibnamefont {Kornjaca}}, \bibinfo {author}
  {\bibfnamefont {A.}~\bibnamefont {Lukin}}, \bibinfo {author} {\bibfnamefont
  {S.~H.}\ \bibnamefont {Cantu}}, \bibinfo {author} {\bibfnamefont
  {F.}~\bibnamefont {Liu}}, \bibinfo {author} {\bibfnamefont {S.}~\bibnamefont
  {Wang}}, \bibinfo {author} {\bibfnamefont {K.}~\bibnamefont {A.}}, \bibinfo
  {author} {\bibfnamefont {M.~D.}\ \bibnamefont {Lukin}}, \bibinfo {author}
  {\bibfnamefont {P.}~\bibnamefont {Zoller}},\ and\ \bibinfo {author}
  {\bibfnamefont {A.}~\bibnamefont {Bylinskii}},\ }\href
  {https://arxiv.org/abs/2410.16558} {\bibinfo {title} {Observation of string
  breaking on a (2 + 1)d rydberg quantum simulator}} (\bibinfo {year} {2024}),\
  \Eprint {https://arxiv.org/abs/2410.16558} {arXiv:2410.16558 [quant-ph]}
  \BibitemShut {NoStop}%
\bibitem [{\citenamefont {Surace}\ \emph {et~al.}(2020)\citenamefont {Surace},
  \citenamefont {Mazza}, \citenamefont {Giudici}, \citenamefont {Lerose},
  \citenamefont {Gambassi},\ and\ \citenamefont
  {Dalmonte}}]{surace2020lattice}%
  \BibitemOpen
  \bibfield  {author} {\bibinfo {author} {\bibfnamefont {F.~M.}\ \bibnamefont
  {Surace}}, \bibinfo {author} {\bibfnamefont {P.~P.}\ \bibnamefont {Mazza}},
  \bibinfo {author} {\bibfnamefont {G.}~\bibnamefont {Giudici}}, \bibinfo
  {author} {\bibfnamefont {A.}~\bibnamefont {Lerose}}, \bibinfo {author}
  {\bibfnamefont {A.}~\bibnamefont {Gambassi}},\ and\ \bibinfo {author}
  {\bibfnamefont {M.}~\bibnamefont {Dalmonte}},\ }\bibfield  {title} {\bibinfo
  {title} {Lattice gauge theories and string dynamics in rydberg atom quantum
  simulators},\ }\bibfield  {journal} {\bibinfo  {journal} {Physical Review X}\
  }\textbf {\bibinfo {volume} {10}},\ \href
  {https://doi.org/10.1103/physrevx.10.021041} {10.1103/physrevx.10.021041}
  (\bibinfo {year} {2020})\BibitemShut {NoStop}%
\bibitem [{\citenamefont {Mildenberger}\ \emph {et~al.}(2025)\citenamefont
  {Mildenberger}, \citenamefont {Mruczkiewicz}, \citenamefont {Halimeh},
  \citenamefont {Jiang},\ and\ \citenamefont
  {Hauke}}]{mildenberger2025confinement}%
  \BibitemOpen
  \bibfield  {author} {\bibinfo {author} {\bibfnamefont {J.}~\bibnamefont
  {Mildenberger}}, \bibinfo {author} {\bibfnamefont {W.}~\bibnamefont
  {Mruczkiewicz}}, \bibinfo {author} {\bibfnamefont {J.~C.}\ \bibnamefont
  {Halimeh}}, \bibinfo {author} {\bibfnamefont {Z.}~\bibnamefont {Jiang}},\
  and\ \bibinfo {author} {\bibfnamefont {P.}~\bibnamefont {Hauke}},\ }\bibfield
   {title} {\bibinfo {title} {Confinement in a {$Z_2$} lattice gauge theory on
  a quantum computer},\ }\href {https://doi.org/10.1038/s41567-024-02723-6}
  {\bibfield  {journal} {\bibinfo  {journal} {Nature Physics}\ }\textbf
  {\bibinfo {volume} {21}},\ \bibinfo {pages} {312–317} (\bibinfo {year}
  {2025})}\BibitemShut {NoStop}%
\bibitem [{\citenamefont {Meth}\ \emph {et~al.}(2025)\citenamefont {Meth},
  \citenamefont {Zhang}, \citenamefont {Haase}, \citenamefont {Edmunds},
  \citenamefont {Postler}, \citenamefont {Jena}, \citenamefont {Steiner},
  \citenamefont {Dellantonio}, \citenamefont {Blatt}, \citenamefont {Zoller},
  \citenamefont {Monz}, \citenamefont {Schindler}, \citenamefont {Muschik},\
  and\ \citenamefont {Ringbauer}}]{meth2025simulating}%
  \BibitemOpen
  \bibfield  {author} {\bibinfo {author} {\bibfnamefont {M.}~\bibnamefont
  {Meth}}, \bibinfo {author} {\bibfnamefont {J.}~\bibnamefont {Zhang}},
  \bibinfo {author} {\bibfnamefont {J.~F.}\ \bibnamefont {Haase}}, \bibinfo
  {author} {\bibfnamefont {C.}~\bibnamefont {Edmunds}}, \bibinfo {author}
  {\bibfnamefont {L.}~\bibnamefont {Postler}}, \bibinfo {author} {\bibfnamefont
  {A.~J.}\ \bibnamefont {Jena}}, \bibinfo {author} {\bibfnamefont
  {A.}~\bibnamefont {Steiner}}, \bibinfo {author} {\bibfnamefont
  {L.}~\bibnamefont {Dellantonio}}, \bibinfo {author} {\bibfnamefont
  {R.}~\bibnamefont {Blatt}}, \bibinfo {author} {\bibfnamefont
  {P.}~\bibnamefont {Zoller}}, \bibinfo {author} {\bibfnamefont
  {T.}~\bibnamefont {Monz}}, \bibinfo {author} {\bibfnamefont {P.}~\bibnamefont
  {Schindler}}, \bibinfo {author} {\bibfnamefont {C.}~\bibnamefont {Muschik}},\
  and\ \bibinfo {author} {\bibfnamefont {M.}~\bibnamefont {Ringbauer}},\
  }\bibfield  {title} {\bibinfo {title} {Simulating two-dimensional lattice
  gauge theories on a qudit quantum computer},\ }\href
  {https://doi.org/10.1038/s41567-025-02797-w} {\bibfield  {journal} {\bibinfo
  {journal} {Nature Physics}\ }\textbf {\bibinfo {volume} {21}},\ \bibinfo
  {pages} {570–576} (\bibinfo {year} {2025})}\BibitemShut {NoStop}%
\bibitem [{\citenamefont {Verstraete}\ \emph {et~al.}(2008)\citenamefont
  {Verstraete}, \citenamefont {Murg},\ and\ \citenamefont
  {Cirac}}]{verstraete2008matrix}%
  \BibitemOpen
  \bibfield  {author} {\bibinfo {author} {\bibfnamefont {F.}~\bibnamefont
  {Verstraete}}, \bibinfo {author} {\bibfnamefont {V.}~\bibnamefont {Murg}},\
  and\ \bibinfo {author} {\bibfnamefont {J.}~\bibnamefont {Cirac}},\ }\bibfield
   {title} {\bibinfo {title} {Matrix product states, projected entangled pair
  states, and variational renormalization group methods for quantum spin
  systems},\ }\href {https://doi.org/10.1080/14789940801912366} {\bibfield
  {journal} {\bibinfo  {journal} {Advances in Physics}\ }\textbf {\bibinfo
  {volume} {57}},\ \bibinfo {pages} {143–224} (\bibinfo {year}
  {2008})}\BibitemShut {NoStop}%
\bibitem [{\citenamefont {Schollwöck}(2011)}]{schollwock2011dmrg}%
  \BibitemOpen
  \bibfield  {author} {\bibinfo {author} {\bibfnamefont {U.}~\bibnamefont
  {Schollwöck}},\ }\bibfield  {title} {\bibinfo {title} {The density-matrix
  renormalization group in the age of matrix product states},\ }\href
  {https://doi.org/https://doi.org/10.1016/j.aop.2010.09.012} {\bibfield
  {journal} {\bibinfo  {journal} {Annals of Physics}\ }\textbf {\bibinfo
  {volume} {326}},\ \bibinfo {pages} {96} (\bibinfo {year} {2011})},\ \bibinfo
  {note} {january 2011 Special Issue}\BibitemShut {NoStop}%
\bibitem [{\citenamefont {Orús}(2014)}]{orus2014practical}%
  \BibitemOpen
  \bibfield  {author} {\bibinfo {author} {\bibfnamefont {R.}~\bibnamefont
  {Orús}},\ }\bibfield  {title} {\bibinfo {title} {A practical introduction to
  tensor networks: Matrix product states and projected entangled pair states},\
  }\href {https://doi.org/10.1016/j.aop.2014.06.013} {\bibfield  {journal}
  {\bibinfo  {journal} {Annals of Physics}\ }\textbf {\bibinfo {volume}
  {349}},\ \bibinfo {pages} {117–158} (\bibinfo {year} {2014})}\BibitemShut
  {NoStop}%
\bibitem [{\citenamefont {Bridgeman}\ and\ \citenamefont
  {Chubb}(2017)}]{Bridgeman_2017}%
  \BibitemOpen
  \bibfield  {author} {\bibinfo {author} {\bibfnamefont {J.~C.}\ \bibnamefont
  {Bridgeman}}\ and\ \bibinfo {author} {\bibfnamefont {C.~T.}\ \bibnamefont
  {Chubb}},\ }\bibfield  {title} {\bibinfo {title} {Hand-waving and
  interpretive dance: an introductory course on tensor networks},\ }\href
  {https://doi.org/10.1088/1751-8121/aa6dc3} {\bibfield  {journal} {\bibinfo
  {journal} {Journal of Physics A: Mathematical and Theoretical}\ }\textbf
  {\bibinfo {volume} {50}},\ \bibinfo {pages} {223001} (\bibinfo {year}
  {2017})},\ \Eprint {https://arxiv.org/abs/1603.03039} {arXiv:1603.03039
  [quant-ph]} \BibitemShut {NoStop}%
\bibitem [{\citenamefont {Silvi}\ \emph {et~al.}(2019)\citenamefont {Silvi},
  \citenamefont {Tschirsich}, \citenamefont {Gerster}, \citenamefont
  {Jünemann}, \citenamefont {Jaschke}, \citenamefont {Rizzi},\ and\
  \citenamefont {Montangero}}]{silvi2019tensor}%
  \BibitemOpen
  \bibfield  {author} {\bibinfo {author} {\bibfnamefont {P.}~\bibnamefont
  {Silvi}}, \bibinfo {author} {\bibfnamefont {F.}~\bibnamefont {Tschirsich}},
  \bibinfo {author} {\bibfnamefont {M.}~\bibnamefont {Gerster}}, \bibinfo
  {author} {\bibfnamefont {J.}~\bibnamefont {Jünemann}}, \bibinfo {author}
  {\bibfnamefont {D.}~\bibnamefont {Jaschke}}, \bibinfo {author} {\bibfnamefont
  {M.}~\bibnamefont {Rizzi}},\ and\ \bibinfo {author} {\bibfnamefont
  {S.}~\bibnamefont {Montangero}},\ }\bibfield  {title} {\bibinfo {title} {The
  tensor networks anthology: Simulation techniques for many-body quantum
  lattice systems},\ }\bibfield  {journal} {\bibinfo  {journal} {SciPost
  Physics Lecture Notes}\ }\href
  {https://doi.org/10.21468/scipostphyslectnotes.8}
  {10.21468/scipostphyslectnotes.8} (\bibinfo {year} {2019})\BibitemShut
  {NoStop}%
\bibitem [{\citenamefont {Ran}\ \emph {et~al.}(2020)\citenamefont {Ran},
  \citenamefont {Tirrito}, \citenamefont {Peng}, \citenamefont {Chen},
  \citenamefont {Tagliacozzo}, \citenamefont {Su},\ and\ \citenamefont
  {Lewenstein}}]{Ran2020tncontr}%
  \BibitemOpen
  \bibfield  {author} {\bibinfo {author} {\bibfnamefont {S.-J.}\ \bibnamefont
  {Ran}}, \bibinfo {author} {\bibfnamefont {E.}~\bibnamefont {Tirrito}},
  \bibinfo {author} {\bibfnamefont {C.}~\bibnamefont {Peng}}, \bibinfo {author}
  {\bibfnamefont {X.}~\bibnamefont {Chen}}, \bibinfo {author} {\bibfnamefont
  {L.}~\bibnamefont {Tagliacozzo}}, \bibinfo {author} {\bibfnamefont
  {G.}~\bibnamefont {Su}},\ and\ \bibinfo {author} {\bibfnamefont
  {M.}~\bibnamefont {Lewenstein}},\ }\href
  {https://doi.org/10.1007/978-3-030-34489-4} {\emph {\bibinfo {title} {Tensor
  Network Contractions}}},\ Lecture Notes in Physics\ (\bibinfo  {publisher}
  {Springer International Publishing},\ \bibinfo {year} {2020})\BibitemShut
  {NoStop}%
\bibitem [{\citenamefont {Cirac}\ \emph {et~al.}(2021)\citenamefont {Cirac},
  \citenamefont {P\'erez-Garc\'{\i}a}, \citenamefont {Schuch},\ and\
  \citenamefont {Verstraete}}]{Cirac2021rmp}%
  \BibitemOpen
  \bibfield  {author} {\bibinfo {author} {\bibfnamefont {J.~I.}\ \bibnamefont
  {Cirac}}, \bibinfo {author} {\bibfnamefont {D.}~\bibnamefont
  {P\'erez-Garc\'{\i}a}}, \bibinfo {author} {\bibfnamefont {N.}~\bibnamefont
  {Schuch}},\ and\ \bibinfo {author} {\bibfnamefont {F.}~\bibnamefont
  {Verstraete}},\ }\bibfield  {title} {\bibinfo {title} {Matrix product states
  and projected entangled pair states: Concepts, symmetries, theorems},\ }\href
  {https://doi.org/10.1103/RevModPhys.93.045003} {\bibfield  {journal}
  {\bibinfo  {journal} {Rev. Mod. Phys.}\ }\textbf {\bibinfo {volume} {93}},\
  \bibinfo {pages} {045003} (\bibinfo {year} {2021})},\ \Eprint
  {https://arxiv.org/abs/2011.12127} {arXiv:2011.12127 [quant-ph]} \BibitemShut
  {NoStop}%
\bibitem [{\citenamefont {Okunishi}\ \emph {et~al.}(2022)\citenamefont
  {Okunishi}, \citenamefont {Nishino},\ and\ \citenamefont
  {Hiroshi}}]{Okunishi2022}%
  \BibitemOpen
  \bibfield  {author} {\bibinfo {author} {\bibfnamefont {K.}~\bibnamefont
  {Okunishi}}, \bibinfo {author} {\bibfnamefont {T.}~\bibnamefont {Nishino}},\
  and\ \bibinfo {author} {\bibfnamefont {U.}~\bibnamefont {Hiroshi}},\
  }\bibfield  {title} {\bibinfo {title} {{Developments in the Tensor Network
  --- from Statistical Mechanics to Quantum Entanglement}},\ }\href
  {https://doi.org/https://doi.org/10.7566/JPSJ.91.062001} {\bibfield
  {journal} {\bibinfo  {journal} {J. Phys. Soc. Jpn}\ }\textbf {\bibinfo
  {volume} {91}},\ \bibinfo {pages} {062001} (\bibinfo {year}
  {2022})}\BibitemShut {NoStop}%
\bibitem [{\citenamefont {Ba\~{n}uls}(2023)}]{Banuls2023}%
  \BibitemOpen
  \bibfield  {author} {\bibinfo {author} {\bibfnamefont {M.~C.}\ \bibnamefont
  {Ba\~{n}uls}},\ }\bibfield  {title} {\bibinfo {title} {{Tensor Network
  Algorithms: A Route Map}},\ }\bibfield  {journal} {\bibinfo  {journal} {Annu.
  Rev. Condens. Matter Phys}\ }\textbf {\bibinfo {volume} {14}},\ \href
  {https://doi.org/10.1146/annurev-conmatphys-040721-022705}
  {10.1146/annurev-conmatphys-040721-022705} (\bibinfo {year}
  {2023})\BibitemShut {NoStop}%
\bibitem [{\citenamefont {Carmen~Bañuls}\ and\ \citenamefont
  {Cichy}(2020)}]{banuls2020review}%
  \BibitemOpen
  \bibfield  {author} {\bibinfo {author} {\bibfnamefont {M.}~\bibnamefont
  {Carmen~Bañuls}}\ and\ \bibinfo {author} {\bibfnamefont {K.}~\bibnamefont
  {Cichy}},\ }\bibfield  {title} {\bibinfo {title} {Review on novel methods for
  lattice gauge theories},\ }\href {https://doi.org/10.1088/1361-6633/ab6311}
  {\bibfield  {journal} {\bibinfo  {journal} {Reports on Progress in Physics}\
  }\textbf {\bibinfo {volume} {83}},\ \bibinfo {pages} {024401} (\bibinfo
  {year} {2020})}\BibitemShut {NoStop}%
\bibitem [{\citenamefont {Meurice}\ \emph {et~al.}(2022)\citenamefont
  {Meurice}, \citenamefont {Sakai},\ and\ \citenamefont
  {Unmuth-Yockey}}]{meurice2022trg}%
  \BibitemOpen
  \bibfield  {author} {\bibinfo {author} {\bibfnamefont {Y.}~\bibnamefont
  {Meurice}}, \bibinfo {author} {\bibfnamefont {R.}~\bibnamefont {Sakai}},\
  and\ \bibinfo {author} {\bibfnamefont {J.}~\bibnamefont {Unmuth-Yockey}},\
  }\bibfield  {title} {\bibinfo {title} {Tensor lattice field theory for
  renormalization and quantum computing},\ }\href
  {https://doi.org/10.1103/RevModPhys.94.025005} {\bibfield  {journal}
  {\bibinfo  {journal} {Rev. Mod. Phys.}\ }\textbf {\bibinfo {volume} {94}},\
  \bibinfo {pages} {025005} (\bibinfo {year} {2022})},\ \Eprint
  {https://arxiv.org/abs/2010.06539} {arXiv:2010.06539 [hep-lat]} \BibitemShut
  {NoStop}%
\bibitem [{\citenamefont {Kadoh}(2022)}]{Kadoh:2022Ia}%
  \BibitemOpen
  \bibfield  {author} {\bibinfo {author} {\bibfnamefont {D.}~\bibnamefont
  {Kadoh}},\ }\bibfield  {title} {\bibinfo {title} {{Recent progress in the
  tensor renormalization group}},\ }\href {https://doi.org/10.22323/1.396.0633}
  {\bibfield  {journal} {\bibinfo  {journal} {Proceedings of The 38th
  International Symposium on Lattice Field Theory {\textemdash}
  PoS(LATTICE2021)}\ }\textbf {\bibinfo {volume} {396}},\ \bibinfo {pages}
  {633} (\bibinfo {year} {2022})}\BibitemShut {NoStop}%
\bibitem [{\citenamefont {Felser}\ \emph {et~al.}(2020)\citenamefont {Felser},
  \citenamefont {Silvi}, \citenamefont {Collura},\ and\ \citenamefont
  {Montangero}}]{felser2020u1}%
  \BibitemOpen
  \bibfield  {author} {\bibinfo {author} {\bibfnamefont {T.}~\bibnamefont
  {Felser}}, \bibinfo {author} {\bibfnamefont {P.}~\bibnamefont {Silvi}},
  \bibinfo {author} {\bibfnamefont {M.}~\bibnamefont {Collura}},\ and\ \bibinfo
  {author} {\bibfnamefont {S.}~\bibnamefont {Montangero}},\ }\bibfield  {title}
  {\bibinfo {title} {{Two-Dimensional Quantum-Link Lattice Quantum
  Electrodynamics at Finite Density}},\ }\href
  {https://doi.org/10.1103/PhysRevX.10.041040} {\bibfield  {journal} {\bibinfo
  {journal} {Phys. Rev. X}\ }\textbf {\bibinfo {volume} {10}},\ \bibinfo
  {pages} {041040} (\bibinfo {year} {2020})},\ \Eprint
  {https://arxiv.org/abs/1911.09693} {arXiv:1911.09693 [quant-ph]} \BibitemShut
  {NoStop}%
\bibitem [{\citenamefont {Robaina}\ \emph {et~al.}(2021)\citenamefont
  {Robaina}, \citenamefont {Bañuls},\ and\ \citenamefont
  {Cirac}}]{robaina2021simulating}%
  \BibitemOpen
  \bibfield  {author} {\bibinfo {author} {\bibfnamefont {D.}~\bibnamefont
  {Robaina}}, \bibinfo {author} {\bibfnamefont {M.~C.}\ \bibnamefont
  {Bañuls}},\ and\ \bibinfo {author} {\bibfnamefont {J.~I.}\ \bibnamefont
  {Cirac}},\ }\bibfield  {title} {\bibinfo {title} {Simulating {2+1D} {$Z_3$}
  lattice gauge theory with an infinite projected entangled-pair state},\
  }\bibfield  {journal} {\bibinfo  {journal} {Physical Review Letters}\
  }\textbf {\bibinfo {volume} {126}},\ \href
  {https://doi.org/10.1103/physrevlett.126.050401}
  {10.1103/physrevlett.126.050401} (\bibinfo {year} {2021})\BibitemShut
  {NoStop}%
\bibitem [{\citenamefont {Magnifico}\ \emph {et~al.}(2021)\citenamefont
  {Magnifico}, \citenamefont {Felser}, \citenamefont {Silvi},\ and\
  \citenamefont {Montangero}}]{magnifico2021qed3d}%
  \BibitemOpen
  \bibfield  {author} {\bibinfo {author} {\bibfnamefont {G.}~\bibnamefont
  {Magnifico}}, \bibinfo {author} {\bibfnamefont {T.}~\bibnamefont {Felser}},
  \bibinfo {author} {\bibfnamefont {P.}~\bibnamefont {Silvi}},\ and\ \bibinfo
  {author} {\bibfnamefont {S.}~\bibnamefont {Montangero}},\ }\bibfield  {title}
  {\bibinfo {title} {Lattice quantum electrodynamics in (3+1)-dimensions at
  finite density with tensor networks},\ }\href
  {https://doi.org/10.1038/s41467-021-23646-3} {\bibfield  {journal} {\bibinfo
  {journal} {Nature Communications}\ }\textbf {\bibinfo {volume} {12}},\
  \bibinfo {pages} {3600} (\bibinfo {year} {2021})},\ \Eprint
  {https://arxiv.org/abs/2011.10658} {arXiv:2011.10658 [hep-lat]} \BibitemShut
  {NoStop}%
\bibitem [{\citenamefont {Emonts}\ \emph {et~al.}(2023)\citenamefont {Emonts},
  \citenamefont {Kelman}, \citenamefont {Borla}, \citenamefont {Moroz},
  \citenamefont {Gazit},\ and\ \citenamefont {Zohar}}]{emonts2023z2}%
  \BibitemOpen
  \bibfield  {author} {\bibinfo {author} {\bibfnamefont {P.}~\bibnamefont
  {Emonts}}, \bibinfo {author} {\bibfnamefont {A.}~\bibnamefont {Kelman}},
  \bibinfo {author} {\bibfnamefont {U.}~\bibnamefont {Borla}}, \bibinfo
  {author} {\bibfnamefont {S.}~\bibnamefont {Moroz}}, \bibinfo {author}
  {\bibfnamefont {S.}~\bibnamefont {Gazit}},\ and\ \bibinfo {author}
  {\bibfnamefont {E.}~\bibnamefont {Zohar}},\ }\bibfield  {title} {\bibinfo
  {title} {Finding the ground state of a lattice gauge theory with fermionic
  tensor networks: A {$2+1\mathrm{D}$} {$Z_2$} demonstration},\ }\href
  {https://doi.org/10.1103/PhysRevD.107.014505} {\bibfield  {journal} {\bibinfo
   {journal} {Phys. Rev. D}\ }\textbf {\bibinfo {volume} {107}},\ \bibinfo
  {pages} {014505} (\bibinfo {year} {2023})},\ \Eprint
  {https://arxiv.org/abs/2211.00023} {arXiv:2211.00023 [quant-ph]} \BibitemShut
  {NoStop}%
\bibitem [{\citenamefont {Kelman}\ \emph {et~al.}(2024)\citenamefont {Kelman},
  \citenamefont {Borla}, \citenamefont {Gomelski}, \citenamefont {Elyovich},
  \citenamefont {Roose}, \citenamefont {Emonts},\ and\ \citenamefont
  {Zohar}}]{kelman2024gpeps}%
  \BibitemOpen
  \bibfield  {author} {\bibinfo {author} {\bibfnamefont {A.}~\bibnamefont
  {Kelman}}, \bibinfo {author} {\bibfnamefont {U.}~\bibnamefont {Borla}},
  \bibinfo {author} {\bibfnamefont {I.}~\bibnamefont {Gomelski}}, \bibinfo
  {author} {\bibfnamefont {J.}~\bibnamefont {Elyovich}}, \bibinfo {author}
  {\bibfnamefont {G.}~\bibnamefont {Roose}}, \bibinfo {author} {\bibfnamefont
  {P.}~\bibnamefont {Emonts}},\ and\ \bibinfo {author} {\bibfnamefont
  {E.}~\bibnamefont {Zohar}},\ }\bibfield  {title} {\bibinfo {title} {Gauged
  gaussian projected entangled pair states: A high dimensional tensor network
  formulation for lattice gauge theories},\ }\href
  {https://doi.org/10.1103/PhysRevD.110.054511} {\bibfield  {journal} {\bibinfo
   {journal} {Phys. Rev. D}\ }\textbf {\bibinfo {volume} {110}},\ \bibinfo
  {pages} {054511} (\bibinfo {year} {2024})},\ \Eprint
  {https://arxiv.org/abs/2404.13123} {arXiv:2404.13123 [hep-lat]} \BibitemShut
  {NoStop}%
\bibitem [{\citenamefont {Akiyama}\ \emph {et~al.}(2024)\citenamefont
  {Akiyama}, \citenamefont {Meurice},\ and\ \citenamefont
  {Sakai}}]{Akiyama_2024}%
  \BibitemOpen
  \bibfield  {author} {\bibinfo {author} {\bibfnamefont {S.}~\bibnamefont
  {Akiyama}}, \bibinfo {author} {\bibfnamefont {Y.}~\bibnamefont {Meurice}},\
  and\ \bibinfo {author} {\bibfnamefont {R.}~\bibnamefont {Sakai}},\ }\bibfield
   {title} {\bibinfo {title} {Tensor renormalization group for fermions},\
  }\href {https://doi.org/10.1088/1361-648X/ad4760} {\bibfield  {journal}
  {\bibinfo  {journal} {Journal of Physics: Condensed Matter}\ }\textbf
  {\bibinfo {volume} {36}},\ \bibinfo {pages} {343002} (\bibinfo {year}
  {2024})},\ \Eprint {https://arxiv.org/abs/2401.08542} {arXiv:2401.08542
  [hep-lat]} \BibitemShut {NoStop}%
\bibitem [{\citenamefont {Pichler}\ \emph {et~al.}(2016)\citenamefont
  {Pichler}, \citenamefont {Dalmonte}, \citenamefont {Rico}, \citenamefont
  {Zoller},\ and\ \citenamefont {Montangero}}]{pichler2016real-time}%
  \BibitemOpen
  \bibfield  {author} {\bibinfo {author} {\bibfnamefont {T.}~\bibnamefont
  {Pichler}}, \bibinfo {author} {\bibfnamefont {M.}~\bibnamefont {Dalmonte}},
  \bibinfo {author} {\bibfnamefont {E.}~\bibnamefont {Rico}}, \bibinfo {author}
  {\bibfnamefont {P.}~\bibnamefont {Zoller}},\ and\ \bibinfo {author}
  {\bibfnamefont {S.}~\bibnamefont {Montangero}},\ }\bibfield  {title}
  {\bibinfo {title} {Real-time dynamics in u(1) lattice gauge theories with
  tensor networks},\ }\bibfield  {journal} {\bibinfo  {journal} {Physical
  Review X}\ }\textbf {\bibinfo {volume} {6}},\ \href
  {https://doi.org/10.1103/physrevx.6.011023} {10.1103/physrevx.6.011023}
  (\bibinfo {year} {2016})\BibitemShut {NoStop}%
\bibitem [{\citenamefont {Chanda}\ \emph {et~al.}(2020)\citenamefont {Chanda},
  \citenamefont {Zakrzewski}, \citenamefont {Lewenstein},\ and\ \citenamefont
  {Tagliacozzo}}]{chanda2020quenches}%
  \BibitemOpen
  \bibfield  {author} {\bibinfo {author} {\bibfnamefont {T.}~\bibnamefont
  {Chanda}}, \bibinfo {author} {\bibfnamefont {J.}~\bibnamefont {Zakrzewski}},
  \bibinfo {author} {\bibfnamefont {M.}~\bibnamefont {Lewenstein}},\ and\
  \bibinfo {author} {\bibfnamefont {L.}~\bibnamefont {Tagliacozzo}},\
  }\bibfield  {title} {\bibinfo {title} {{Confinement and Lack of
  Thermalization after Quenches in the Bosonic Schwinger Model}},\ }\href
  {https://doi.org/10.1103/PhysRevLett.124.180602} {\bibfield  {journal}
  {\bibinfo  {journal} {Phys. Rev. Lett.}\ }\textbf {\bibinfo {volume} {124}},\
  \bibinfo {pages} {180602} (\bibinfo {year} {2020})}\BibitemShut {NoStop}%
\bibitem [{\citenamefont {Notarnicola}\ \emph {et~al.}(2020)\citenamefont
  {Notarnicola}, \citenamefont {Collura},\ and\ \citenamefont
  {Montangero}}]{notarnicola2020ryd}%
  \BibitemOpen
  \bibfield  {author} {\bibinfo {author} {\bibfnamefont {S.}~\bibnamefont
  {Notarnicola}}, \bibinfo {author} {\bibfnamefont {M.}~\bibnamefont
  {Collura}},\ and\ \bibinfo {author} {\bibfnamefont {S.}~\bibnamefont
  {Montangero}},\ }\bibfield  {title} {\bibinfo {title} {{Real-time-dynamics
  quantum simulation of $(1+1)\text{-dimensional}$ lattice QED with Rydberg
  atoms}},\ }\href {https://doi.org/10.1103/PhysRevResearch.2.013288}
  {\bibfield  {journal} {\bibinfo  {journal} {Phys. Rev. Res.}\ }\textbf
  {\bibinfo {volume} {2}},\ \bibinfo {pages} {013288} (\bibinfo {year}
  {2020})}\BibitemShut {NoStop}%
\bibitem [{\citenamefont {Rigobello}\ \emph {et~al.}(2021)\citenamefont
  {Rigobello}, \citenamefont {Notarnicola}, \citenamefont {Magnifico},\ and\
  \citenamefont {Montangero}}]{rigobello2021entanglement}%
  \BibitemOpen
  \bibfield  {author} {\bibinfo {author} {\bibfnamefont {M.}~\bibnamefont
  {Rigobello}}, \bibinfo {author} {\bibfnamefont {S.}~\bibnamefont
  {Notarnicola}}, \bibinfo {author} {\bibfnamefont {G.}~\bibnamefont
  {Magnifico}},\ and\ \bibinfo {author} {\bibfnamefont {S.}~\bibnamefont
  {Montangero}},\ }\bibfield  {title} {\bibinfo {title} {{Entanglement
  generation in $(1+1)\mathrm{D}$ QED scattering processes}},\ }\href
  {https://doi.org/10.1103/PhysRevD.104.114501} {\bibfield  {journal} {\bibinfo
   {journal} {Phys. Rev. D}\ }\textbf {\bibinfo {volume} {104}},\ \bibinfo
  {pages} {114501} (\bibinfo {year} {2021})}\BibitemShut {NoStop}%
\bibitem [{\citenamefont {Belyansky}\ \emph {et~al.}(2023)\citenamefont
  {Belyansky}, \citenamefont {Whitsitt}, \citenamefont {Mueller}, \citenamefont
  {Fahimniya}, \citenamefont {Bennewitz}, \citenamefont {Davoudi},\ and\
  \citenamefont {Gorshkov}}]{belyansky2023high}%
  \BibitemOpen
  \bibfield  {author} {\bibinfo {author} {\bibfnamefont {R.}~\bibnamefont
  {Belyansky}}, \bibinfo {author} {\bibfnamefont {S.}~\bibnamefont {Whitsitt}},
  \bibinfo {author} {\bibfnamefont {N.}~\bibnamefont {Mueller}}, \bibinfo
  {author} {\bibfnamefont {A.}~\bibnamefont {Fahimniya}}, \bibinfo {author}
  {\bibfnamefont {E.~R.}\ \bibnamefont {Bennewitz}}, \bibinfo {author}
  {\bibfnamefont {Z.}~\bibnamefont {Davoudi}},\ and\ \bibinfo {author}
  {\bibfnamefont {A.~V.}\ \bibnamefont {Gorshkov}},\ }\bibfield  {title}
  {\bibinfo {title} {{High-Energy Collision of Quarks and Hadrons in the
  Schwinger Model: From Tensor Networks to Circuit QED}},\ }\bibfield
  {journal} {\bibinfo  {journal} {arXiv preprint}\ }\href
  {https://doi.org/10.48550/arXiv.2307.02522} {10.48550/arXiv.2307.02522}
  (\bibinfo {year} {2023})\BibitemShut {NoStop}%
\bibitem [{\citenamefont {Barata}\ and\ \citenamefont
  {Rico}(2025)}]{barata2025real}%
  \BibitemOpen
  \bibfield  {author} {\bibinfo {author} {\bibfnamefont {J.}~\bibnamefont
  {Barata}}\ and\ \bibinfo {author} {\bibfnamefont {E.}~\bibnamefont {Rico}},\
  }\bibfield  {title} {\bibinfo {title} {Real-time simulation of jet energy
  loss and entropy production in high-energy scattering with matter},\
  }\href@noop {} {\bibfield  {journal} {\bibinfo  {journal} {arXiv preprint
  arXiv:2502.17558}\ } (\bibinfo {year} {2025})}\BibitemShut {NoStop}%
\bibitem [{\citenamefont {Papaefstathiou}\ \emph {et~al.}(2025)\citenamefont
  {Papaefstathiou}, \citenamefont {Knolle},\ and\ \citenamefont
  {Ba\~nuls}}]{papaefstathiou2025}%
  \BibitemOpen
  \bibfield  {author} {\bibinfo {author} {\bibfnamefont {I.}~\bibnamefont
  {Papaefstathiou}}, \bibinfo {author} {\bibfnamefont {J.}~\bibnamefont
  {Knolle}},\ and\ \bibinfo {author} {\bibfnamefont {M.~C.}\ \bibnamefont
  {Ba\~nuls}},\ }\bibfield  {title} {\bibinfo {title} {Real-time scattering in
  the lattice schwinger model},\ }\href
  {https://doi.org/10.1103/PhysRevD.111.014504} {\bibfield  {journal} {\bibinfo
   {journal} {Phys. Rev. D}\ }\textbf {\bibinfo {volume} {111}},\ \bibinfo
  {pages} {014504} (\bibinfo {year} {2025})}\BibitemShut {NoStop}%
\bibitem [{\citenamefont {Bañuls}\ \emph {et~al.}(2025)\citenamefont
  {Bañuls}, \citenamefont {Cichy}, \citenamefont {Lin},\ and\ \citenamefont
  {Schneider}}]{banuls2025pdf}%
  \BibitemOpen
  \bibfield  {author} {\bibinfo {author} {\bibfnamefont {M.~C.}\ \bibnamefont
  {Bañuls}}, \bibinfo {author} {\bibfnamefont {K.}~\bibnamefont {Cichy}},
  \bibinfo {author} {\bibfnamefont {C.~J.~D.}\ \bibnamefont {Lin}},\ and\
  \bibinfo {author} {\bibfnamefont {M.}~\bibnamefont {Schneider}},\ }\href
  {https://arxiv.org/abs/2504.07508} {\bibinfo {title} {Parton distribution
  functions in the schwinger model from tensor network states}} (\bibinfo
  {year} {2025}),\ \Eprint {https://arxiv.org/abs/2504.07508} {arXiv:2504.07508
  [hep-lat]} \BibitemShut {NoStop}%
\bibitem [{\citenamefont {Kogut}\ \emph {et~al.}(1981)\citenamefont {Kogut},
  \citenamefont {Sinclair}, \citenamefont {Pearson}, \citenamefont
  {Richardson},\ and\ \citenamefont {Shigemitsu}}]{kogut1981string}%
  \BibitemOpen
  \bibfield  {author} {\bibinfo {author} {\bibfnamefont {J.~B.}\ \bibnamefont
  {Kogut}}, \bibinfo {author} {\bibfnamefont {D.~K.}\ \bibnamefont {Sinclair}},
  \bibinfo {author} {\bibfnamefont {R.~B.}\ \bibnamefont {Pearson}}, \bibinfo
  {author} {\bibfnamefont {J.~L.}\ \bibnamefont {Richardson}},\ and\ \bibinfo
  {author} {\bibfnamefont {J.}~\bibnamefont {Shigemitsu}},\ }\bibfield  {title}
  {\bibinfo {title} {Fluctuating string of lattice gauge theory: The
  heavy-quark potential, the restoration of rotational symmetry, and
  roughening},\ }\href {https://doi.org/10.1103/physrevd.23.2945} {\bibfield
  {journal} {\bibinfo  {journal} {Physical Review D}\ }\textbf {\bibinfo
  {volume} {23}},\ \bibinfo {pages} {2945–2961} (\bibinfo {year}
  {1981})}\BibitemShut {NoStop}%
\bibitem [{\citenamefont {Xu}\ \emph {et~al.}(2025)\citenamefont {Xu},
  \citenamefont {Knap},\ and\ \citenamefont {Pollmann}}]{xu2025tensor}%
  \BibitemOpen
  \bibfield  {author} {\bibinfo {author} {\bibfnamefont {W.-T.}\ \bibnamefont
  {Xu}}, \bibinfo {author} {\bibfnamefont {M.}~\bibnamefont {Knap}},\ and\
  \bibinfo {author} {\bibfnamefont {F.}~\bibnamefont {Pollmann}},\ }\href
  {https://arxiv.org/abs/2503.19027} {\bibinfo {title} {Tensor-network study of
  the roughening transition in (2 + 1)d lattice gauge theories}} (\bibinfo
  {year} {2025}),\ \Eprint {https://arxiv.org/abs/2503.19027} {arXiv:2503.19027
  [cond-mat.str-el]} \BibitemShut {NoStop}%
\bibitem [{\citenamefont {Sugihara}(2005)}]{sugihara2005matrix}%
  \BibitemOpen
  \bibfield  {author} {\bibinfo {author} {\bibfnamefont {T.}~\bibnamefont
  {Sugihara}},\ }\bibfield  {title} {\bibinfo {title} {Matrix product
  representation of gauge invariant states in a {$Z_2$} lattice gauge theory},\
  }\href {https://doi.org/10.1088/1126-6708/2005/07/022} {\bibfield  {journal}
  {\bibinfo  {journal} {Journal of High Energy Physics}\ }\textbf {\bibinfo
  {volume} {2005}},\ \bibinfo {pages} {022–022} (\bibinfo {year}
  {2005})}\BibitemShut {NoStop}%
\bibitem [{\citenamefont {Tagliacozzo}\ and\ \citenamefont
  {Vidal}(2011)}]{tagliacozzo2011entanglement}%
  \BibitemOpen
  \bibfield  {author} {\bibinfo {author} {\bibfnamefont {L.}~\bibnamefont
  {Tagliacozzo}}\ and\ \bibinfo {author} {\bibfnamefont {G.}~\bibnamefont
  {Vidal}},\ }\bibfield  {title} {\bibinfo {title} {Entanglement
  renormalization and gauge symmetry},\ }\href
  {https://doi.org/10.1103/PhysRevB.83.115127} {\bibfield  {journal} {\bibinfo
  {journal} {Phys. Rev. B}\ }\textbf {\bibinfo {volume} {83}},\ \bibinfo
  {pages} {115127} (\bibinfo {year} {2011})}\BibitemShut {NoStop}%
\bibitem [{\citenamefont {Wu}\ and\ \citenamefont
  {Liu}(2025)}]{wu2025gauge_invariant_tn}%
  \BibitemOpen
  \bibfield  {author} {\bibinfo {author} {\bibfnamefont {Y.}~\bibnamefont
  {Wu}}\ and\ \bibinfo {author} {\bibfnamefont {W.-Y.}\ \bibnamefont {Liu}},\
  }\href {https://arxiv.org/abs/2503.20566} {\bibinfo {title} {Accurate
  gauge-invariant tensor network simulations for abelian lattice gauge theory
  in (2+1)d}} (\bibinfo {year} {2025}),\ \Eprint
  {https://arxiv.org/abs/2503.20566} {arXiv:2503.20566 [cond-mat.str-el]}
  \BibitemShut {NoStop}%
\bibitem [{\citenamefont {Liu}\ \emph {et~al.}(2013)\citenamefont {Liu},
  \citenamefont {Meurice}, \citenamefont {Qin}, \citenamefont {Unmuth-Yockey},
  \citenamefont {Xiang}, \citenamefont {Xie}, \citenamefont {Yu},\ and\
  \citenamefont {Zou}}]{liu2013blocking_formulas}%
  \BibitemOpen
  \bibfield  {author} {\bibinfo {author} {\bibfnamefont {Y.}~\bibnamefont
  {Liu}}, \bibinfo {author} {\bibfnamefont {Y.}~\bibnamefont {Meurice}},
  \bibinfo {author} {\bibfnamefont {M.~P.}\ \bibnamefont {Qin}}, \bibinfo
  {author} {\bibfnamefont {J.}~\bibnamefont {Unmuth-Yockey}}, \bibinfo {author}
  {\bibfnamefont {T.}~\bibnamefont {Xiang}}, \bibinfo {author} {\bibfnamefont
  {Z.~Y.}\ \bibnamefont {Xie}}, \bibinfo {author} {\bibfnamefont {J.~F.}\
  \bibnamefont {Yu}},\ and\ \bibinfo {author} {\bibfnamefont {H.}~\bibnamefont
  {Zou}},\ }\bibfield  {title} {\bibinfo {title} {Exact blocking formulas for
  spin and gauge models},\ }\href {https://doi.org/10.1103/PhysRevD.88.056005}
  {\bibfield  {journal} {\bibinfo  {journal} {Phys. Rev. D}\ }\textbf {\bibinfo
  {volume} {88}},\ \bibinfo {pages} {056005} (\bibinfo {year}
  {2013})}\BibitemShut {NoStop}%
\bibitem [{\citenamefont {Borla}\ \emph {et~al.}(2025)\citenamefont {Borla},
  \citenamefont {Osborne}, \citenamefont {Moroz},\ and\ \citenamefont
  {Halimeh}}]{borla2025breaking}%
  \BibitemOpen
  \bibfield  {author} {\bibinfo {author} {\bibfnamefont {U.}~\bibnamefont
  {Borla}}, \bibinfo {author} {\bibfnamefont {J.~J.}\ \bibnamefont {Osborne}},
  \bibinfo {author} {\bibfnamefont {S.}~\bibnamefont {Moroz}},\ and\ \bibinfo
  {author} {\bibfnamefont {J.~C.}\ \bibnamefont {Halimeh}},\ }\href
  {https://arxiv.org/abs/2501.17929} {\bibinfo {title} {String breaking in a
  2+1d {$Z_2$} lattice gauge theory}} (\bibinfo {year} {2025}),\ \Eprint
  {https://arxiv.org/abs/2501.17929} {arXiv:2501.17929 [quant-ph]} \BibitemShut
  {NoStop}%
\bibitem [{\citenamefont {Kogut}\ and\ \citenamefont
  {Susskind}(1975)}]{kogut1975hamiltonian}%
  \BibitemOpen
  \bibfield  {author} {\bibinfo {author} {\bibfnamefont {J.}~\bibnamefont
  {Kogut}}\ and\ \bibinfo {author} {\bibfnamefont {L.}~\bibnamefont
  {Susskind}},\ }\bibfield  {title} {\bibinfo {title} {Hamiltonian formulation
  of wilson's lattice gauge theories},\ }\href
  {https://doi.org/10.1103/PhysRevD.11.395} {\bibfield  {journal} {\bibinfo
  {journal} {Phys. Rev. D}\ }\textbf {\bibinfo {volume} {11}},\ \bibinfo
  {pages} {395} (\bibinfo {year} {1975})}\BibitemShut {NoStop}%
\bibitem [{\citenamefont {Horn}\ \emph {et~al.}(1979)\citenamefont {Horn},
  \citenamefont {Weinstein},\ and\ \citenamefont
  {Yankielowicz}}]{horn1979hamiltonian}%
  \BibitemOpen
  \bibfield  {author} {\bibinfo {author} {\bibfnamefont {D.}~\bibnamefont
  {Horn}}, \bibinfo {author} {\bibfnamefont {M.}~\bibnamefont {Weinstein}},\
  and\ \bibinfo {author} {\bibfnamefont {S.}~\bibnamefont {Yankielowicz}},\
  }\bibfield  {title} {\bibinfo {title} {Hamiltonian approach to $z(n)$ lattice
  gauge theories},\ }\href {https://doi.org/10.1103/physrevd.19.3715}
  {\bibfield  {journal} {\bibinfo  {journal} {Physical Review D}\ }\textbf
  {\bibinfo {volume} {19}},\ \bibinfo {pages} {3715–3731} (\bibinfo {year}
  {1979})}\BibitemShut {NoStop}%
\bibitem [{\citenamefont {P{\'e}rez-Garc{\'\i}a}\ \emph
  {et~al.}(2007)\citenamefont {P{\'e}rez-Garc{\'\i}a}, \citenamefont
  {Verstraete}, \citenamefont {Wolf},\ and\ \citenamefont
  {Cirac}}]{PerezGarcia2007}%
  \BibitemOpen
  \bibfield  {author} {\bibinfo {author} {\bibfnamefont {D.}~\bibnamefont
  {P{\'e}rez-Garc{\'\i}a}}, \bibinfo {author} {\bibfnamefont {F.}~\bibnamefont
  {Verstraete}}, \bibinfo {author} {\bibfnamefont {M.~M.}\ \bibnamefont
  {Wolf}},\ and\ \bibinfo {author} {\bibfnamefont {J.~I.}\ \bibnamefont
  {Cirac}},\ }\bibfield  {title} {\bibinfo {title} {Matrix product state
  representations},\ }\href
  {http://www.rintonpress.com/xqic7/qic-7-56/401-430.pdf} {\bibfield  {journal}
  {\bibinfo  {journal} {Quantum Inf. Comput.}\ }\textbf {\bibinfo {volume}
  {7}},\ \bibinfo {pages} {401} (\bibinfo {year} {2007})}\BibitemShut {NoStop}%
\bibitem [{\citenamefont {Paeckel}\ \emph {et~al.}(2019)\citenamefont
  {Paeckel}, \citenamefont {K{\"o}hler}, \citenamefont {Swoboda}, \citenamefont
  {Manmana}, \citenamefont {Schollw{\"o}ck},\ and\ \citenamefont
  {Hubig}}]{Paeckel2019tevol}%
  \BibitemOpen
  \bibfield  {author} {\bibinfo {author} {\bibfnamefont {S.}~\bibnamefont
  {Paeckel}}, \bibinfo {author} {\bibfnamefont {T.}~\bibnamefont {K{\"o}hler}},
  \bibinfo {author} {\bibfnamefont {A.}~\bibnamefont {Swoboda}}, \bibinfo
  {author} {\bibfnamefont {S.~R.}\ \bibnamefont {Manmana}}, \bibinfo {author}
  {\bibfnamefont {U.}~\bibnamefont {Schollw{\"o}ck}},\ and\ \bibinfo {author}
  {\bibfnamefont {C.}~\bibnamefont {Hubig}},\ }\bibfield  {title} {\bibinfo
  {title} {Time-evolution methods for matrix-product states},\ }\href
  {https://doi.org/https://doi.org/10.1016/j.aop.2019.167998} {\bibfield
  {journal} {\bibinfo  {journal} {Annals Phys.}\ }\textbf {\bibinfo {volume}
  {411}},\ \bibinfo {pages} {167998} (\bibinfo {year} {2019})}\BibitemShut
  {NoStop}%
\bibitem [{\citenamefont {Krinitsin}\ \emph {et~al.}(2024)\citenamefont
  {Krinitsin}, \citenamefont {Tausendpfund}, \citenamefont {Rizzi},
  \citenamefont {Heyl},\ and\ \citenamefont
  {Schmitt}}]{krinitsin2024roughening}%
  \BibitemOpen
  \bibfield  {author} {\bibinfo {author} {\bibfnamefont {W.}~\bibnamefont
  {Krinitsin}}, \bibinfo {author} {\bibfnamefont {N.}~\bibnamefont
  {Tausendpfund}}, \bibinfo {author} {\bibfnamefont {M.}~\bibnamefont {Rizzi}},
  \bibinfo {author} {\bibfnamefont {M.}~\bibnamefont {Heyl}},\ and\ \bibinfo
  {author} {\bibfnamefont {M.}~\bibnamefont {Schmitt}},\ }\href
  {https://arxiv.org/abs/2412.10145} {\bibinfo {title} {Roughening dynamics of
  interfaces in two-dimensional quantum matter}} (\bibinfo {year} {2024}),\
  \Eprint {https://arxiv.org/abs/2412.10145} {arXiv:2412.10145 [quant-ph]}
  \BibitemShut {NoStop}%
\bibitem [{\citenamefont {Fradkin}\ and\ \citenamefont
  {Susskind}(1978)}]{fradkin1978order}%
  \BibitemOpen
  \bibfield  {author} {\bibinfo {author} {\bibfnamefont {E.}~\bibnamefont
  {Fradkin}}\ and\ \bibinfo {author} {\bibfnamefont {L.}~\bibnamefont
  {Susskind}},\ }\bibfield  {title} {\bibinfo {title} {Order and disorder in
  gauge systems and magnets},\ }\href
  {https://doi.org/10.1103/PhysRevD.17.2637} {\bibfield  {journal} {\bibinfo
  {journal} {Phys. Rev. D}\ }\textbf {\bibinfo {volume} {17}},\ \bibinfo
  {pages} {2637} (\bibinfo {year} {1978})}\BibitemShut {NoStop}%
\bibitem [{\citenamefont {Kogut}(1979)}]{kogut1979introduction}%
  \BibitemOpen
  \bibfield  {author} {\bibinfo {author} {\bibfnamefont {J.~B.}\ \bibnamefont
  {Kogut}},\ }\bibfield  {title} {\bibinfo {title} {An introduction to lattice
  gauge theory and spin systems},\ }\href
  {https://doi.org/10.1103/RevModPhys.51.659} {\bibfield  {journal} {\bibinfo
  {journal} {Rev. Mod. Phys.}\ }\textbf {\bibinfo {volume} {51}},\ \bibinfo
  {pages} {659} (\bibinfo {year} {1979})}\BibitemShut {NoStop}%
\bibitem [{\citenamefont {Wegner}(1971)}]{wegner1971duality}%
  \BibitemOpen
  \bibfield  {author} {\bibinfo {author} {\bibfnamefont {F.~J.}\ \bibnamefont
  {Wegner}},\ }\bibfield  {title} {\bibinfo {title} {Duality in generalized
  ising models and phase transitions without local order parameters},\ }\href
  {https://doi.org/10.1063/1.1665530} {\bibfield  {journal} {\bibinfo
  {journal} {Journal of Mathematical Physics}\ }\textbf {\bibinfo {volume}
  {12}},\ \bibinfo {pages} {2259} (\bibinfo {year} {1971})}\BibitemShut
  {NoStop}%
\bibitem [{\citenamefont {Cobanera}\ \emph {et~al.}(2011)\citenamefont
  {Cobanera}, \citenamefont {Ortiz},\ and\ \citenamefont
  {Nussinov}}]{cobanera2011bond}%
  \BibitemOpen
  \bibfield  {author} {\bibinfo {author} {\bibfnamefont {E.}~\bibnamefont
  {Cobanera}}, \bibinfo {author} {\bibfnamefont {G.}~\bibnamefont {Ortiz}},\
  and\ \bibinfo {author} {\bibfnamefont {Z.}~\bibnamefont {Nussinov}},\
  }\bibfield  {title} {\bibinfo {title} {The bond-algebraic approach to
  dualities},\ }\href {https://doi.org/10.1080/00018732.2011.619814} {\bibfield
   {journal} {\bibinfo  {journal} {Advances in physics}\ }\textbf {\bibinfo
  {volume} {60}},\ \bibinfo {pages} {679} (\bibinfo {year} {2011})}\BibitemShut
  {NoStop}%
\bibitem [{\citenamefont {Berezinskii}(1971)}]{berezinskii1971long-range}%
  \BibitemOpen
  \bibfield  {author} {\bibinfo {author} {\bibfnamefont {V.}~\bibnamefont
  {Berezinskii}},\ }\bibfield  {title} {\bibinfo {title} {Destruction of
  long-range order in one-dimensional and two-dimensional systems having a
  continuous symmetry group i. classical systems},\ }\href@noop {} {\bibfield
  {journal} {\bibinfo  {journal} {Sov. Phys. JETP}\ }\textbf {\bibinfo {volume}
  {32}},\ \bibinfo {pages} {493} (\bibinfo {year} {1971})}\BibitemShut
  {NoStop}%
\bibitem [{\citenamefont {Kosterlitz}\ and\ \citenamefont
  {Thouless}(1973)}]{kosterlitz1973ordering}%
  \BibitemOpen
  \bibfield  {author} {\bibinfo {author} {\bibfnamefont {J.~M.}\ \bibnamefont
  {Kosterlitz}}\ and\ \bibinfo {author} {\bibfnamefont {D.~J.}\ \bibnamefont
  {Thouless}},\ }\bibfield  {title} {\bibinfo {title} {Ordering, metastability
  and phase transitions in two-dimensional systems},\ }\href
  {https://doi.org/10.1088/0022-3719/6/7/010} {\bibfield  {journal} {\bibinfo
  {journal} {Journal of Physics C: Solid State Physics}\ }\textbf {\bibinfo
  {volume} {6}},\ \bibinfo {pages} {1181–1203} (\bibinfo {year}
  {1973})}\BibitemShut {NoStop}%
\bibitem [{\citenamefont {Kosterlitz}(1974)}]{kosterlitz1974critical}%
  \BibitemOpen
  \bibfield  {author} {\bibinfo {author} {\bibfnamefont {J.~M.}\ \bibnamefont
  {Kosterlitz}},\ }\bibfield  {title} {\bibinfo {title} {The critical
  properties of the two-dimensional xy model},\ }\href
  {https://doi.org/10.1088/0022-3719/7/6/005} {\bibfield  {journal} {\bibinfo
  {journal} {Journal of Physics C: Solid State Physics}\ }\textbf {\bibinfo
  {volume} {7}},\ \bibinfo {pages} {1046–1060} (\bibinfo {year}
  {1974})}\BibitemShut {NoStop}%
\bibitem [{\citenamefont {Fröhlich}\ and\ \citenamefont
  {Spencer}(1981)}]{frohlich1981kosterlitz-thouless}%
  \BibitemOpen
  \bibfield  {author} {\bibinfo {author} {\bibfnamefont {J.}~\bibnamefont
  {Fröhlich}}\ and\ \bibinfo {author} {\bibfnamefont {T.}~\bibnamefont
  {Spencer}},\ }\bibfield  {title} {\bibinfo {title} {The kosterlitz-thouless
  transition in two-dimensional abelian spin systems and the coulomb gas},\
  }\href {https://doi.org/10.1007/bf01208273} {\bibfield  {journal} {\bibinfo
  {journal} {Communications in Mathematical Physics}\ }\textbf {\bibinfo
  {volume} {81}},\ \bibinfo {pages} {527–602} (\bibinfo {year}
  {1981})}\BibitemShut {NoStop}%
\bibitem [{\citenamefont {Blöte}\ and\ \citenamefont
  {Deng}(2002)}]{blote2002cluster}%
  \BibitemOpen
  \bibfield  {author} {\bibinfo {author} {\bibfnamefont {H.~W.~J.}\
  \bibnamefont {Blöte}}\ and\ \bibinfo {author} {\bibfnamefont
  {Y.}~\bibnamefont {Deng}},\ }\bibfield  {title} {\bibinfo {title} {Cluster
  monte carlo simulation of the transverse ising model},\ }\bibfield  {journal}
  {\bibinfo  {journal} {Physical Review E}\ }\textbf {\bibinfo {volume} {66}},\
  \href {https://doi.org/10.1103/physreve.66.066110}
  {10.1103/physreve.66.066110} (\bibinfo {year} {2002})\BibitemShut {NoStop}%
\bibitem [{Note1()}]{Note1}%
  \BibitemOpen
  \bibinfo {note} {The saturation values will be $w^2 \rightarrow (2k^2 + 1)/6$
  if $N = 2k$ and $w^2 \rightarrow (k(k + 1))/3$ if $N = 2k+1$.}\BibitemShut
  {Stop}%
\bibitem [{\citenamefont {Amico}\ \emph {et~al.}(2008)\citenamefont {Amico},
  \citenamefont {Fazio}, \citenamefont {Osterloh},\ and\ \citenamefont
  {Vedral}}]{amico2008entanglement}%
  \BibitemOpen
  \bibfield  {author} {\bibinfo {author} {\bibfnamefont {L.}~\bibnamefont
  {Amico}}, \bibinfo {author} {\bibfnamefont {R.}~\bibnamefont {Fazio}},
  \bibinfo {author} {\bibfnamefont {A.}~\bibnamefont {Osterloh}},\ and\
  \bibinfo {author} {\bibfnamefont {V.}~\bibnamefont {Vedral}},\ }\bibfield
  {title} {\bibinfo {title} {Entanglement in many-body systems},\ }\href
  {https://doi.org/10.1103/revmodphys.80.517} {\bibfield  {journal} {\bibinfo
  {journal} {Reviews of Modern Physics}\ }\textbf {\bibinfo {volume} {80}},\
  \bibinfo {pages} {517–576} (\bibinfo {year} {2008})}\BibitemShut {NoStop}%
\bibitem [{\citenamefont {Holzhey}\ \emph {et~al.}(1994)\citenamefont
  {Holzhey}, \citenamefont {Larsen},\ and\ \citenamefont
  {Wilczek}}]{holzhey1994entropy}%
  \BibitemOpen
  \bibfield  {author} {\bibinfo {author} {\bibfnamefont {C.}~\bibnamefont
  {Holzhey}}, \bibinfo {author} {\bibfnamefont {F.}~\bibnamefont {Larsen}},\
  and\ \bibinfo {author} {\bibfnamefont {F.}~\bibnamefont {Wilczek}},\
  }\bibfield  {title} {\bibinfo {title} {Geometric and renormalized entropy in
  conformal field theory},\ }\href
  {https://doi.org/https://doi.org/10.1016/0550-3213(94)90402-2} {\bibfield
  {journal} {\bibinfo  {journal} {Nuclear Physics B}\ }\textbf {\bibinfo
  {volume} {424}},\ \bibinfo {pages} {443} (\bibinfo {year}
  {1994})}\BibitemShut {NoStop}%
\bibitem [{\citenamefont {Calabrese}\ and\ \citenamefont
  {Cardy}(2004)}]{calabrese2004entanglement}%
  \BibitemOpen
  \bibfield  {author} {\bibinfo {author} {\bibfnamefont {P.}~\bibnamefont
  {Calabrese}}\ and\ \bibinfo {author} {\bibfnamefont {J.}~\bibnamefont
  {Cardy}},\ }\bibfield  {title} {\bibinfo {title} {Entanglement entropy and
  quantum field theory},\ }\href
  {https://doi.org/10.1088/1742-5468/2004/06/P06002} {\bibfield  {journal}
  {\bibinfo  {journal} {Journal of Statistical Mechanics: Theory and
  Experiment}\ }\textbf {\bibinfo {volume} {2004}},\ \bibinfo {pages} {P06002}
  (\bibinfo {year} {2004})}\BibitemShut {NoStop}%
\bibitem [{\citenamefont {Korepin}(2004)}]{korepin2004universality}%
  \BibitemOpen
  \bibfield  {author} {\bibinfo {author} {\bibfnamefont {V.~E.}\ \bibnamefont
  {Korepin}},\ }\bibfield  {title} {\bibinfo {title} {Universality of entropy
  scaling in one dimensional gapless models},\ }\bibfield  {journal} {\bibinfo
  {journal} {Physical Review Letters}\ }\textbf {\bibinfo {volume} {92}},\
  \href {https://doi.org/10.1103/physrevlett.92.096402}
  {10.1103/physrevlett.92.096402} (\bibinfo {year} {2004})\BibitemShut
  {NoStop}%
\bibitem [{\citenamefont {Vidal}\ \emph {et~al.}(2003)\citenamefont {Vidal},
  \citenamefont {Latorre}, \citenamefont {Rico},\ and\ \citenamefont
  {Kitaev}}]{vidal2003entanglement}%
  \BibitemOpen
  \bibfield  {author} {\bibinfo {author} {\bibfnamefont {G.}~\bibnamefont
  {Vidal}}, \bibinfo {author} {\bibfnamefont {J.~I.}\ \bibnamefont {Latorre}},
  \bibinfo {author} {\bibfnamefont {E.}~\bibnamefont {Rico}},\ and\ \bibinfo
  {author} {\bibfnamefont {A.}~\bibnamefont {Kitaev}},\ }\bibfield  {title}
  {\bibinfo {title} {Entanglement in quantum critical phenomena},\ }\href
  {https://doi.org/10.1103/PhysRevLett.90.227902} {\bibfield  {journal}
  {\bibinfo  {journal} {Phys. Rev. Lett.}\ }\textbf {\bibinfo {volume} {90}},\
  \bibinfo {pages} {227902} (\bibinfo {year} {2003})}\BibitemShut {NoStop}%
\bibitem [{\citenamefont {Lüscher}\ \emph {et~al.}(1980)\citenamefont
  {Lüscher}, \citenamefont {Symanzik},\ and\ \citenamefont
  {Weisz}}]{luscher1980anomalies}%
  \BibitemOpen
  \bibfield  {author} {\bibinfo {author} {\bibfnamefont {M.}~\bibnamefont
  {Lüscher}}, \bibinfo {author} {\bibfnamefont {K.}~\bibnamefont {Symanzik}},\
  and\ \bibinfo {author} {\bibfnamefont {P.}~\bibnamefont {Weisz}},\ }\bibfield
   {title} {\bibinfo {title} {Anomalies of the free loop wave equation in the
  wkb approximation},\ }\href {https://doi.org/10.1016/0550-3213(80)90009-7}
  {\bibfield  {journal} {\bibinfo  {journal} {Nuclear Physics B}\ }\textbf
  {\bibinfo {volume} {173}},\ \bibinfo {pages} {365–396} (\bibinfo {year}
  {1980})}\BibitemShut {NoStop}%
\bibitem [{\citenamefont {Kuti}(2006)}]{kuti2006lattice}%
  \BibitemOpen
  \bibfield  {author} {\bibinfo {author} {\bibfnamefont {J.}~\bibnamefont
  {Kuti}},\ }\bibfield  {title} {\bibinfo {title} {Lattice qcd and string
  theory},\ }\href {https://doi.org/10.1142/s0217751x06031910} {\bibfield
  {journal} {\bibinfo  {journal} {International Journal of Modern Physics A}\
  }\textbf {\bibinfo {volume} {21}},\ \bibinfo {pages} {699–706} (\bibinfo
  {year} {2006})}\BibitemShut {NoStop}%
\bibitem [{\citenamefont {Caselle}\ \emph {et~al.}(2013)\citenamefont
  {Caselle}, \citenamefont {Fioravanti}, \citenamefont {Gliozzi},\ and\
  \citenamefont {Tateo}}]{caselle2013quantisation}%
  \BibitemOpen
  \bibfield  {author} {\bibinfo {author} {\bibfnamefont {M.}~\bibnamefont
  {Caselle}}, \bibinfo {author} {\bibfnamefont {D.}~\bibnamefont {Fioravanti}},
  \bibinfo {author} {\bibfnamefont {F.}~\bibnamefont {Gliozzi}},\ and\ \bibinfo
  {author} {\bibfnamefont {R.}~\bibnamefont {Tateo}},\ }\bibfield  {title}
  {\bibinfo {title} {Quantisation of the effective string with tba},\
  }\href@noop {} {\bibfield  {journal} {\bibinfo  {journal} {Journal of High
  Energy Physics}\ }\textbf {\bibinfo {volume} {2013}},\ \bibinfo {pages} {1}
  (\bibinfo {year} {2013})}\BibitemShut {NoStop}%
\bibitem [{\citenamefont {Blöte}\ \emph {et~al.}(1986)\citenamefont {Blöte},
  \citenamefont {Cardy},\ and\ \citenamefont
  {Nightingale}}]{blote1986conformal}%
  \BibitemOpen
  \bibfield  {author} {\bibinfo {author} {\bibfnamefont {H.~W.~J.}\
  \bibnamefont {Blöte}}, \bibinfo {author} {\bibfnamefont {J.~L.}\
  \bibnamefont {Cardy}},\ and\ \bibinfo {author} {\bibfnamefont {M.~P.}\
  \bibnamefont {Nightingale}},\ }\bibfield  {title} {\bibinfo {title}
  {Conformal invariance, the central charge, and universal finite-size
  amplitudes at criticality},\ }\href
  {https://doi.org/10.1103/physrevlett.56.742} {\bibfield  {journal} {\bibinfo
  {journal} {Physical Review Letters}\ }\textbf {\bibinfo {volume} {56}},\
  \bibinfo {pages} {742–745} (\bibinfo {year} {1986})}\BibitemShut {NoStop}%
\bibitem [{\citenamefont {Affleck}(1988)}]{affleck1988universal}%
  \BibitemOpen
  \bibfield  {author} {\bibinfo {author} {\bibfnamefont {I.}~\bibnamefont
  {Affleck}},\ }\bibinfo {title} {Universal term in the free energy at a
  critical point and the conformal anomaly},\ in\ \href
  {https://doi.org/10.1016/b978-0-444-87109-1.50034-0} {\emph {\bibinfo
  {booktitle} {Finite-Size Scaling}}}\ (\bibinfo  {publisher} {Elsevier},\
  \bibinfo {year} {1988})\ p.\ \bibinfo {pages} {347–349}\BibitemShut
  {NoStop}%
\bibitem [{\citenamefont {Francesco}\ \emph {et~al.}(2012)\citenamefont
  {Francesco}, \citenamefont {Mathieu},\ and\ \citenamefont
  {S{\'e}n{\'e}chal}}]{difrancesco2012conformal}%
  \BibitemOpen
  \bibfield  {author} {\bibinfo {author} {\bibfnamefont {P.}~\bibnamefont
  {Francesco}}, \bibinfo {author} {\bibfnamefont {P.}~\bibnamefont {Mathieu}},\
  and\ \bibinfo {author} {\bibfnamefont {D.}~\bibnamefont {S{\'e}n{\'e}chal}},\
  }\href@noop {} {\emph {\bibinfo {title} {Conformal field theory}}}\ (\bibinfo
   {publisher} {Springer Science \& Business Media},\ \bibinfo {year}
  {2012})\BibitemShut {NoStop}%
\bibitem [{Note2()}]{Note2}%
  \BibitemOpen
  \bibinfo {note} {Lorentz-invariance on a lattice has to be understood as
  isotropy between the time and space directions}\BibitemShut {NoStop}%
\bibitem [{\citenamefont {Kogut}\ and\ \citenamefont
  {Sinclair}(1981)}]{kogut1981analyticity}%
  \BibitemOpen
  \bibfield  {author} {\bibinfo {author} {\bibfnamefont {J.~B.}\ \bibnamefont
  {Kogut}}\ and\ \bibinfo {author} {\bibfnamefont {D.~K.}\ \bibnamefont
  {Sinclair}},\ }\bibfield  {title} {\bibinfo {title} {Analyticity of the
  off-axis string tension and the restoration of rotational symmetry in lattice
  systems},\ }\href {https://doi.org/10.1103/physrevd.24.1610} {\bibfield
  {journal} {\bibinfo  {journal} {Physical Review D}\ }\textbf {\bibinfo
  {volume} {24}},\ \bibinfo {pages} {1610–1616} (\bibinfo {year}
  {1981})}\BibitemShut {NoStop}%
\bibitem [{\citenamefont {Calabrese}\ and\ \citenamefont
  {Cardy}(2005)}]{calabrese2005evolution}%
  \BibitemOpen
  \bibfield  {author} {\bibinfo {author} {\bibfnamefont {P.}~\bibnamefont
  {Calabrese}}\ and\ \bibinfo {author} {\bibfnamefont {J.}~\bibnamefont
  {Cardy}},\ }\bibfield  {title} {\bibinfo {title} {Evolution of entanglement
  entropy in one-dimensional systems},\ }\href
  {https://doi.org/10.1088/1742-5468/2005/04/P04010} {\bibfield  {journal}
  {\bibinfo  {journal} {Journal of Statistical Mechanics: Theory and
  Experiment}\ }\textbf {\bibinfo {volume} {2005}},\ \bibinfo {pages} {P04010}
  (\bibinfo {year} {2005})}\BibitemShut {NoStop}%
\bibitem [{\citenamefont {De~Chiara}\ \emph {et~al.}(2006)\citenamefont
  {De~Chiara}, \citenamefont {Montangero}, \citenamefont {Calabrese},\ and\
  \citenamefont {Rosario}}]{dechiara2006heisenberg}%
  \BibitemOpen
  \bibfield  {author} {\bibinfo {author} {\bibfnamefont {G.}~\bibnamefont
  {De~Chiara}}, \bibinfo {author} {\bibfnamefont {S.}~\bibnamefont
  {Montangero}}, \bibinfo {author} {\bibfnamefont {P.}~\bibnamefont
  {Calabrese}},\ and\ \bibinfo {author} {\bibfnamefont {F.}~\bibnamefont
  {Rosario}},\ }\bibfield  {title} {\bibinfo {title} {Entanglement entropy
  dynamics of heisenberg chains},\ }\href
  {https://doi.org/10.1088/1742-5468/2006/03/P03001} {\bibfield  {journal}
  {\bibinfo  {journal} {Journal of Statistical Mechanics: Theory and
  Experiment}\ }\textbf {\bibinfo {volume} {2006}},\ \bibinfo {pages} {P03001}
  (\bibinfo {year} {2006})}\BibitemShut {NoStop}%
\bibitem [{\citenamefont {Alba}\ and\ \citenamefont
  {Calabrese}(2017)}]{Alba2017}%
  \BibitemOpen
  \bibfield  {author} {\bibinfo {author} {\bibfnamefont {V.}~\bibnamefont
  {Alba}}\ and\ \bibinfo {author} {\bibfnamefont {P.}~\bibnamefont
  {Calabrese}},\ }\bibfield  {title} {\bibinfo {title} {Entanglement and
  thermodynamics after a quantum quench in integrable systems},\ }\href
  {https://doi.org/10.1073/pnas.1703516114} {\bibfield  {journal} {\bibinfo
  {journal} {Proceedings of the National Academy of Sciences}\ }\textbf
  {\bibinfo {volume} {114}},\ \bibinfo {pages} {7947} (\bibinfo {year}
  {2017})},\ \Eprint
  {https://arxiv.org/abs/https://www.pnas.org/doi/pdf/10.1073/pnas.1703516114}
  {https://www.pnas.org/doi/pdf/10.1073/pnas.1703516114} \BibitemShut {NoStop}%
\bibitem [{\citenamefont {Kim}\ and\ \citenamefont
  {Huse}(2013)}]{Kim2013ballistic}%
  \BibitemOpen
  \bibfield  {author} {\bibinfo {author} {\bibfnamefont {H.}~\bibnamefont
  {Kim}}\ and\ \bibinfo {author} {\bibfnamefont {D.~A.}\ \bibnamefont {Huse}},\
  }\bibfield  {title} {\bibinfo {title} {Ballistic spreading of entanglement in
  a diffusive nonintegrable system},\ }\href
  {https://doi.org/10.1103/PhysRevLett.111.127205} {\bibfield  {journal}
  {\bibinfo  {journal} {Phys. Rev. Lett.}\ }\textbf {\bibinfo {volume} {111}},\
  \bibinfo {pages} {127205} (\bibinfo {year} {2013})}\BibitemShut {NoStop}%
\bibitem [{\citenamefont {Ho}\ and\ \citenamefont {Abanin}(2017)}]{Ho2017ent}%
  \BibitemOpen
  \bibfield  {author} {\bibinfo {author} {\bibfnamefont {W.~W.}\ \bibnamefont
  {Ho}}\ and\ \bibinfo {author} {\bibfnamefont {D.~A.}\ \bibnamefont
  {Abanin}},\ }\bibfield  {title} {\bibinfo {title} {Entanglement dynamics in
  quantum many-body systems},\ }\href
  {https://doi.org/10.1103/PhysRevB.95.094302} {\bibfield  {journal} {\bibinfo
  {journal} {Phys. Rev. B}\ }\textbf {\bibinfo {volume} {95}},\ \bibinfo
  {pages} {094302} (\bibinfo {year} {2017})}\BibitemShut {NoStop}%
\bibitem [{\citenamefont {Nahum}\ \emph {et~al.}(2017)\citenamefont {Nahum},
  \citenamefont {Ruhman}, \citenamefont {Vijay},\ and\ \citenamefont
  {Haah}}]{Nahum2017rand}%
  \BibitemOpen
  \bibfield  {author} {\bibinfo {author} {\bibfnamefont {A.}~\bibnamefont
  {Nahum}}, \bibinfo {author} {\bibfnamefont {J.}~\bibnamefont {Ruhman}},
  \bibinfo {author} {\bibfnamefont {S.}~\bibnamefont {Vijay}},\ and\ \bibinfo
  {author} {\bibfnamefont {J.}~\bibnamefont {Haah}},\ }\bibfield  {title}
  {\bibinfo {title} {Quantum entanglement growth under random unitary
  dynamics},\ }\href {https://doi.org/10.1103/PhysRevX.7.031016} {\bibfield
  {journal} {\bibinfo  {journal} {Phys. Rev. X}\ }\textbf {\bibinfo {volume}
  {7}},\ \bibinfo {pages} {031016} (\bibinfo {year} {2017})}\BibitemShut
  {NoStop}%
\bibitem [{\citenamefont {\ifmmode \check{Z}\else
  \v{Z}\fi{}nidari\ifmmode~\check{c}\else \v{c}\fi{}}\ \emph
  {et~al.}(2008)\citenamefont {\ifmmode \check{Z}\else
  \v{Z}\fi{}nidari\ifmmode~\check{c}\else \v{c}\fi{}}, \citenamefont {Prosen},\
  and\ \citenamefont {Prelov\ifmmode~\check{s}\else
  \v{s}\fi{}ek}}]{Znidaric2008mbl}%
  \BibitemOpen
  \bibfield  {author} {\bibinfo {author} {\bibfnamefont {M.}~\bibnamefont
  {\ifmmode \check{Z}\else \v{Z}\fi{}nidari\ifmmode~\check{c}\else
  \v{c}\fi{}}}, \bibinfo {author} {\bibfnamefont {T.~c.~v.}\ \bibnamefont
  {Prosen}},\ and\ \bibinfo {author} {\bibfnamefont {P.}~\bibnamefont
  {Prelov\ifmmode~\check{s}\else \v{s}\fi{}ek}},\ }\bibfield  {title} {\bibinfo
  {title} {Many-body localization in the heisenberg $xxz$ magnet in a random
  field},\ }\href {https://doi.org/10.1103/PhysRevB.77.064426} {\bibfield
  {journal} {\bibinfo  {journal} {Phys. Rev. B}\ }\textbf {\bibinfo {volume}
  {77}},\ \bibinfo {pages} {064426} (\bibinfo {year} {2008})}\BibitemShut
  {NoStop}%
\bibitem [{\citenamefont {Bardarson}\ \emph {et~al.}(2012)\citenamefont
  {Bardarson}, \citenamefont {Pollmann},\ and\ \citenamefont
  {Moore}}]{Bardarson2012mbl}%
  \BibitemOpen
  \bibfield  {author} {\bibinfo {author} {\bibfnamefont {J.~H.}\ \bibnamefont
  {Bardarson}}, \bibinfo {author} {\bibfnamefont {F.}~\bibnamefont
  {Pollmann}},\ and\ \bibinfo {author} {\bibfnamefont {J.~E.}\ \bibnamefont
  {Moore}},\ }\bibfield  {title} {\bibinfo {title} {Unbounded growth of
  entanglement in models of many-body localization},\ }\href
  {https://doi.org/10.1103/PhysRevLett.109.017202} {\bibfield  {journal}
  {\bibinfo  {journal} {Phys. Rev. Lett.}\ }\textbf {\bibinfo {volume} {109}},\
  \bibinfo {pages} {017202} (\bibinfo {year} {2012})}\BibitemShut {NoStop}%
\bibitem [{\citenamefont {Fradm98}(2025)}]{qs-mps-repo}%
  \BibitemOpen
  \bibfield  {author} {\bibinfo {author} {\bibnamefont {Fradm98}},\ }\href@noop
  {} {\bibinfo {title} {Quantum simulation package with {MPS} in python}}
  (\bibinfo {year} {2025}),\ \bibinfo {note}
  {\url{https://github.com/Fradm98/qs-mps} [GitHub repository for
  {\texttt{qs-mps}}]}\BibitemShut {NoStop}%
\bibitem [{\citenamefont {Nambu}(1979)}]{nambu1979qcd}%
  \BibitemOpen
  \bibfield  {author} {\bibinfo {author} {\bibfnamefont {Y.}~\bibnamefont
  {Nambu}},\ }\bibfield  {title} {\bibinfo {title} {Qcd and the string model},\
  }\href {https://doi.org/10.1016/0370-2693(79)91193-6} {\bibfield  {journal}
  {\bibinfo  {journal} {Physics Letters B}\ }\textbf {\bibinfo {volume} {80}},\
  \bibinfo {pages} {372–376} (\bibinfo {year} {1979})}\BibitemShut {NoStop}%
\bibitem [{\citenamefont {Goddard}\ \emph {et~al.}(1973)\citenamefont
  {Goddard}, \citenamefont {Goldstone}, \citenamefont {Rebbi},\ and\
  \citenamefont {Thorn}}]{goddard1973quantum}%
  \BibitemOpen
  \bibfield  {author} {\bibinfo {author} {\bibfnamefont {P.}~\bibnamefont
  {Goddard}}, \bibinfo {author} {\bibfnamefont {J.}~\bibnamefont {Goldstone}},
  \bibinfo {author} {\bibfnamefont {C.}~\bibnamefont {Rebbi}},\ and\ \bibinfo
  {author} {\bibfnamefont {C.}~\bibnamefont {Thorn}},\ }\bibfield  {title}
  {\bibinfo {title} {Quantum dynamics of a massless relativistic string},\
  }\href {https://doi.org/10.1016/0550-3213(73)90223-x} {\bibfield  {journal}
  {\bibinfo  {journal} {Nuclear Physics B}\ }\textbf {\bibinfo {volume} {56}},\
  \bibinfo {pages} {109–135} (\bibinfo {year} {1973})}\BibitemShut {NoStop}%
\bibitem [{\citenamefont {Polyakov}(1986)}]{polyakov1986string}%
  \BibitemOpen
  \bibfield  {author} {\bibinfo {author} {\bibfnamefont {A.}~\bibnamefont
  {Polyakov}},\ }\bibfield  {title} {\bibinfo {title} {Fine structure of
  strings},\ }\href {https://doi.org/10.1016/0550-3213(86)90162-8} {\bibfield
  {journal} {\bibinfo  {journal} {Nuclear Physics B}\ }\textbf {\bibinfo
  {volume} {268}},\ \bibinfo {pages} {406–412} (\bibinfo {year}
  {1986})}\BibitemShut {NoStop}%
\bibitem [{\citenamefont {Polchinski}\ and\ \citenamefont
  {Strominger}(1991)}]{polchinski1991effective}%
  \BibitemOpen
  \bibfield  {author} {\bibinfo {author} {\bibfnamefont {J.}~\bibnamefont
  {Polchinski}}\ and\ \bibinfo {author} {\bibfnamefont {A.}~\bibnamefont
  {Strominger}},\ }\bibfield  {title} {\bibinfo {title} {Effective string
  theory},\ }\href {https://doi.org/10.1103/physrevlett.67.1681} {\bibfield
  {journal} {\bibinfo  {journal} {Physical Review Letters}\ }\textbf {\bibinfo
  {volume} {67}},\ \bibinfo {pages} {1681–1684} (\bibinfo {year}
  {1991})}\BibitemShut {NoStop}%
\bibitem [{\citenamefont {Mishra}\ \emph {et~al.}(2011)\citenamefont {Mishra},
  \citenamefont {Carrasquilla},\ and\ \citenamefont {Rigol}}]{mishra2011phase}%
  \BibitemOpen
  \bibfield  {author} {\bibinfo {author} {\bibfnamefont {T.}~\bibnamefont
  {Mishra}}, \bibinfo {author} {\bibfnamefont {J.}~\bibnamefont
  {Carrasquilla}},\ and\ \bibinfo {author} {\bibfnamefont {M.}~\bibnamefont
  {Rigol}},\ }\bibfield  {title} {\bibinfo {title} {Phase diagram of the
  half-filled one-dimensional {$t-V-V^{\prime}$} model},\ }\bibfield  {journal}
  {\bibinfo  {journal} {Physical Review B}\ }\textbf {\bibinfo {volume} {84}},\
  \href {https://doi.org/10.1103/physrevb.84.115135}
  {10.1103/physrevb.84.115135} (\bibinfo {year} {2011})\BibitemShut {NoStop}%
\bibitem [{\citenamefont {Dalmonte}\ \emph {et~al.}(2015)\citenamefont
  {Dalmonte}, \citenamefont {Carrasquilla}, \citenamefont {Taddia},
  \citenamefont {Ercolessi},\ and\ \citenamefont {Rigol}}]{dalmonte2015gap}%
  \BibitemOpen
  \bibfield  {author} {\bibinfo {author} {\bibfnamefont {M.}~\bibnamefont
  {Dalmonte}}, \bibinfo {author} {\bibfnamefont {J.}~\bibnamefont
  {Carrasquilla}}, \bibinfo {author} {\bibfnamefont {L.}~\bibnamefont
  {Taddia}}, \bibinfo {author} {\bibfnamefont {E.}~\bibnamefont {Ercolessi}},\
  and\ \bibinfo {author} {\bibfnamefont {M.}~\bibnamefont {Rigol}},\ }\bibfield
   {title} {\bibinfo {title} {Gap scaling at berezinskii-kosterlitz-thouless
  quantum critical points in one-dimensional hubbard and heisenberg models},\
  }\bibfield  {journal} {\bibinfo  {journal} {Physical Review B}\ }\textbf
  {\bibinfo {volume} {91}},\ \href {https://doi.org/10.1103/physrevb.91.165136}
  {10.1103/physrevb.91.165136} (\bibinfo {year} {2015})\BibitemShut {NoStop}%
\bibitem [{\citenamefont {Woynarovich}\ and\ \citenamefont
  {Eckle}(1987)}]{woynarovich1987finite-size}%
  \BibitemOpen
  \bibfield  {author} {\bibinfo {author} {\bibfnamefont {F.}~\bibnamefont
  {Woynarovich}}\ and\ \bibinfo {author} {\bibfnamefont {H.~P.}\ \bibnamefont
  {Eckle}},\ }\bibfield  {title} {\bibinfo {title} {Finite-size corrections for
  the low lying states of a half-filled hubbard chain},\ }\href
  {https://doi.org/10.1088/0305-4470/20/7/005} {\bibfield  {journal} {\bibinfo
  {journal} {Journal of Physics A: Mathematical and General}\ }\textbf
  {\bibinfo {volume} {20}},\ \bibinfo {pages} {L443–L449} (\bibinfo {year}
  {1987})}\BibitemShut {NoStop}%
\end{thebibliography}
\end{document}